\documentclass[12pt]{article}
\usepackage{graphics}
\usepackage{graphicx}
\usepackage{amsmath,amssymb,amsthm,amsfonts}
\usepackage{algorithm,algorithmic}
\usepackage{url}
\usepackage{bbold}
\usepackage[utf8]{inputenc}
\usepackage[english]{babel}
\usepackage{float}
\usepackage{listings}
\usepackage[letterpaper, margin=1in]{geometry}
\usepackage{mathtools}
\usepackage{multirow}
\usepackage{authblk}
\usepackage[super,sort&compress]{natbib}
\setcitestyle{super,citesep={,}} 
\usepackage{xcolor}
\usepackage{booktabs}
\usepackage{threeparttable}
\usepackage{caption}
\usepackage{subcaption}
\usepackage{rotating}
\usepackage{bm}
\usepackage{hyperref}
\usepackage{indentfirst}
\usepackage{setspace}

\title{\Large Causal Inference with Multiple Misclassified Exposures: A Control Variate-Adjusted Calibration Weighting Approach}
\author[1]{Nandini Murali}
\author[2]{Keith Barnatchez}
\author[3]{Jordana E. Hoppe}
\author[1]{Brandie D. Wagner}
\author[4]{Kayleigh P. Keller}
\author[1]{Kevin P. Josey\thanks{Corresponding author: Kevin P. Josey, Department of Biostatistics \& Informatics, Colorado School of Public Health, 13001 East 17th Place, Aurora, CO 80045, USA. Email: kevin.josey@cuanschutz.edu}}
\affil[1]{Department of Biostatistics \& Informatics, Colorado School of Public Health, Aurora, CO}
\affil[2]{Department of Biostatistics, Johns Hopkins Bloomberg School of Public Health, Baltimore, MD}
\affil[3]{Department of Pediatrics, University of Colorado School of Medicine, Aurora, CO}
\affil[4]{Department of Statistics, Colorado State University, Fort Collins, CO}

\newtheorem{assumption}{Assumption}

\newcommand{\bX}{\mathbf{X}}

\newcommand{\bA}{\mathbf{A}}
\newcommand{\bO}{\mathbf{O}}
\newcommand{\ba}{\mathbf{a}}
\newcommand{\bc}{\mathbf{c}}

\DeclareMathOperator*{\argmin}{arg\,min}
\newcommand{\E}{\mathbb{E}}
\newcommand{\var}{\text{Var}}
\newcommand{\cov}{\text{Cov}}
\newcommand{\ind}{\mathbf{1}}

\newcommand{\gs}{\mathrm{gold}}
\newcommand{\val}{\mathrm{val}}

\newcommand{\beginsupplement}{%
        \setcounter{table}{0}
        \renewcommand{\thetable}{S\arabic{table}}%
        \setcounter{figure}{0}
        \renewcommand{\thefigure}{S\arabic{figure}}%
        \setcounter{section}{0}
        \renewcommand\thesection{{\bf S.}\arabic{section}}
     }

\begin{document}

\singlespacing
\maketitle

\begin{abstract}
\noindent Exposure misclassification is a common issue prevalent in studies of respiratory infections in cystic fibrosis. Throat swabs are frequently used in place of expectorated or induced sputum cultures, although they have imperfect sensitivity and specificity to detect \textit{Pseudomonas aeruginosa} and \textit{Staphylococcus aureus}. We develop calibration weighting and control variate estimators for causal inference with multiple misclassified binary exposures and clustered observations. The calibration approach treats misclassification as a missing data problem, achieving consistency without modeling the misclassification mechanism. The control variate adjustment integrates information from error-prone observations to reduce variance while preserving the consistency of the gold-standard estimator. We show that the resulting estimator inherits double robustness from its component estimators. We also characterize a structural ceiling on efficiency gains in the bivariate setting, where joint correct classification of both exposures limits the variance reduction achievable relative to univariate applications. Simulation studies confirm the consistency and double robustness of the proposed estimators under model misspecification. We then apply these methods to a cohort of $651$ cystic fibrosis patients ages $6$--$21$. Swab-based estimates attenuate the effect of \textit{P.~aeruginosa} on percent predicted FEV$_1$ by approximately $79\%$ relative to sputum-based estimates ($-1.68$ vs.\ $-7.88$ percentage points; $95\%$ confidence interval for sputum: $-10.08$, $-5.67$). These findings suggest that relying on throat swabs may lead to under-treatment of \textit{P.~aeruginosa} infections. More broadly, the methods provide a framework for causal inference with multiple misclassified exposures.
\end{abstract}

\vspace{1em}

\doublespacing
\section{Introduction}

Cystic fibrosis (CF) is a genetic disorder that results in abnormal mucus production, thereby predisposing patients to chronic bacterial colonization in the respiratory tract.\cite{yu2024} Two opportunistic pathogens, \textit{Pseudomonas aeruginosa} and \textit{Staphylococcus aureus} (the latter of which includes methicillin-sensitive and -resistant strains), contribute to progressive lung damage and impaired pulmonary function. Clinical management relies on periodic microbiological surveillance to guide antibiotic selection. However, the magnitude of the independent and joint effects of these pathogens on respiratory outcomes remains uncertain.\cite{hauser2011}

Sputum samples represent a gold standard for detecting airway pathogens, but obtaining sputum poses practical challenges, particularly for patients who are younger, receiving mucus-thinning therapies,\cite{ward2023} or undergoing highly effective modulator therapies.\cite{nichols2023, fajac2024} Alternatively, throat swabs can be used to detect bacterial colonization, but validation studies suggest they are prone to substantial misclassification,\cite{hoppe2015} with negative predictive values of $50$--$60\%$ for \textit{P.~aeruginosa} and positive predictive values of $41$--$57\%$ for \textit{S.~aureus}.\cite{eyns2018} Molecular methods such as quantitative polymerase chain reaction detect both pathogens at higher rates than culture, with the largest discrepancies for \textit{P.~aeruginosa} in pediatric patients.\cite{gavillet2022} However, these gains in analytic sensitivity do not resolve the underlying sample-type problem. Throat swabs sample the upper airway and remain less sensitive than sputum for lower-airway pathogens, regardless of the assay applied.\cite{hoppe2015} In estimating causal effects, such measurement error can bias estimates. In particular, nondifferential misclassification commonly produces attenuation towards the null.\cite{carroll2006, josey2023}

Much of the existing literature at the intersection of measurement error and causal inference has relied on parametric models for either the misclassification mechanism, the treatment assignment, or both.\cite{carroll2006, braun2017} Work on misclassified binary exposures has primarily focused on adjustments to the propensity scores with extensions to continuous\cite{josey2023} and categorical\cite{wu2019} settings. Although a multivariate generalized propensity score has been developed for joint, continuous exposures without measurement error,\cite{williams2020} extensions to mismeasured multivariate exposures remain unaddressed. Multiple imputation offers a potential solution,\cite{webbvargas2017} but the method relies upon correctly specifying the imputation model. More recently, methods that present measurement error as a missing data problem have minimized assumptions in estimation by using tools from semiparametric efficiency theory.\cite{kennedy2020} Relatedly, calibration weighting estimators correspond to semiparametric models for incomplete data,\cite{lumley2011} which enables their application to measurement error problems without explicit modeling of the misclassification mechanism.\cite{edwards2015} The calibration approach is attractive because neither the sensitivity nor the specificity enters the estimator directly. The constructed weights balance covariate distributions across subsamples and consistency follows from the resulting balance rather than a correctly specified error model. This calibration step has been combined with augmented inverse probability weighting (AIPW) to obtain a doubly robust estimator for transporting trial effects to an observational target population.\cite{lee2023} We adopt the same calibrated AIPW construction to transport effects across measurement subsamples.

Building on this calibration framework, we extend the control variates method\cite{yang2020, barnatchez2025flexible} to the scenario of multiple misclassified binary exposures with clustered observations. The control variates framework, which was originally developed for unmeasured confounding and later adapted to univariate measurement error, augments a consistent but inefficient gold-standard estimator with a mean-zero term constructed from error-prone data to minimize variance. The degree of variance reduction depends on the correlation between the gold-standard and error-prone estimators, which is identifiable through a validation sample where both measurement types are observed.

Our data structure motivates the approach. We partition observations into three mutually exclusive subsamples: a gold-standard sample with only sputum cultures; an error-prone sample with only throat swabs; and a small validation sample with paired sputum and swab measurements collected during the same visit. These splits reflect the reality of many CF studies in which the majority of patients contribute observations based on only one measurement type. By applying calibration-weighted AIPW estimators to the gold-standard and error-prone samples separately, we can isolate and quantify the bias attributable to measurement error. The validation sample then provides the empirical covariance needed to construct the control variate adjustment.

The contributions of this paper are fourfold. First, we quantify the bias in swab-based causal effect estimates for \textit{P.~aeruginosa} and \textit{S.~aureus} by comparing them against sputum-based estimates, thus demonstrating the clinical consequences of exposure misclassification. Second, we outline calibration weighting estimators that treat misclassification as missing data and apply them separately to gold-standard and error-prone samples without modeling the misclassification mechanism. Third, we develop a control variate adjustment for bivariate binary exposures that integrates information from both estimators via the validation sample, and in doing so characterize the structural ceiling on efficiency gains in the bivariate setting. Fourth, we evaluate these methods through simulation under model misspecification and apply them to data from a cohort of CF patients.

The remainder of this paper is organized as follows. Section~\ref{sec:prelim} defines the data structure, causal estimands, and identification assumptions. Section~\ref{sec:methods} introduces the calibrated AIPW estimator and the control variate adjustment. Section~\ref{sec:simulation} presents a simulation study that evaluates the proposed estimators under bivariate measurement error and model misspecification. Section~\ref{sec:illustrate} applies the methods to the CF cohort. Section~\ref{sec:discussion} concludes with a discussion of findings, limitations, and directions for future work.

\section{Preliminaries}
\label{sec:prelim}

\subsection{Notation and Data Structure}

We examine the causal effects of two binary exposures, the presence of \textit{P.~aeruginosa} and \textit{S.~aureus} in respiratory samples, on lung function in cystic fibrosis patients. For individual $i = 1,\ldots,n$ and observation $j = 1,\ldots,m_i$, with $N = \sum_{i = 1}^n m_i$ total observations, we denote the bivariate true exposure vector as $\bA_{ij} = (A_{ij0}, A_{ij1})$, where $A_{ij0} = 1$ indicates the presence of \textit{P.~aeruginosa} and $A_{ij1} = 1$ indicates the presence of \textit{S.~aureus}. The outcome $Y_{ij}$ represents percent predicted FEV$_1$ (forced expiratory volume in one second) and $\bX_{ij}$ is a $(p \times 1)$ vector of baseline covariates that includes an intercept term.

Let $Z_{ij} = 1$ indicate that observation $(i,j)$ has a gold-standard sputum culture where the true exposure $\bA_{ij}$ is observed directly. $Z_{ij} = 0$ denotes a throat swab measurement in which we observe the error-prone exposure $\bA^*_{ij} = (A^*_{ij0}, A^*_{ij1})$ instead of $\bA_{ij}$. We also define $S_{ij} = 1$ to indicate that the observation is from the validation set in which both sputum and swab measurements were collected simultaneously from the same individual at the same visit. Then, $S_{ij} = 1$ implies $Z_{ij} = 1$.

The data are partitioned into three mutually exclusive subsamples based on available measurements: the gold-standard-only subsample ($Z_{ij} = 1$, $S_{ij} = 0$), where only sputum is available; the error-prone-only subsample ($Z_{ij} = 0$, $S_{ij} = 0$), where only the throat swab is available; and the validation subsample ($S_{ij} = 1$), where both measurement types are observed. In the validation sample, we observe both $\bA_{ij}$ and $\bA^*_{ij}$ for the same observation, which allows direct assessment of misclassification rates and, as we show in Section~\ref{sec:cv}, the identification of the control variate adjustment. We write $\bO_{ij} = (Y_{ij}, \bA_{ij}, \bA^*_{ij}, \bX_{ij}, Z_{ij}, S_{ij})$ for the complete observation, with the convention that $\bA_{ij}$ is recorded only when $Z_{ij} = 1$ and $\bA^*_{ij}$ only for observations based on a throat swab.

The misclassification mechanism is characterized by pathogen-specific sensitivity $\Pr(A^*_{ijk} = 1 \mid A_{ijk} = 1)$ and specificity $\Pr(A^*_{ijk} = 0 \mid A_{ijk} = 0)$ for $k = 0$ (\textit{P.~aeruginosa}) and $k = 1$ (\textit{S.~aureus}). However, we note that neither the sensitivity nor specificity directly enters our estimators. The calibration weighting approach described in Section~\ref{sec:methods} treats misclassification as a missing data problem, and therefore achieves consistency without modeling the misclassification mechanism.

\subsection{Causal Estimands}

We adopt the potential outcomes framework\cite{rubin1974} to construct causal estimands. We define $Y_{ij}^{\ba}$ as the counterfactual lung function that would be observed under the exposure combination $\ba = (a_0, a_1) \in \{0,1\}^2$. For example, $Y_{ij}^{(1,0)}$ represents the lung function that would arise under \textit{P.~aeruginosa} infection alone. Potential outcomes are defined in terms of the true exposure $\bA_{ij}$, not the error-prone measurement $\bA^*_{ij}$. Throughout, $\bA_{ij}$ denotes a point-in-time exposure measured at visit $j$, and $Y_{ij}^{\ba}$  indicates the contemporaneous counterfactual outcome under a single-visit assignment to $\ba$. Identification relies on cross-sectional exchangeability conditions rather than the sequential exchangeability required in longitudinal settings since we do not consider sustained exposure regimes or treatment trajectories.

For each exposure combination $\ba \in \{(1,0), (0,1), (1,1)\}$, we define the average treatment effect as $\tau^{\ba} \coloneqq \E\!\left[Y_{ij}^{\ba} - Y_{ij}^{\mathbf{0}}\right]$, where $\mathbf{0} = (0,0)$ indicates that neither pathogen is present. Exposure may vary across a patient's visits, and the expectation averages over the person-visit population, so with unit base weights each visit contributes equally. These parameters quantify the average change in percent predicted FEV$_1$ that is attributable to \textit{P.~aeruginosa} alone ($\ba = (1,0)$), \textit{S.~aureus} alone ($\ba = (0,1)$), or co-infection ($\ba = (1,1)$). The interaction effect, which captures synergism or antagonism between the two pathogens, parallels the factorial interaction contrast studied in experimental causal inference literature\cite{egami2019} and is given by $\Delta \coloneqq \tau^{(1,1)} - \tau^{(1,0)} - \tau^{(0,1)}$.

\subsection{Identification Assumptions}

Under Assumptions~\ref{as:sutva}--\ref{as:exch}, the causal parameters $\tau^{\ba}$ and $\Delta$ are identifiable from the gold-standard data. Assumption~\ref{as:nondiff-mis} allows the use of error-prone data to improve efficiency without introducing bias.

\begin{assumption}[Stable Unit Treatment Value]
\label{as:sutva}
No interference between units and consistency: $Y_{ij} = Y_{ij}^{\bA_{ij}}$.
\end{assumption}

Assumption \ref{as:sutva}  requires that a patient's infection status does not affect another patient's lung function, and that the observed outcome equals the potential outcome under the true exposure. The unit is the person-visit, so no interference asks that the exposure at one visit not affect the outcome at another; this is plausible whether visits are grouped within patients or within other clusters, such as clinical sites, provided distinct units do not influence one another. Using $\bA^*_{ij}$ in place of $\bA_{ij}$ would violate consistency, since $Y_{ij}^{\bA_{ij}} \neq Y_{ij}^{\bA^*_{ij}}$ when misclassification occurs.

\begin{assumption}[Ignorable Exposure Assignment]
\label{as:ignorability}
(i) \textit{Unconfoundedness:} $Y_{ij}^{\ba} \perp \bA_{ij} \mid \bX_{ij}, Z_{ij} = 1$ for all $\ba \in \{0,1\}^2$. (ii) \textit{Positivity:} $\Pr\{\bA_{ij} = \ba \mid \bX_{ij}, Z_{ij} = 1\} > 0$ for all $\ba \in \{0, 1\}^2$ and $\bX_{ij}$ in the support.
\end{assumption}

Unconfoundedness requires that all common causes of exposures and the outcome are captured in $\bX_{ij}$. Positivity ensures a nonzero probability of each exposure combination at every covariate pattern. Both conditions apply to  gold-standard measurements ($Z_{ij} = 1$) since the true exposure $\bA_{ij}$ is only observed in sputum samples.

The following assumptions concern the relationship between gold-standard and error-prone observations and are necessary to incorporate error-prone data into the analysis.

\begin{assumption}[Exchangeable Measurement Selection]
\label{as:exch}
(i) \textit{Effect transportability:}
\[
\E\!\left[Y_{ij}^{\ba} - Y_{ij}^{\mathbf{0}} \mid Z_{ij} = 1, \bX_{ij}\right] = \E\!\left[Y_{ij}^{\ba} - Y_{ij}^{\mathbf{0}} \mid Z_{ij} = 0, \bX_{ij}\right] \quad \text{for all } \ba \in \{0,1\}^2.
\]
(ii) \textit{Measurement positivity:} $0 < \Pr\{Z_{ij} = 1 \mid \bX_{ij}\} < 1$ for all $\bX_{ij}$ in the support.
\end{assumption}

Effect transportability ensures that the causal effects estimated in the sputum sample generalize to patients measured with swabs when conditioned on the covariates. This assumption is plausible if all effect modifiers associated with measurement type are included in $\bX_{ij}$. Measurement positivity ensures that for every covariate pattern, both sputum and swab measurements are possible.

\begin{assumption}[Exchangeable Error-Prone Sampling]
\label{as:nondiff-mis}
(i) \textit{Mean exchangeability:} the conditional mean outcome given the error-prone exposure and covariates agrees across the error-prone-only ($Z_{ij} = 0$) and validation ($S_{ij} = 1$) subsamples,
\[
\E\!\left[Y_{ij} \mid \bA^*_{ij} = \ba, \bX_{ij}, Z_{ij} = 0\right] = \E\!\left[Y_{ij} \mid \bA^*_{ij} = \ba, \bX_{ij}, S_{ij} = 1\right] \quad \text{for all } \ba \in \{0,1\}^2.
\]
(ii) \textit{Error-prone positivity:} $0 < \Pr\{S_{ij} = 1 \mid \bX_{ij}\} < 1$, $0 < \Pr\{\bA^*_{ij} = \ba \mid \bX_{ij}, Z_{ij} = 0\}$, and $0 < \Pr\{\bA^*_{ij} = \ba \mid \bX_{ij}, S_{ij} = 1\}$ for all $\ba \in \{0,1\}^2$ and $\bX_{ij}$ in the support.
\end{assumption}

Part (i) states that, among the observations from the throat swab, the conditional mean outcome for a given error-prone exposure and covariate value is the same regardless of whether a paired sputum culture was also obtained. This holds when the gold-standard measurement is selected at random, given that the validation subsample is a conditionally random subset of the gold-standard sample. The simulation design of Section~\ref{sec:simulation} and, approximately, the application have this structure. Unlike classical nondifferential misclassification, which limits the error mechanism $\Pr\{\bA^*_{ij} \mid \bA_{ij}\}$, Part (i) constrains only the comparability of the subsamples that carry $\bA^*_{ij}$. The error mechanism itself may be arbitrary and even differential, provided that it is common to the error-prone-only and validation subsamples. Part (ii) is the positivity counterpart, which ensures the calibration weights of Section~\ref{sec:methods} are well defined within each error-prone exposure cell in both subsamples.

Assumption~\ref{as:nondiff-mis} has a direct consequence in constructing the control variate in Section~\ref{sec:cv}. Because the error-prone exposure and outcome share a common conditional law across the error-prone-only and validation subsamples, any covariate-standardized contrast built from $\bA^*_{ij}$ has the same population limit whether it is evaluated using the validation sample or the error-prone-only sample. The difference between the two estimators therefore has expectation of zero, and creates the mean-zero property that the control variate adjustment relies upon. This common limit is generally a biased proxy for $\tau^{\ba}$ since it is defined through the error-prone exposure, but the control variate exploits its cancellation, not its value.

\section{Methods}
\label{sec:methods}

\subsection{Calibrated Augmented Inverse Probability Weighting}

Calibration constructs sample weights such that weighted covariate totals reproduce a target population margin.\cite{deville1992} Calibration estimators correspond to semiparametric models for incomplete data,\cite{lumley2011} an observation that extends naturally to measurement error problems where the true exposure is ``missing'' in error-prone observations.\cite{edwards2015} Combining calibration weights with AIPW yields an estimator that is doubly robust against the misspecification of either component.\cite{robins1995, bang2005, funk2011} The calibrated AIPW estimator that we implement here was developed in the context of generalizing trial findings to an observational target population.\cite{lee2023} We use the same construction to transport effects across measurement subsamples and achieve consistency without modeling the misclassification mechanism.

Let $q_{ij} > 0$ denote a baseline weight assigned to each observation. In our analysis, we take $q_{ij} = 1$ for all $(i,j)$. For each exposure combination $\ba \in \{0, 1\}^2$, we construct calibration weights $\hat{\gamma}^{\ba}_1(\mathbf{X})$ on the gold-standard subsample as solutions to
\begin{equation}
\label{eq:calibration}
\sum_{i,j} \ind(Z_{ij} = 1, \bA_{ij} = \ba)\, \hat{\gamma}^{\ba}_1(\mathbf{X}_{ij})\,\mathbf{X}_{ij} \;=\; \sum_{i,j} q_{ij}\,\mathbf{X}_{ij}.
\end{equation}
Because $\mathbf{X}_{ij}$ contains an intercept, the leading row of \eqref{eq:calibration} forces $\sum_{i,j} \ind(Z_{ij}=1,\bA_{ij}=\ba)\,\hat{\gamma}^{\ba}_1 = \sum_{i,j} q_{ij}$, so that the calibrated weights match the total baseline mass on the $(Z=1, \bA=\ba)$ cell. This scaling places the inverse-probability and outcome-regression components of the AIPW pseudo-outcome on a common per-observation scale, without an explicit factor of $N$. We parameterize the weights using exponential tilting, $\hat{\gamma}^{\ba}_1(\mathbf{X}_{ij}) \propto q_{ij}\exp(\mathbf{X}_{ij}^T\hat{\bm{\lambda}}_{\ba})$, with $\hat{\bm{\lambda}}_{\ba}$ obtained as the dual solution to the Lagrangian for \eqref{eq:calibration}. The resulting weights are positive and minimize the Kullback--Leibler divergence to the baseline weights $q_{ij}$, connecting the construction to empirical likelihood\cite{han2021} and to the generalized raking estimator of survey sampling.\cite{deville1993} Exponential tilting is one member of the Bregman-distance family of balancing weights\cite{josey2021a} that has been used to create balance across subsamples in transportability and data-fusion problems.\cite{josey2021b, josey2022} Section~S.1 of the Supplementary Materials sketches a more general formulation that admits arbitrary calibration bases $\bc(\bX_{ij})$ and a wider range of baseline weights $q_{ij}$.

We pair the calibration weights with an outcome regression $\hat{\mu}_1(\ba, \bX_{ij}) = \hat{\mathbb{E}}[Y_{ij} \mid \bA_{ij} = \ba, \bX_{ij}, Z_{ij} = 1]$ estimated by a flexible, data-adaptive regression method. This accommodates nonlinear structure without committing to a fixed functional form. The AIPW pseudo-outcome combines the calibration-weighted residual with the outcome model,
\begin{equation}
\label{eq:aipw-gold}
\hat{\psi}^{\ba}_1(\bO_{ij}) \;=\; \ind(\bA_{ij} = \ba,\, Z_{ij} = 1)\,\hat{\gamma}^{\ba}_1(\bX_{ij})\left[Y_{ij} - \hat{\mu}_1(\ba, \bX_{ij})\right] + \hat{\mu}_1(\ba, \bX_{ij}).
\end{equation}
The gold-standard average treatment effect estimator is
\begin{equation}
\label{eq:ate-gold}
\hat{\tau}^{\ba}_1 \;=\; \frac{1}{N}\sum_{i,j}\left[\hat{\psi}^{\ba}_1(\bO_{ij}) - \hat{\psi}^{\mathbf{0}}_1(\bO_{ij})\right].
\end{equation}
The estimator is consistent if the calibration weights balance the covariates in the expectation, or if the outcome model correctly specifies $\E[Y \mid \bA, \bX]$; in either case the estimator is asymptotically linear with influence function determined by the AIPW pseudo-outcome.\cite{naimi2017}

We construct an analogous estimator on the error-prone subsample. Define calibration weights $\tilde{\gamma}^{\ba}_0(\mathbf{X})$ on observations with $Z_{ij} = 0$ that balance the same target $\sum_{i,j} q_{ij}\,\bX_{ij}$ within each $(Z=0, \bA^*=\ba)$ cell, and an outcome model $\tilde{\mu}_0(\ba, \bX_{ij}) = \hat{\E}[Y_{ij} \mid \bA^*_{ij} = \ba, \bX_{ij}, Z_{ij} = 0]$ fit on the error-prone observations. The error-prone pseudo-outcome takes the same form as the gold-standard one with $\bA^*_{ij}$ in place of $\bA_{ij}$ and the error-prone nuisance functions replacing their gold-standard counterparts,
\[
\tilde{\psi}^{\ba}_0(\bO_{ij}) \;=\; \ind(\bA^*_{ij} = \ba,\, Z_{ij} = 0)\,\tilde{\gamma}^{\ba}_0(\bX_{ij})\left[Y_{ij} - \tilde{\mu}_0(\ba, \bX_{ij})\right] + \tilde{\mu}_0(\ba, \bX_{ij}).
\]
Averaging the pseudo-outcome contrasts against the reference cell $\ba = \mathbf{0}$ yields the error-prone estimator,
\[
\tilde{\tau}^{\ba}_0 = \frac{1}{N}\sum_{i,j}\left[\tilde{\psi}^{\ba}_0(\bO_{ij}) - \tilde{\psi}^{\mathbf{0}}_0(\bO_{ij})\right].
\]
Because $\bA^*_{ij}$ enters in place of $\bA_{ij}$, $\tilde{\tau}^{\ba}_0$ no longer estimates $\tau^{\ba}$; the discrepancy $\hat{\tau}^{\ba}_1 - \tilde{\tau}^{\ba}_0$ summarizes the resulting bias. However, it remains a well-defined statistical quantity. It targets the covariate-standardized error-prone contrast $\theta^{\ba} \coloneqq \E_{\bX}\!\left[m^*(\ba, \bX) - m^*(\mathbf{0}, \bX)\right]$, where $m^*(\ba, \bX) = \E[Y_{ij} \mid \bA^*_{ij} = \ba, \bX_{ij} = \bX, Z_{ij} = 0]$ is the error-prone outcome regression. Assumption~\ref{as:nondiff-mis} makes $m^*$ common to the error-prone-only and validation subsamples. The full-sample error-prone estimator constructed in Section~\ref{sec:cv} targets the same $\theta^{\ba}$, which makes their difference mean-zero.

The interaction estimator for gold-standard data, which captures synergism or antagonism between the two exposures, is
\[
\hat{\Delta}_1 = \frac{1}{N} \sum_{i,j} \left[\hat{\psi}^{(1,1)}_1 - \hat{\psi}^{(1,0)}_1 - \hat{\psi}^{(0,1)}_1 + \hat{\psi}^{\mathbf{0}}_1\right],
\]
with an analogous expression $\tilde{\Delta}_0$ for the error-prone subsample. Inference proceeds from the asymptotic linearity of $\hat{\tau}^{\ba}_1$. Repeated observations on the same individual induce within-cluster correlation in the pseudo-outcomes, which we accommodate through a cluster-robust score-squared variance estimator,
\[
\widehat{\var}\left(\hat{\tau}^{\ba}_1\right) = \frac{n}{(n-1)N^2} \sum_{i=1}^{n} \left\{\sum_{j=1}^{m_i} \left[\hat{\psi}^{\ba}_1(\bO_{ij}) - \hat{\psi}^{\mathbf{0}}_1(\bO_{ij}) - \hat{\tau}^{\ba}_1\right]\right\}^2,
\]
which can be computed using the CR1 small-sample correction within the \texttt{sandwich} package in \texttt{R}.\cite{zeileis2006}

\subsection{Control Variate Adjustment}
\label{sec:cv}

The gold-standard estimator $\hat{\tau}^{\ba}_1$ is consistent but inefficient when gold-standard observations are limited. The control variates approach, originally developed for unmeasured confounding\cite{yang2020} and later applied to measurement error settings,\cite{barnatchez2025flexible} uses information from error-prone observations to reduce variance while preserving consistency. The construction borrows tools from the Monte Carlo literature.\cite{rubinstein1985} With a mean-zero auxiliary quantity whose realization is correlated with the estimator of interest, a suitable linear combination of the two has lower variance.\cite{casella2002} In our setting, the validation subsample, on which both measurement types are observed, provides the link needed to construct such an auxiliary quantity.

We build the auxiliary quantity from two error-prone AIPW estimators that share a population limit but use different observations. The first is the error-prone-only estimator $\tilde{\tau}^{\ba}_0$ defined above, which uses only observations with $Z_{ij}=0$. The second applies the same AIPW construction to the entire error-prone sample $\{(1-Z_{ij})+S_{ij}=1\}$, the error-prone-only and validation subsamples combined. Its calibration weights $\tilde{\gamma}^{\ba}(\mathbf{X})$ solve
\[
\sum_{i,j} \ind\{(1-Z_{ij})+S_{ij}=1,\, \bA^*_{ij}=\ba\}\, \tilde{\gamma}^{\ba}(\mathbf{X}_{ij})\,\mathbf{X}_{ij} \;=\; \sum_{i,j} q_{ij}\,\mathbf{X}_{ij},
\]
its outcome model is $\tilde{\mu}(\ba, \bX_{ij}) = \hat{\E}[Y_{ij} \mid \bA^*_{ij} = \ba, \bX_{ij}, (1-Z_{ij})+S_{ij}=1]$, and its pseudo-outcome and estimator are
\[
\tilde{\psi}^{\ba}(\bO_{ij}) \;=\; \ind\{\bA^*_{ij} = \ba,\, (1-Z_{ij})+S_{ij}=1\}\,\tilde{\gamma}^{\ba}(\bX_{ij})\left[Y_{ij} - \tilde{\mu}(\ba, \bX_{ij})\right] + \tilde{\mu}(\ba, \bX_{ij}),
\]
\[
\tilde{\tau}^{\ba} = \frac{1}{N}\sum_{i,j}\left[\tilde{\psi}^{\ba}(\bO_{ij}) - \tilde{\psi}^{\mathbf{0}}(\bO_{ij})\right].
\]
Both estimators use the misclassified exposure $\bA^*_{ij}$, and Assumption~\ref{as:nondiff-mis} implies that they share the same population limit $\theta^{\ba}$. We define the control variate as their difference, $\hat{\zeta}^{\ba} = \tilde{\tau}^{\ba} - \tilde{\tau}^{\ba}_0$ and $\E[\hat{\zeta}^{\ba}] = 0.$ The two estimators coincide except on the validation observations, which enter $\tilde{\tau}^{\ba}$ but not $\tilde{\tau}^{\ba}_0$, so $\hat{\zeta}^{\ba}$ isolates the error-prone contribution of the validation subsample. Because the gold-standard estimator $\hat{\tau}^{\ba}_1$ also draws on those observations, $\hat{\zeta}^{\ba}$ is correlated with $\hat{\tau}^{\ba}_1$ while adding no asymptotic bias. Both $\tilde{\tau}^{\ba}_0$ and $\tilde{\tau}^{\ba}$ fit their nuisance functions on the large error-prone samples, so the construction never estimates an outcome model or calibration weights on the sparse validation subsample.

The estimator adjusted by control variates combines the gold-standard estimator with the mean-zero auxiliary as $\hat{\xi}^{\ba} = \hat{\tau}^{\ba}_1 + b\,\hat{\zeta}^{\ba}$, where the variance-minimizing coefficient is
\begin{equation}
\label{eq:bopt}
b \;=\; -\frac{\cov\left(\hat{\tau}^{\ba}_1,\; \hat{\zeta}^{\ba}\right)}{\var\left(\hat{\zeta}^{\ba}\right)}.
\end{equation}
With this, $\var\left(\hat{\xi}^{\ba}\right)/\var\left(\hat{\tau}^{\ba}_1\right) = 1 - \rho^2$, where $\rho$ is the asymptotic correlation between $\hat{\tau}^{\ba}_1$ and $\hat{\zeta}^{\ba}$. The reduction in variance is determined by how strongly $\hat{\tau}^{\ba}_1$ and $\hat{\zeta}^{\ba}$ co-vary, which is in turn driven by the rate at which the error-prone measurement agrees with the true exposure.

The covariance term in \eqref{eq:bopt} is identified through the validation subsample. On observations with $S_{ij} = 1$, the gold-standard pseudo-outcome and the full-sample error-prone pseudo-outcome $\tilde{\psi}^{\ba}$ simultaneously carry residual-weighted contributions tied to the same $Y_{ij}$, producing the within-observation correlation that drives the adjustment. Without these observations, $\tilde{\tau}^{\ba}$ and $\tilde{\tau}^{\ba}_0$ coincide, $\hat{\zeta}^{\ba}$ vanishes, and no variance reduction is achieved.

In practice, we obtain $\hat{\xi}^{\ba}$ from a joint moment regression. We fit a multivariate intercept-only model with response $\bigl(\hat{\psi}^{\ba}_1 - \hat{\psi}^{\mathbf{0}}_1,\; (\tilde{\psi}^{\ba} - \tilde{\psi}^{\mathbf{0}}) - (\tilde{\psi}^{\ba}_0 - \tilde{\psi}^{\mathbf{0}}_0)\bigr)$ over all observations and extract the cluster-robust covariance matrix $\hat{\bm{\Sigma}}$ of the two intercepts via the \texttt{sandwich} package in \texttt{R}. By writing $\hat{\Gamma} = \hat{\Sigma}_{12}$ and $\hat{V}_\zeta = \hat{\Sigma}_{22}$, the control variate estimator and its variance are $\hat{\xi}^{\ba} = \hat{\tau}^{\ba}_1 - (\hat{\Gamma}/\hat{V}_\zeta)\, \hat{\zeta}^{\ba}$ and $\widehat{\var}\left(\hat{\xi}^{\ba}\right) = \hat{\Omega} - \hat{\Gamma}^2/\hat{V}_\zeta$, with Wald confidence intervals applied at the nominal level. The numerator $\hat{\Gamma}$ is identified by the validation observations through the mechanism described above, while $\hat{V}_\zeta$ aggregates the variance of $\hat{\zeta}^{\ba}$ over all observations entering either error-prone estimator.

Each of the three constituent AIPW estimators is doubly robust and asymptotically linear, and $\hat{\xi}^{\ba}$ inherits both properties, with asymptotic variance $\Omega - \Gamma^2/V_\zeta$ under standard regularity conditions. Consistency for $\tau^{\ba}$ rests on the gold-standard estimator. The estimator $\hat{\tau}^{\ba}_1$ is consistent whenever its calibration weights or its outcome model is correctly specified, and $\hat{\zeta}^{\ba}$ adds no bias because it targets zero under Assumption~\ref{as:nondiff-mis}, so the error-prone estimators contribute only variance reduction. This inherited double robustness mirrors the result established for the univariate setting.\cite{barnatchez2025flexible}

It is worth noting that there are other mean-zero contrasts among the error-prone estimators fitting the same template. A natural alternative replaces the full-error-prone estimator $\tilde{\tau}^{\ba}$ with $\tilde{\tau}^{\ba}_{\val}$, an error-prone AIPW estimator restricted to the validation observations $\{S_{ij} = 1\}$. This estimator also targets $\theta^{\ba}$ under Assumption~\ref{as:nondiff-mis}, so the contrast $\tilde{\tau}^{\ba}_{\val} - \tilde{\tau}^{\ba}_0$ remains mean-zero. This validation-based construction was used to build the univariate control variate in previous work.\cite{barnatchez2025flexible} Concentrating the validation signal in a single estimator can raise the correlation between the resulting control variate and $\hat{\tau}^{\ba}_1$, producing larger efficiency gains than the full-error-prone contrast we primarily utilize. The tradeoff is that $\tilde{\tau}^{\ba}_{\val}$ fits its calibration weights and outcome regression on the validation observations alone, which requires a validation sample that is sufficiently large enough to support flexible, data-adaptive nuisance estimation. When the validation sample is small, as in our application, the model fits become unreliable. We therefore adopt $\hat{\zeta}^{\ba} = \tilde{\tau}^{\ba} - \tilde{\tau}^{\ba}_0$, which estimates its nuisances on the two full-size error-prone samples and never fits on the sparse validation subsample.

\section{Simulation Study}
\label{sec:simulation}

We conduct a simulation study to evaluate the proposed estimators under bivariate exposures, clustered observations, and model misspecification. We compare bias, root mean squared error (RMSE), coverage probability, and relative efficiency across three modeling scenarios: (1) correctly specified models for both exposures and outcome; (2) a misspecified propensity score model with a correctly specified outcome model; and (3) correctly specified propensity score models with a misspecified outcome model.

\subsection{Simulation Design}

We generate $N = 2{,}000$ observations distributed across $n = 500$ individuals, with each observation assigned uniformly at random to an individual. Thus, the resulting average cluster size is four. The individual-specific random intercepts induce within-person correlation in exposures and outcomes. We vary the proportion of units with observed gold-standard measurements, $\eta \in \{0.2, 0.3, 0.4, 0.5, 0.6, 0.7, 0.8\}$, with gold-standard status assigned deterministically to the first $\lceil N \times \eta \rceil$ observations. Among gold-standard observations, half are randomly assigned to the validation sample ($S_{ij} = 1$), where both sputum and swab measurements are observed. The remaining gold-standard observations have only sputum, and all non-gold-standard observations have only swab measurements. For example, at $\eta=0.2$ the validation, gold-standard-only, and error-prone-only subsamples constitute approximately $10\%$, $10\%$, and $80\%$ of the data, respectively.

Two baseline covariates capture patient heterogeneity. The first covariate is generated as $X_{ij0} \mid Z_{ij}=1 \sim N(0,2^2)$ and $X_{ij0} \mid Z_{ij}=0 \sim N(-1,1)$, reflecting differences between patients who can and cannot produce sputum. The second covariate is generated as $X_{ij1} \mid Z_{ij}=1 \sim N(1,1)$ and $X_{ij1} \mid Z_{ij}=0 \sim N(0,2^2)$. To evaluate robustness, we generate transformed covariates $U_{ij0} = \exp\{-X_{ij0}/4\}$ and $U_{ij1} = |X_{ij0} - X_{ij1}|$ (each standardized to mean zero and unit variance) that violate linearity assumptions when substituted for $X_{ij0}$ and $X_{ij1}$.

Binary exposures follow a sequential generation process that does not depend on measurement type given covariates, so that $\bA_{ij} \perp Z_{ij} \mid \bX_{ij}$ aligns with Assumption~\ref{as:nondiff-mis}. The first exposure has $\Pr(A_{ij0} = 1 \mid \bX_{ij}) = \text{expit}(\alpha_{i0} + 0.5X_{ij0} - 0.5X_{ij1})$ with $\alpha_{i0} \sim N(0,1)$. The second exposure depends on the first through $\Pr(A_{ij1} = 1 \mid A_{ij0}, \bX_{ij}) = \text{expit}(\alpha_{i1} - 0.75X_{ij0} - 0.25X_{ij1} + 0.75A_{ij0}X_{ij1})$ with $\alpha_{i1} \sim N(0,1)$. Generating the second exposure conditional on the first induces dependence within the joint exposure $\bA_{ij}$ but leaves identification intact, since unconfoundedness and positivity concern the joint exposure given $\bX_{ij}$, not the order in which its components are drawn. Both exposures are measured at the same visit, so this is a bivariate contemporaneous exposure rather than a sequence of exposures, and it requires no sequential g-methods. The covariate distributions continue to differ by measurement type, so the gold-standard and error-prone subsamples still require calibration. However, relative to a design with measurement-type-dependent exposures the selection into the gold standard is uninformative about the exposures given $\bX_{ij}$. The continuous outcome follows
\[ Y_{ij} = \beta_i - 0.25A_{ij0} + 0.75A_{ij1} - 0.75X_{ij0} + 0.25X_{ij1} + 0.5A_{ij0}X_{ij0} - 0.5A_{ij0}A_{ij1} + \epsilon_{ij},\]
where $\beta_i \sim N(3, 1)$ and $\epsilon_{ij} \sim N(0,2^2)$. Because $X_{ij0}$ has a covariate-mixture distribution that depends on the gold-standard proportion, the marginal causal contrasts vary slightly with $\eta$: the average treatment effects are $\tau^{(1,0)} = -0.25 + 0.5\,\E[X_{ij0}]$; $\tau^{(0,1)} = 0.75$; $\tau^{(1,1)} = 0.5\,\E[X_{ij0}]$; and the interaction is $\Delta = -0.5$. Across the proportions we examine, $\E[X_{ij0}]$ ranges from $-0.8$ to $-0.2$, yielding $\tau^{(1,0)}$ between $-0.65$ and $-0.35$ and $\tau^{(1,1)}$ between $-0.4$ and $-0.1$. Bias and RMSE are reported relative to the realized true contrast in each replication.

We examine five misclassification configurations, specified as $(\text{sens}_0, \text{spec}_0, \text{sens}_1, \text{spec}_1)$: $(0.7, 0.7, 0.9, 0.9)$ for a case where measurement error is concentrated in the first exposure; $(0.7, 0.7, 0.7, 0.7)$ for measurement error in both exposures; $(0.9, 0.7, 0.9, 0.7)$ for uniformly low specificity; $(0.7, 0.9, 0.7, 0.9)$ for uniformly low sensitivity; and $(0.7, 0.9, 0.9, 0.7)$ for mixed error patterns across exposures. Misclassification occurs independently across exposures conditional on true exposure status.

We evaluate robustness under three scenarios, with the estimators fitting their nuisance models in the true covariates $\mathbf{X}$ throughout: the first generates the propensity and outcome surfaces from $\mathbf{X}$ as well; the second generates the propensity from the transformed covariates $\mathbf{U}$; and the third generates the outcome surface from $\mathbf{U}$, so that the fitted propensity and outcome models are misspecified in turn.

Five estimators are compared across $5{,}000$ replications: an oracle that uses true exposures for all observations; a naive estimator that treats all observed exposures, either gold-standard when available or error-prone, as correct; an error-prone estimator that uses only swab data; a gold-standard calibration estimator; and a control variate estimator. For validation observations, where the true and error-prone observations exist for the same point in time, the oracle and naive estimators both use the gold-standard exposure $\bA_{ij}$. The two estimators differ only on the error-prone-only subsample, where the oracle has access to the (counterfactual) true exposure while the naive uses $\bA^*_{ij}$ instead. Summary statistics (bias, RMSE, variance) are computed on estimates winsorized at the $1$st and $99$th percentiles, which limits the influence of occasional extreme calibration weights. Implementation uses \texttt{R}, with all outcome models fit by a SuperLearner ensemble\cite{vanderLaan2007} of a sample mean, $\ell_1$-penalized regression, and degree-two multivariate adaptive regression splines\cite{friedman1991}, calibration weights obtained through exponential tilting, and cluster-robust standard errors computed with the \texttt{sandwich} package. Additional implementation details, including the form of the calibration optimization, appear in Section~S.1 of the Supplementary Materials.

\subsection{Simulation Results}

\begin{figure}[ht]
    \centering
    \includegraphics[width=1\linewidth]{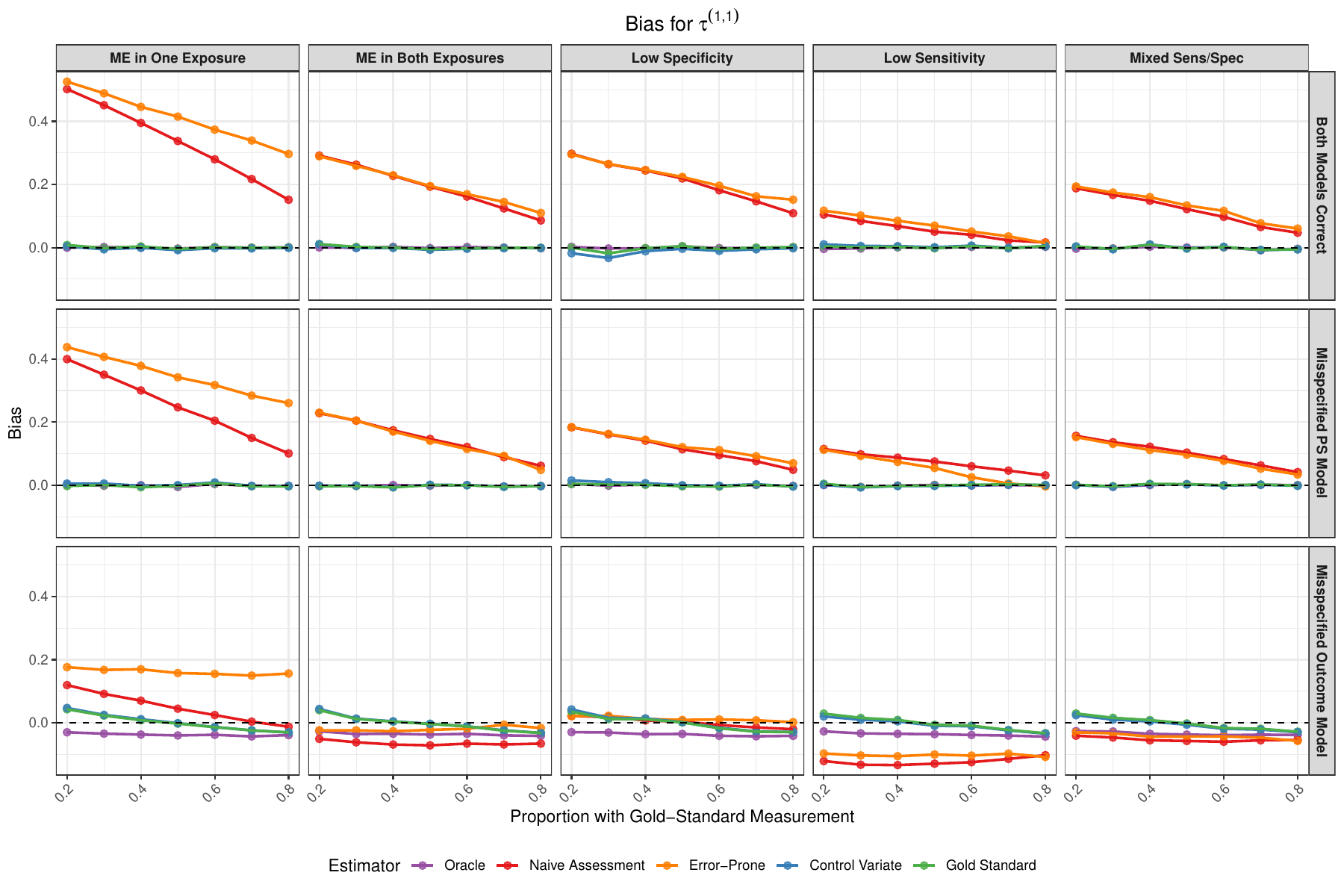}
    \caption{Bias for the joint exposure effect $\tau^{(1,1)}$ across measurement error configurations and model specifications.}
    \label{fig:bias}
\end{figure}

\begin{figure}[ht]
    \centering
    \includegraphics[width=1\linewidth]{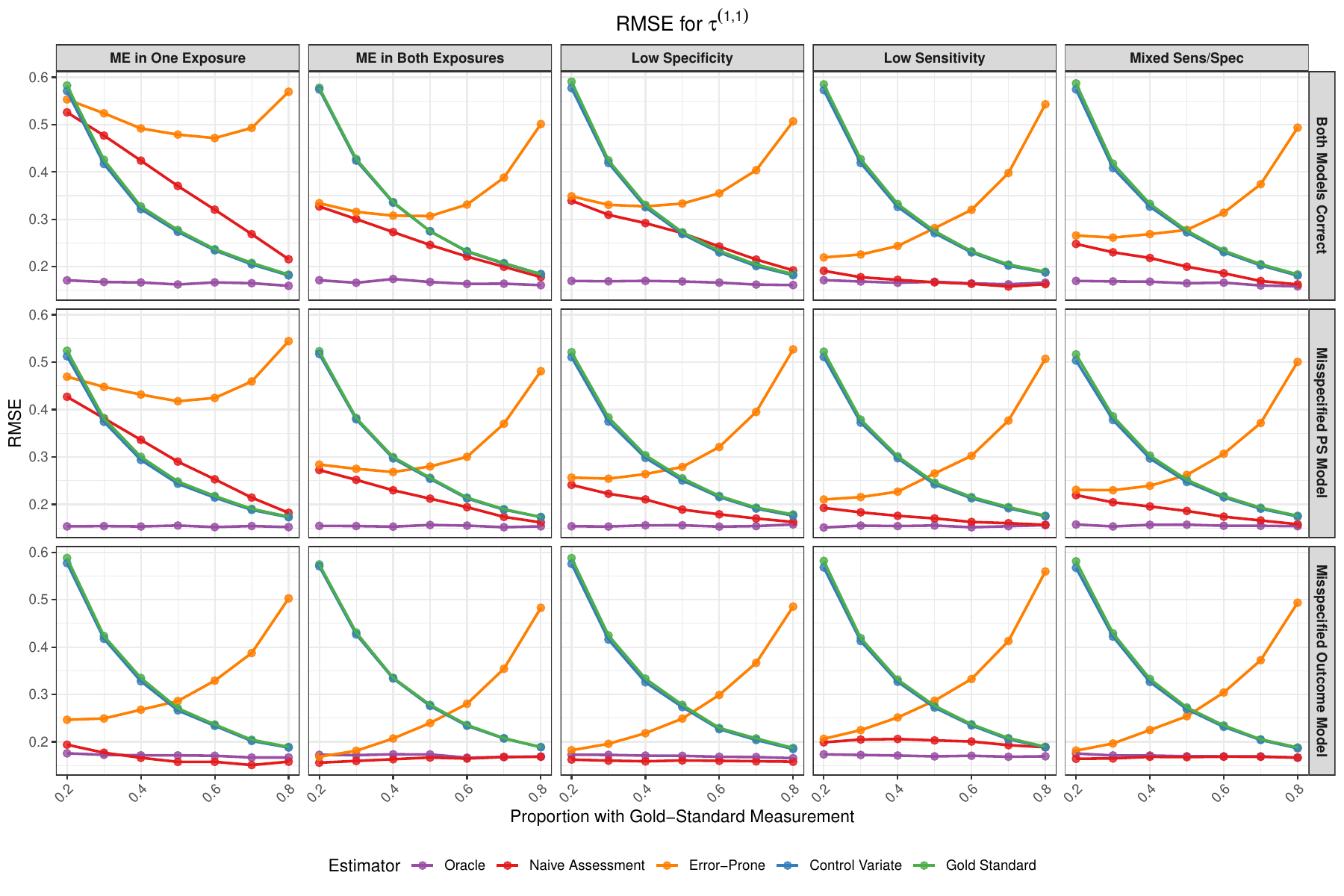}
    \caption{Root mean squared error (RMSE) for the joint exposure effect $\tau^{(1,1)}$ across measurement error configurations and model specifications.}
    \label{fig:rmse}
\end{figure}

\begin{figure}[ht]
    \centering
    \includegraphics[width=1\linewidth]{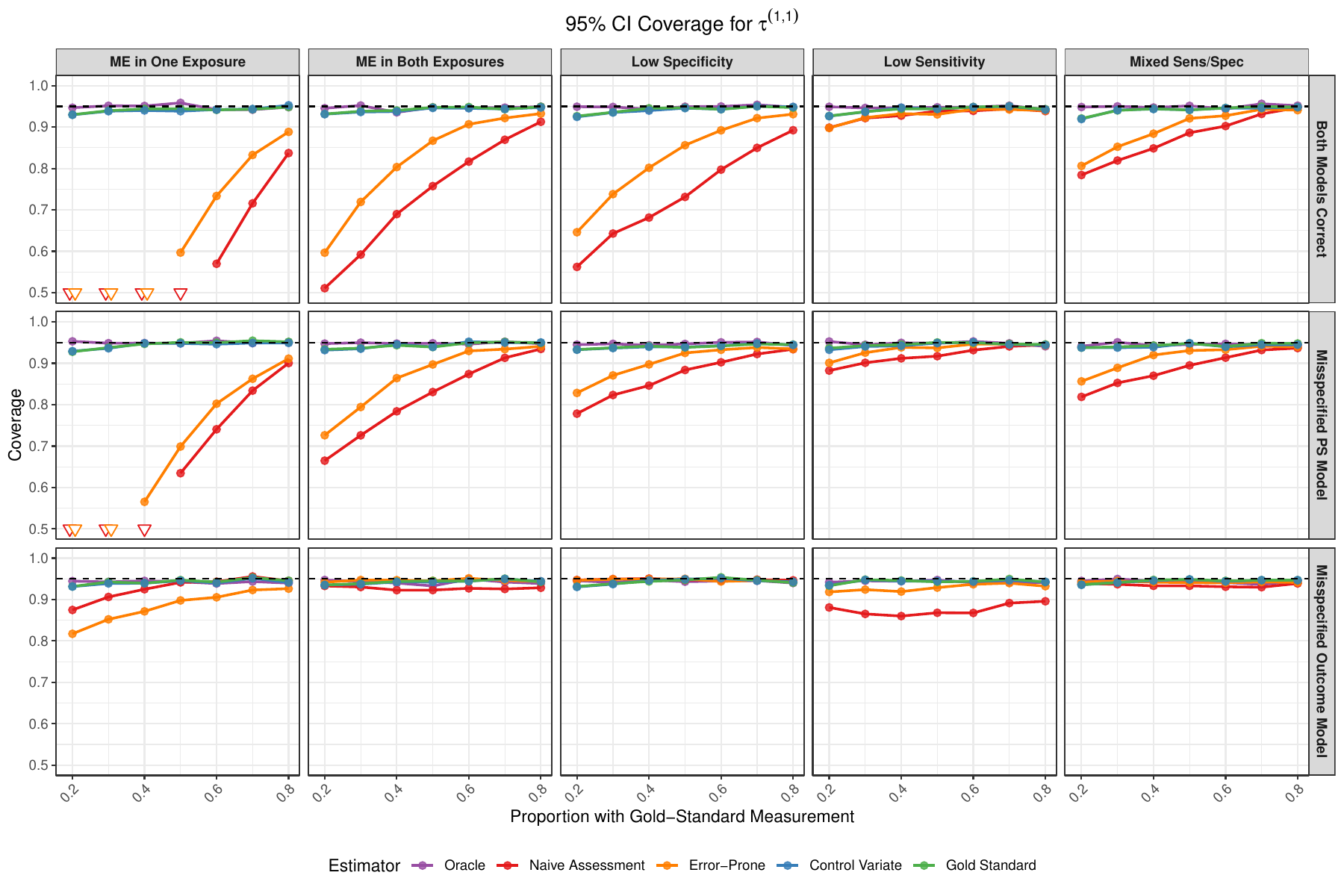}
    \caption{Empirical coverage probability of nominal 95\% confidence intervals. The dashed line indicates the target 95\% level.}
    \label{fig:coverage}
\end{figure}

\begin{figure}[ht]
    \centering
    \includegraphics[width=1\linewidth]{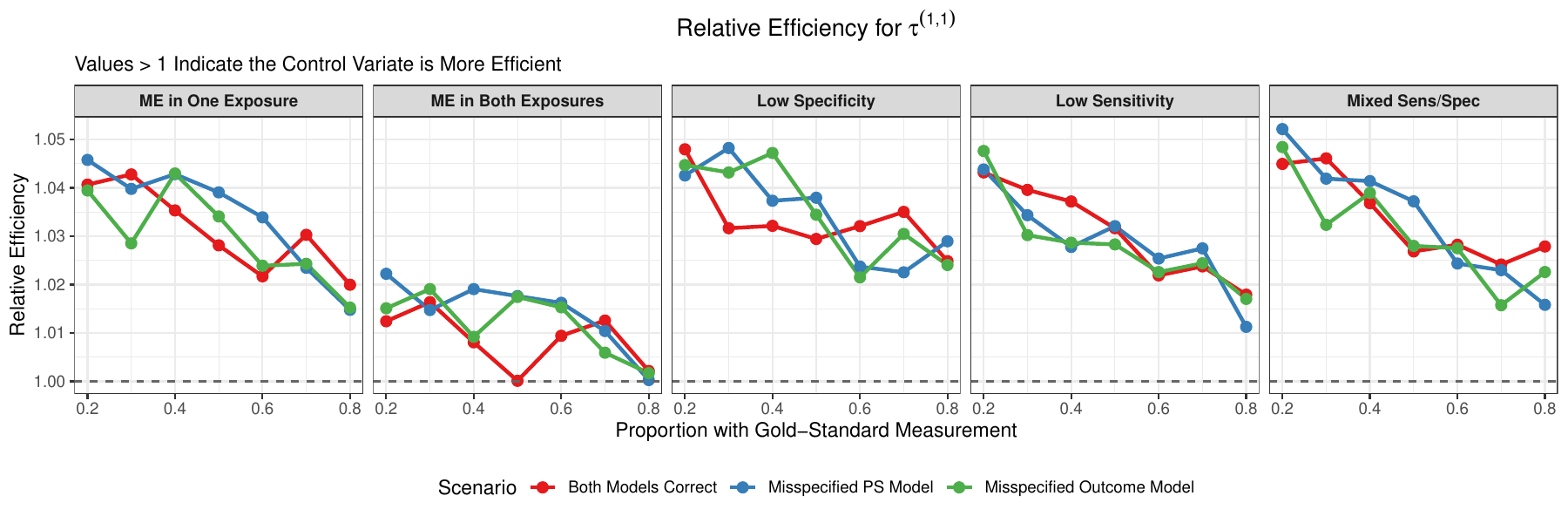}
    \caption{Relative efficiency of the control variate-adjusted joint estimator $\hat{\xi}^{(1,1)}$ compared to the gold-standard estimator $\hat{\tau}_1^{(1,1)}$, computed as $\text{Var}\left(\hat{\tau}_1^{(1,1)}\right)/\text{Var}\left(\hat{\xi}^{(1,1)}\right)$. Values above 1 indicate efficiency gains from the control variate adjustment.}
    \label{fig:efficiency}
\end{figure}

Figures~\ref{fig:bias}--\ref{fig:efficiency} present results for the joint exposure effect $\tau^{(1,1)}$. The results for the marginal effects and interaction are similar and appear in the Supplementary Materials.

Figure~\ref{fig:bias} illustrates the bias across measurement error configurations and different model specifications. The oracle, gold-standard, and control variate estimators maintain bias close to zero in all scenarios. The three model specification settings produce similar results that are consistent with the double robustness of the calibration-weighted AIPW estimator. In contrast, the naive and error-prone estimators exhibit substantial bias that persists regardless of the validation proportion. The bias on $\tau^{(1,1)}$ is largest in the ``ME in One Exposure'' configuration, where measurement error is concentrated on the first exposure, and smallest in the cases with uniformly low sensitivity or mixed error patterns.

Figure~\ref{fig:rmse} shows root mean squared error across scenarios. The oracle estimator has the lowest RMSE because it uses correctly measured exposures for all observations. The gold-standard and control variate estimators have RMSE that decreases as $\eta$ increases, converging on the oracle as expected. When $\eta$ is small, the RMSE of these estimators is elevated because the calibration weights are calculated using few observations, which results in a higher variance that more than offsets the near-zero bias. The naive and error-prone estimators have RMSE that is dominated by their bias in most configurations, particularly when measurement error is present in both exposures.

Figure~\ref{fig:coverage} shows the empirical coverage probabilities of the nominal 95\% confidence intervals. The oracle estimator achieves approximately 95\% coverage across all configurations, as expected. The gold-standard and control variate estimators have mild undercoverage that is largest at low $\eta$. At $\eta=0.2$, coverage falls to approximately $92$--$94\%$ across configurations and improves to roughly $94$--$95\%$ at $\eta=0.8$. This undercoverage reflects finite-sample instability in both the calibration weights and the resulting cluster-robust variance estimates when the gold-standard subsample is small. Performance under the three model specification scenarios is similar for these estimators, demonstrating consistency with double robustness. The naive and error-prone estimators have poor coverage due to bias, and both estimators fall below $20\%$ in the worst cases.

Figure~\ref{fig:efficiency} compares the variance of the gold-standard and control variate estimators for $\tau^{(1,1)}$. The relative efficiency, calculated as $\widehat{\var}\left(\hat{\tau}^{(1,1)}_1\right)/\widehat{\var}\left(\hat{\xi}^{(1,1)}\right)$, ranges from approximately $1.00$ to $1.05$ across scenarios, corresponding to variance reductions of up to $5\%$. Improvements are larger at lower values of $\eta$, where the error-prone subsample is larger relative to the gold-standard subsample, and decline as gold-standard data become more abundant. Performance is similar across the three model specification scenarios. Similar efficiency results for $\tau^{(0,1)}$, $\tau^{(1,0)}$, and the interaction $\Delta$ appear in the Supplementary Materials. These follow the same pattern, with maximum efficiency gains of approximately $4\%$, $4\%$, and $3\%$, respectively.

The efficiency gains in the bivariate setting are more modest than those seen in univariate applications.\cite{barnatchez2025flexible} The variance reduction that results from control variates depends on the squared correlation $\rho^2$ between the gold-standard estimator and the control variate, with variance reduced by a factor of $1 - \rho^2$. However, in the univariate setting this correlation is typically high because the influence functions coincide whenever the exposure is correctly classified. In the bivariate setting the joint exposure $(A_0, A_1)$ is correctly classified only when both components are modeled effectively, so the joint correct-classification rate is no larger than either marginal rate. This heuristically weakens the correlation between the estimators and shrinks the gain, and the effect compounds as the number of exposures grows. Despite these moderate gains, the control variate estimator nearly always improves efficiency compared to the gold-standard estimator without introducing additional bias. The validation-only construction of Section~\ref{sec:cv} achieves substantially larger gains. Concentrating the validation signal in a single error-prone estimator raises the correlation between the control variate and the gold-standard estimator and increases the joint-effect gains to as much as $17\%$ when the validation sample is large enough to fit the nuisance models. Section~S.6 of the Supplementary Materials reports these gains in full.

\section[Analyzing Percent Predicted FEV1 Changes from Bacterial Infections]{Analyzing Percent Predicted FEV$_1$ Changes from Bacterial Infections}
\label{sec:illustrate}

\subsection{Cystic Fibrosis Dataset}

Our analysis uses longitudinal data from Children's Hospital Colorado, consisting of bacterial cultures collected from cystic fibrosis patients at varying ages. We observe $55$ paired observations from $24$ participants that are usable as validation observations. Both sources contain information on sampling types, indicators of microorganism infection, and clinical covariates. We restrict the analysis to observations on \textit{P.~aeruginosa} and \textit{S.~aureus} infections detected through sputum or swab samples among patients ages $6$--$21$ with complete information on age, sex, height, and weight. We also limit the data to observations where percent predicted FEV$_1$ takes on values between $20$ and $150$ in accordance with plausible bounds reported for FEV$_1$ in cystic fibrosis epidemiologic studies.\cite{szczesniak2017} Data cleaning corrected $12$ height or weight data entry errors, marked $9$ additional height or weight observations as missing due to implausible values, and excluded observations with percent predicted FEV$_1$ values outside $[20, 150]$. The final cohort consists of $651$ individuals who contribute $12{,}971$ observations, of which $5{,}434$ ($41.9\%$) are sputum-based (primarily expectorated sputum) and $7{,}537$ ($58.1\%$) are swab-based.

Percent predicted FEV$_1$ was not available in the validation sample and was calculated using the Global Lung Function Initiative reference equations\cite{quanjer2012} via the \texttt{rspiro} package (version $0.5$) in \texttt{R}, which derives predicted values from age, height, sex, and observed FEV$_1$.

From the paired observations with complete culture results for both pathogens, swab samples exhibit $78.3\%$ sensitivity and $95.8\%$ specificity to detect \textit{P.~aeruginosa}; the corresponding values for \textit{S.~aureus} are $96.0\%$ and $63.6\%$. The error structure differs by pathogen. \textit{P.~aeruginosa} is prone to false negatives (swabs miss roughly one in five true infections), while \textit{S.~aureus} tends towards false positives (roughly one in three patients without true \textit{S.~aureus} infection is incorrectly classified as positive on the swab).

\subsection{Results}

Tables~\ref{tab:tbl1a} and~\ref{tab:tbl1b} provide descriptive summaries by observation and individual, respectively. Patients who contribute sputum samples are on average older ($14.4$ vs.\ $11.5$ years), taller, heavier, and have a lower percent predicted FEV$_1$ ($79.9$ vs.\ $92.0$). These numbers are consistent with sputum production being more common in older patients with more advanced disease. The prevalence of \textit{P.~aeruginosa} is roughly twice as high in sputum observations ($31.9\%$ vs.\ $15.3\%$), reflecting both true differences in colonization by age and the expected direction of swab misclassification. Across individuals, the median proportion of sputum observations is $34.3\%$, with wide variation.

\begin{figure}[ht]
    \centering
    \includegraphics[width=1\linewidth]{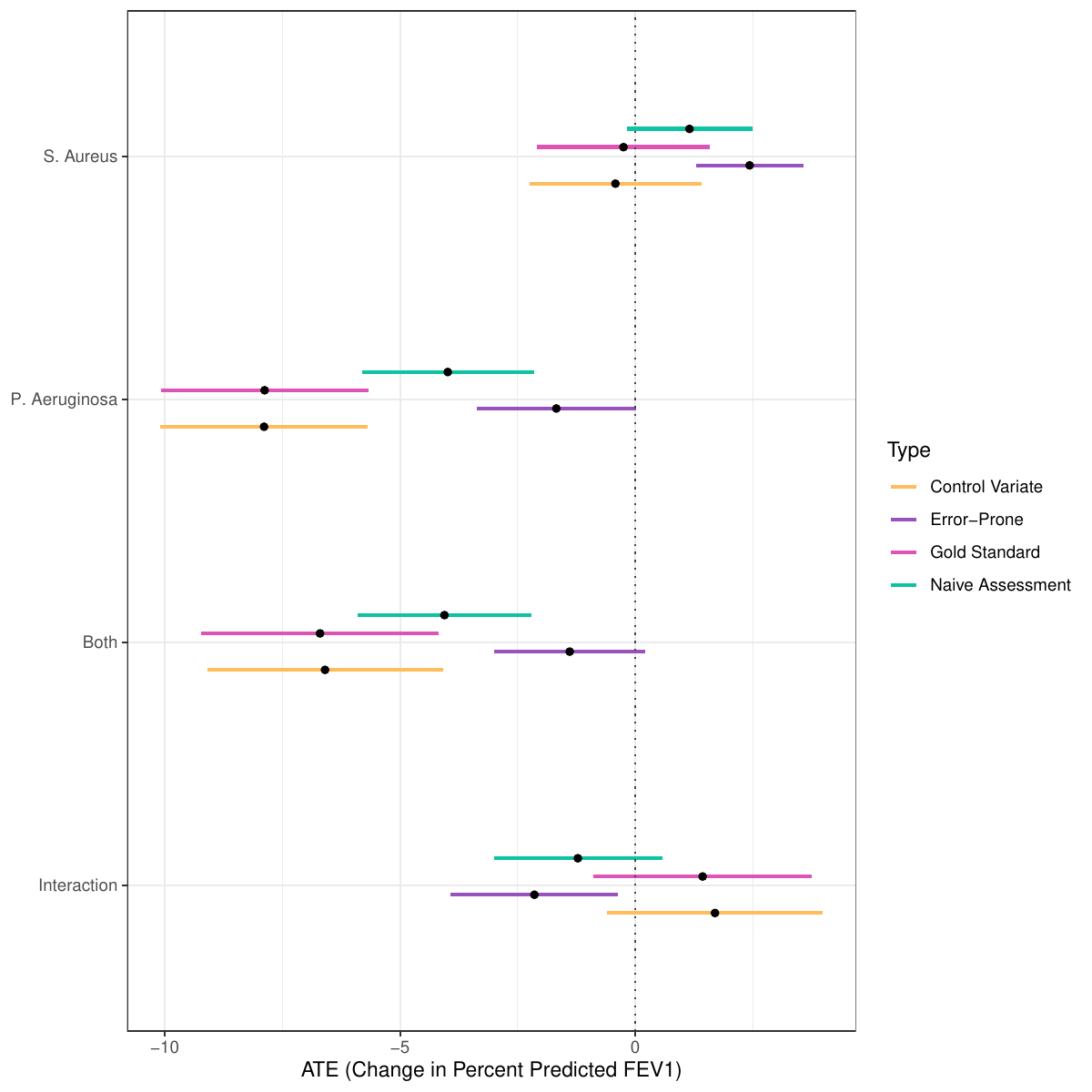}
    \caption{Estimated average treatment effects of pathogen exposures on percent predicted FEV$_1$ with $95\%$ confidence intervals by estimator. The dashed line at zero indicates no effect.}
    \label{fig:resplot}
\end{figure}

Table~\ref{tab:results} and Figure~\ref{fig:resplot} present the estimated average treatment effects, standard errors, and $95\%$ confidence intervals (CI) for each estimator. We fit all nuisance models with a SuperLearner ensemble\cite{vanderLaan2007} of a sample mean, $\ell_1$-penalized regression, and degree-two multivariate adaptive regression splines\cite{friedman1991}, adjusting for age, sex, height, weight, and cystic fibrosis transmembrane conductance regulator (CFTR) F508del genotype (homozygous, heterozygous, neither, or unknown). Cluster-robust standard errors account for repeated observations within individuals.

The largest difference between sample-specific estimators appears for \textit{P.~aeruginosa}. The sputum-based estimator finds that \textit{P.~aeruginosa} alone reduces percent predicted FEV$_1$ by $7.88$ percentage points ($95\%$ CI: $-10.08$, $-5.67$), while the swab-based estimator yields a $1.68$-point reduction with a confidence interval that covers zero ($95\%$ CI: $-3.37$, $0.02$). This attenuation of approximately $79\%$ is consistent with the limited sensitivity of throat swabs for \textit{P.~aeruginosa} ($78.3\%$). False negatives dilute the exposure contrast and pull the estimated effect toward the null. The naive estimator, which treats sputum and swab measures as interchangeable, falls between the two at $-3.99$ ($95\%$ CI: $-5.81$, $-2.16$). The control variate estimator produces an estimate of $-7.89$ ($95\%$ CI: $-10.10$, $-5.68$), close to the sputum-based result.

For \textit{S.~aureus}, the swab-based estimator finds a significant positive effect of $2.43$ percentage points ($95\%$ CI: $1.29$, $3.57$), suggesting that swab-detected \textit{S.~aureus} is associated with better lung function. This finding is likely an artifact of the low specificity of throat swabs for \textit{S.~aureus} ($63.6\%$). False positives label uncolonized patients as colonized, and because these patients are healthier on average, the estimated effect is biased upward. The sputum-based estimator does not find a significant effect of \textit{S.~aureus} ($-0.25$; $95\%$ CI: $-2.09$, $1.58$), and the control variate estimator similarly does not find a significant effect ($-0.42$; $95\%$ CI: $-2.25$, $1.40$).

The joint effect from co-infection follows the pattern for \textit{P.~aeruginosa}. The sputum-based and control variate estimators indicate that co-infection reduces percent predicted FEV$_1$ by $6.70$ ($95\%$ CI: $-9.22$, $-4.18$) and $6.59$ ($95\%$ CI: $-9.10$, $-4.09$) percentage points, respectively. The swab-based estimator finds a smaller, nonsignificant reduction of $1.39$ ($95\%$ CI: $-3.00$, $0.21$). The sputum-based and control variate estimators find no significant interaction between \textit{P.~aeruginosa} and \textit{S.~aureus} ($1.43$ and $1.69$ percentage points, with intervals covering zero), indicating an approximately additive joint effect. The swab-based estimator alone suggests a negative interaction ($-2.14$; $95\%$ CI: $-3.92$, $-0.36$), which most likely reflects misclassification rather than true synergism, mirroring the spurious \textit{S.~aureus} association. This additive pattern is consistent with \textit{P.~aeruginosa} alone driving the decline in lung function and \textit{S.~aureus} contributing little on its own.

The control variate adjustment reduces the standard error for every exposure relative to the sputum-based estimator, but only marginally. These negligible gains echo the modest efficiency improvements in the simulation and reflect the structural ceiling on variance reduction in the bivariate setting, compounded by a small validation sample.

\section{Discussion}
\label{sec:discussion}

We have developed calibration weighting and control variate estimators for causal inference with multiple misclassified binary exposures. When applied to a cohort of $651$ cystic fibrosis patients, these methods reveal that throat swab misclassification attenuates the estimated effect of \textit{P.~aeruginosa} on lung function by approximately $79\%$ compared to the sputum-based estimate ($-1.68$ vs.\ $-7.88$ percentage points of predicted FEV$_1$). The swab-based estimator also produces a spurious positive association between \textit{S.~aureus} and lung function, which can be attributed to the low specificity ($63.6\%$) of throat swabs for this pathogen. The sputum-based and control variate estimators agree that \textit{P.~aeruginosa} primarily reduces pulmonary function, with no evidence of a synergistic interaction between the two pathogens.

The control variate adjustment lowers the standard errors for every exposure in the application relative to the gold-standard estimator, but only marginally. Both theory and simulation agree that the bivariate setting places a structural limitation on variance reduction, because the joint correct-classification rate of both exposures is no larger than either marginal rate.\cite{barnatchez2025flexible} In our data, the validation sample consists of only $24$ individuals who contribute $55$ paired observations, which further limits the precision with which the optimal coefficient $\hat{\Gamma}/\hat{V}_\zeta$ can be estimated. However, we expect efficiency gains to approach their theoretical maximum more closely with a larger validation sample, as the validation-only control variate of Section~\ref{sec:cv} illustrates in simulation (Section~S.6 of the Supplementary Materials).

The simulation study supports the consistency and double robustness of the proposed estimators, but exposes a finite-sample limitation. The gold-standard and control variate estimators undercover the nominal $95\%$ confidence intervals when the proportion of gold-standard observations $\eta$ is small, with coverage falling to roughly $92$--$94\%$ at $\eta = 0.2$ and reaching $94$--$95\%$ at $\eta = 0.8$. This pattern reflects finite-sample volatility in the calibration weights when few gold-standard observations are available. In our application, approximately $42\%$ of observations come from sputum samples; at that gold-standard proportion, simulation coverage improves but has not yet reached the nominal level. Developing variance estimators or bias corrections that perform better in small gold-standard samples is a natural direction for future work.

Several limitations still remain unaddressed. As with any causal analysis, our conclusions depend on assumptions that cannot be verified from the data alone. Our estimands are contemporaneous, visit-level contrasts, so consistency presumes no carryover, that exposure at earlier visits does not affect later outcomes given the current exposure and covariates. Because we adjust for baseline covariates rather than exposure history, the cumulative effect of sustained colonization is a separate longitudinal target that would require g-methods and sequential exchangeability assumptions. Unconfoundedness (Assumption~\ref{as:ignorability}) requires that no unmeasured common causes of pathogen colonization and lung function exist aside from age, sex, height, weight, and F508del genotype. Effect transportability (Assumption~\ref{as:exch}) requires that causal effects estimated in the sputum sample generalize to patients measured with swabs, conditional on covariates. However, because sputum production is associated with age and disease severity, the two populations systematically differ, and the calibration weights must adequately rebalance covariates for transportability to hold. Exchangeable error-prone sampling (Assumption~\ref{as:nondiff-mis}) requires that, in visits where a throat swab is taken, the dependence of lung function on the swab-measured exposure is the same whether or not a paired sputum culture was also collected. This assumption would be violated if the visits selected for paired sampling differed systematically from swab-only visits in ways that alter the swab--outcome relationship beyond $\bX_{ij}$. Finally, all nuisance models are fit by a SuperLearner ensemble of a sample mean, $\ell_1$-penalized regression, and multivariate adaptive regression splines. Broadening the library to include tree-based or kernel learners is straightforward, provided each added learner respects the empirical-process conditions discussed in Section~S.5 of the Supplementary Materials.

Several directions could extend this work. First, the calibration weighting framework allows direct extensions to other patterns of measurement error. Pairing error-prone exposures with a single error-prone covariate can be done with little modification of the control variate construction, as we sketch in Section~S.4 of the Supplementary Materials (albeit where both error-prone measurements are observed on the same subset). Joint measurement error in the outcome and the exposure under the same validation structure used here is the focus of separate work that develops an efficient estimator for that setting.\cite{barnatchez2025twophase} Higher-dimensional combinations of error-prone exposures, outcomes, and covariates remain an active area.  The structural efficiency ceiling described here also tightens further as additional error-prone variables enter the construction. Second, two-phase sampling designs that select validation observations informatively, rather than at random, could improve efficiency by focusing on regions of the covariate space where misclassification is most severe. Third, expansions to more than two exposures would address settings in which multiple pathogens or risk factors are measured with error simultaneously, although the structural ceiling on control variate efficiency would become more binding as the number of exposures grows. Finally, the undercoverage we observe at low gold-standard proportions suggests that finite-sample corrections to the sandwich variance estimator, or bootstrap-based inference, may improve the reliability of confidence intervals when validation data are scarce.

\section*{Acknowledgments}
\noindent The authors thank the patients and clinical staff at Children's Hospital Colorado whose data made this study possible.

\section*{Funding}
\noindent This research received no specific grant from any funding agency in the public, commercial, or not-for-profit sectors.

\section*{Conflict of Interest}
\noindent The authors declare no conflict of interest.

\section*{Data Availability Statement}
\noindent The clinical data that support the findings of this study were collected at Children's Hospital Colorado and contain protected health information from pediatric patients. These data are not publicly available owing to privacy and ethical restrictions and may be available from the authors upon reasonable request and with the permission of Children's Hospital Colorado. The \texttt{R} code implementing the proposed estimators and reproducing the simulation study is openly available at \url{https://github.com/kjosey/multi-me}.

\clearpage

\section*{Tables}

\begin{table}[ht]
\centering
\begin{threeparttable}
\begin{tabular}{lcc} \toprule
& \begin{tabular}{c}Sputum\\($N=5{,}434$)\end{tabular} & \begin{tabular}{c}Throat Swab\\($N=7{,}537$)\end{tabular} \\
\midrule
\textbf{Percent Predicted FEV$_1$} & & \\
\quad Mean (SD) & $79.9$ ($20.2$) & $92.0$ ($16.4$) \\
\quad Median [Range] & $82.7$ [$20.2$, $137.3$] & $93.4$ [$21.0$, $144.9$] \\
\textbf{P. aeruginosa} & & \\
\quad Positive & $1{,}735$ ($31.9\%$) & $1{,}156$ ($15.3\%$) \\
\quad Negative & $3{,}699$ ($68.1\%$) & $6{,}381$ ($84.7\%$) \\
\textbf{S. aureus} & & \\
\quad Positive & $3{,}436$ ($63.2\%$) & $4{,}149$ ($55.0\%$) \\
\quad Negative & $1{,}998$ ($36.8\%$) & $3{,}388$ ($45.0\%$) \\
\textbf{Age (years)} & & \\
\quad Mean (SD) & $14.4$ ($3.45$) & $11.5$ ($3.75$) \\
\quad Median [Range] & $14.7$ [$6.0$, $21.0$] & $11.0$ [$6.0$, $21.0$] \\
\textbf{Height (cm)} & & \\
\quad Mean (SD) & $155.7$ ($15.6$) & $143.2$ ($19.0$) \\
\quad Median [Range] & $158$ [$102$, $197$] & $142$ [$98$, $197$] \\
\textbf{Weight (kg)} & & \\
\quad Mean (SD) & $47.6$ ($14.2$) & $38.4$ ($15.3$) \\
\quad Median [Range] & $48.8$ [$12.78$, $170.2$] & $34.7$ [$12.5$, $105$] \\
\bottomrule
\end{tabular}
\end{threeparttable}
\caption{Descriptive statistics by sampling type across all $12{,}971$ observations from the analytic cohort.}
\label{tab:tbl1a}
\end{table}

\begin{table}[ht]
\centering
\begin{threeparttable}
\begin{tabular}{lc} \toprule
& \begin{tabular}{c}Overall\\($N=651$)\end{tabular} \\
\midrule
\textbf{Female} & $313$ ($48.1\%$) \\
\textbf{Race} & \\
\quad Black & $8$ ($1.2\%$) \\
\quad Caucasian & $618$ ($94.9\%$) \\
\quad Other & $25$ ($3.8\%$) \\
\textbf{Hispanic} & $65$ ($10.0\%$) \\
\textbf{Family History} & $68$ ($10.4\%$) \\
\textbf{F508del Status} & \\
\quad Homozygous & $328$ ($50.4\%$) \\
\quad Heterozygous & $241$ ($37.0\%$) \\
\quad Neither & $62$ ($9.5\%$) \\
\quad Unknown & $20$ ($3.1\%$) \\
\textbf{Malabsorption at Diagnosis} & $53$ ($8.1\%$) \\
\textbf{Meconium Ileus at Diagnosis} & $90$ ($13.8\%$) \\
\textbf{Number of Observations} & \\
\quad Mean (SD) & $19.9$ ($15.1$) \\
\quad Median [Range] & $16.0$ [$1$, $72$] \\
\textbf{Proportion Sputum} & \\
\quad Mean (SD) & $0.412$ ($0.342$) \\
\quad Median [Range] & $0.343$ [$0$, $1.00$] \\
\textbf{Proportion Throat Swab} & \\
\quad Mean (SD) & $0.588$ ($0.342$) \\
\quad Median [Range] & $0.657$ [$0$, $1.00$] \\
\textbf{Number of Sputum Observations} & \\
\quad Mean (SD) & $8.35$ ($10.49$) \\
\quad Median [Range] & $4.0$ [$0$, $64$] \\
\textbf{Number of Swab Observations} & \\
\quad Mean (SD) & $11.58$ ($11.10$) \\
\quad Median [Range] & $8.0$ [$0$, $67$] \\
\bottomrule
\end{tabular}
\end{threeparttable}
\caption{Participant-level summary statistics ($N = 651$ patients ages $6$--$21$ contributing $12{,}971$ observations). For each continuous summary, the per-individual statistic (e.g., proportion sputum, observation count) is summarized across the $651$ patients. Patients with unrecorded CFTR genotype (Unknown) are modeled as a separate category in the adjustment model, not folded into the reference.}
\label{tab:tbl1b}
\end{table}

\begin{table}[ht]
\centering
\begin{threeparttable}
\begin{tabular}{llrrl}
\toprule
\textbf{Estimator} & \textbf{Exposure} & \textbf{ATE} & \textbf{SE} & $\mathbf{95\%}$ \textbf{CI} \\
\midrule
\textbf{Swab (EP)} & & & & \\
& \textit{S. aureus} & $2.43$ & $0.58$ & ($1.29$, $3.57$) \\
& \textit{P. aeruginosa} & $-1.68$ & $0.86$ & ($-3.37$, $0.01$) \\
& Both & $-1.39$ & $0.82$ & ($-3.00$, $0.21$) \\
& Interaction & $-2.14$ & $0.91$ & ($-3.92$, $-0.36$) \\
\midrule
\textbf{Sputum (Gold)} & & & & \\
& \textit{S. aureus} & $-0.25$ & $0.94$ & ($-2.09$, $1.58$) \\
& \textit{P. aeruginosa} & $-7.88$ & $1.13$ & ($-10.08$, $-5.67$) \\
& Both & $-6.70$ & $1.28$ & ($-9.22$, $-4.18$) \\
& Interaction & $1.43$ & $1.18$ & ($-0.89$, $3.75$) \\
\midrule
\textbf{Naive} & & & & \\
& \textit{S. aureus} & $1.15$ & $0.68$ & ($-0.18$, $2.49$) \\
& \textit{P. aeruginosa} & $-3.99$ & $0.93$ & ($-5.81$, $-2.16$) \\
& Both & $-4.06$ & $0.95$ & ($-5.91$, $-2.20$) \\
& Interaction & $-1.22$ & $0.91$ & ($-3.01$, $0.57$) \\
\midrule
\textbf{Control Variate} & & & & \\
& \textit{S. aureus} & $-0.42$ & $0.93$ & ($-2.25$, $1.40$) \\
& \textit{P. aeruginosa} & $-7.89$ & $1.13$ & ($-10.10$, $-5.68$) \\
& Both & $-6.59$ & $1.28$ & ($-9.10$, $-4.09$) \\
& Interaction & $1.69$ & $1.17$ & ($-0.60$, $3.99$) \\
\bottomrule
\end{tabular}
\end{threeparttable}
\caption{Estimated average treatment effects (ATE) on percent predicted FEV$_1$, standard errors (SE), and $95\%$ confidence intervals (CI) by estimator and exposure combination.}
\label{tab:results}
\end{table}

\clearpage

\section*{Figures}

\clearpage
\beginsupplement
\begin{center}
  {\Large\bfseries Supplementary Materials ``Causal Inference with Multiple Misclassified Exposures: A Control Variate-Adjusted Calibration Weighting Approach''}
\end{center}
\bigskip

\section{Calibration Weights}
\label{sec:supp-calibration}

We solve three calibration problems, one for each subsample, using the same entropy-balancing form.  Let $q_{ij}>0$ denote a known base weight on each observation and let $\bc(\bX_{ij})$ denote a fixed vector of calibration basis functions of $\bX_{ij}$ whose first coordinate is identical to one.  The main-text construction is the special case $\bc(\bX_{ij})=\bX_{ij}$, with $\bX_{ij}$ already including a constant; richer choices, such as polynomial expansions, splines, or pairwise interactions, fit the same framework and are useful when the propensity- or outcome-regression surface is plausibly nonlinear in $\bX$.  Common choices for $q_{ij}$ include the following:
\begin{itemize}
    \item $q_{ij}=1$ for all $(i,j)$, which gives every observation an equal baseline mass and is the default used in our analysis;
    \item $q_{ij}=m_i^{-1}$, which equalizes each individual's contribution under clustered data so that subjects with many visits do not dominate the target;
    \item $q_{ij}=w_{ij}$, where $w_{ij}$ is a known design weight inherited from a complex sampling scheme and transports the calibration target to the implied superpopulation;
    \item $q_{ij}=\ind(\bX_{ij}\in\mathcal{R})$ for some region $\mathcal{R}$ of the covariate space, which restricts the target population to a subgroup of interest.
\end{itemize}
For any admissible choice of $q_{ij}$, the gold-standard, error-prone-only, and full-error-prone calibration weights satisfy, respectively,
\begin{equation}\label{eq:primal-cal}
\sum_{i,j} \ind(Z_{ij}=1,\bA_{ij}=\ba)\,\hat\gamma^{\ba}_1(\bX_{ij})\,\bc(\bX_{ij}) \;=\; \sum_{i,j} q_{ij}\,\bc(\bX_{ij}),
\end{equation}
\begin{equation}\label{eq:primal-cal-ep}
\sum_{i,j} \ind(Z_{ij}=0,\bA^*_{ij}=\ba)\,\tilde\gamma^{\ba}_0(\bX_{ij})\,\bc(\bX_{ij}) \;=\; \sum_{i,j} q_{ij}\,\bc(\bX_{ij}),
\end{equation}
\begin{equation}\label{eq:primal-cal-full}
\sum_{i,j} \ind\{(1-Z_{ij})+S_{ij}=1,\bA^*_{ij}=\ba\}\,\tilde\gamma^{\ba}(\bX_{ij})\,\bc(\bX_{ij}) \;=\; \sum_{i,j} q_{ij}\,\bc(\bX_{ij}),
\end{equation}
with each weight obtained by minimizing the relative entropy $\sum \ind(\cdot)\,[\gamma\log(\gamma/q) - \gamma]$ subject to the corresponding constraint, following the entropy-balancing form.\cite{josey2021a} Because $\bc(\bX_{ij})$ contains a constant coordinate, each constraint enforces $\sum \ind(\cdot)\,\gamma = \sum_{i,j} q_{ij}$.  The calibrated weights therefore match the total baseline mass on the relevant subsample-and-exposure stratum, and the IPW component of the AIPW pseudo-outcome lies on the same per-observation scale as the marginal mean of $\hat\mu$ (Section~\ref{sec:supp-aipw}) without an explicit factor of $N$.

Each problem has the same closed form.  Writing the gold-standard case explicitly, the Lagrangian stationarity condition gives the exponential tilt
\begin{equation}\label{eq:gamma-form}
\hat\gamma^{\ba}_1(\bX_{ij}) \;=\; q_{ij}\,\exp\!\bigl[\bc^{T}(\bX_{ij})\hat{\bm\lambda}_{\ba}\bigr],
\end{equation}
with $\hat{\bm\lambda}_{\ba}$ solving the strictly convex dual
\begin{equation}\label{eq:dual-cal}
\hat{\bm\lambda}_{\ba} \;=\; \argmin_{\bm\lambda}\left\{\sum_{i,j} \ind(Z_{ij}=1,\bA_{ij}=\ba)q_{ij}\exp\left[\bc^{T}(\bX_{ij})\bm\lambda\right] \;-\; \bm\lambda^T\sum_{i,j} q_{ij}\,\bc(\bX_{ij})\right\},
\end{equation}
whose gradient vanishes at the optimum, recovering \eqref{eq:primal-cal} exactly.  The same derivation, with $\ind(Z=0,\bA^*=\ba)$ or $\ind\{(1-Z)+S=1,\bA^*=\ba\}$ in place of $\ind(Z=1,\bA=\ba)$, yields the closed-form solutions for $\tilde\gamma^{\ba}_0$ and $\tilde\gamma^{\ba}$.  We solve each dual by BFGS with the analytic gradient.  Convergence is followed through the maximum absolute deviation in the calibration constraint, and weights exceeding ten times the within-group mean trigger a diagnostic warning.  If the dual fails to converge, we fall back to base weights and flag the replication.

\section{AIPW Pseudo-Outcomes and Estimators}
\label{sec:supp-aipw}

Let $\mu_1(\ba,\bX) = \E[Y\mid \bA=\ba,\bX,Z=1]$ and $\pi^{\ba}_1(\bX) = \Pr\{\bA=\ba,Z=1\mid\bX\}$.  Under Assumptions~1--3 of the main text,
\begin{equation}\label{eq:moment}
\E[Y^{\ba}] = \E\!\left[\frac{\ind(\bA=\ba,Z=1)}{\pi^{\ba}_1(\bX)}\bigl(Y - \mu_1(\ba,\bX)\bigr) + \mu_1(\ba,\bX)\right],
\end{equation}
the standard AIPW representation of $\E[Y^{\ba}]$.  Substituting the calibration weight $\hat\gamma^{\ba}_1(\bX_{ij})$ from Section~\ref{sec:supp-calibration} for $q_{ij}/\pi^{\ba}_1(\bX_{ij})$ and $\hat\mu_1$ for $\mu_1$ gives the gold-standard pseudo-outcome, written as a function of the full observation $\bO_{ij}$ defined in the main text,
\begin{equation}\label{eq:psi-gold}
\hat\psi^{\ba}_1(\bO_{ij}) \;=\; \ind(\bA_{ij}=\ba,Z_{ij}=1)\,\hat\gamma^{\ba}_1(\bX_{ij})\bigl(Y_{ij}-\hat\mu_1(\ba,\bX_{ij})\bigr) + \hat\mu_1(\ba,\bX_{ij}),
\end{equation}
and the analogous error-prone-only and full-error-prone pseudo-outcomes $\tilde\psi^{\ba}_0$ and $\tilde\psi^{\ba}$, obtained by substituting $\ind(\bA^*_{ij}=\ba,Z_{ij}=0)$ and $\ind\{\bA^*_{ij}=\ba,(1-Z_{ij})+S_{ij}=1\}$ together with $\tilde\gamma^{\ba}_0$, $\tilde\mu_0$ and $\tilde\gamma^{\ba}$, $\tilde\mu$ for the corresponding gold-standard components.  Because $\hat\gamma$ already aggregates to $\sum_{i,j} q_{ij}$ within each stratum, the IPW component is on the same per-observation scale as $\hat\mu_1$, and the subsequent $1/N$ averaging reproduces a finite weighted mean of residuals over the gold-standard subsample.  Each estimator is the average pseudo-outcome contrast,
\[
\hat\tau^{\ba}_1 = \frac{1}{N}\sum_{i,j}\bigl(\hat\psi^{\ba}_{1,ij} - \hat\psi^{\mathbf{0}}_{1,ij}\bigr),
\]
with $\tilde\tau^{\ba}_0$ and $\tilde\tau^{\ba}$ defined the same way.  Each is doubly robust.  The estimator $\hat\tau^{\ba}_1$ is consistent for $\tau^{\ba}$ if either $\hat\gamma^{\ba}_1$ or $\hat\mu_1$ is consistent, by the standard AIPW argument.  The residual $Y-\hat\mu_1$ has conditional mean zero when $\hat\mu_1\to\mu_1$, and the IPW component already targets $\E[Y^{\ba}]-\E[\mu_1(\ba,\bX)]$ when $\hat\gamma^{\ba}_1$ correctly inverts the joint propensity.

The contrast $\hat\phi^{\ba}_{ij} = \hat\psi^{\ba}_{1,ij} - \hat\psi^{\mathbf{0}}_{1,ij}$ is the influence function of $\hat\tau^{\ba}_1$, so cluster-robust variance follows from aggregating influence-function values within individual,
\begin{equation}\label{eq:vhat}
\widehat{\var}\left(\hat\tau^{\ba}_1\right) = \frac{n}{(n-1)N^{2}}\sum_{i=1}^{n}\left[\sum_{j=1}^{m_i}\bigl(\hat\phi^{\ba}_{ij}-\hat\tau^{\ba}_1\bigr)\right]^{2}.
\end{equation}
We compute \eqref{eq:vhat} via the cluster-robust wrapper in the \texttt{sandwich} package\cite{zeileis2006} and form Wald confidence intervals as $\hat\tau^{\ba}_1\pm1.96\sqrt{\widehat\var\left(\hat\tau^{\ba}_1\right)}$.  The interaction $\hat\Delta_1$ uses the four-way contrast $\hat\phi^{\Delta}_{ij} = \hat\psi^{(1,1)}_{1,ij} - \hat\psi^{(1,0)}_{1,ij} - \hat\psi^{(0,1)}_{1,ij} + \hat\psi^{\mathbf{0}}_{1,ij}$ in place of $\hat\phi^{\ba}_{ij}$.

\section{Control Variate-Adjusted Joint Estimator}
\label{sec:supp-cv}

Recall from Section~3.2 of the main text the control variate $\hat\zeta^{\ba} = \tilde\tau^{\ba} - \tilde\tau^{\ba}_0$ and the joint estimator $\hat\xi^{\ba} = \hat\tau^{\ba}_1 + b\,\hat\zeta^{\ba}$.  Their per-observation influence-function representations are $\hat\phi^{\ba}_{ij}$ for $\hat\tau^{\ba}_1$ and $\tilde\phi^{\ba}_{ij} - \tilde\phi^{\ba}_{0,ij}$ for $\hat\zeta^{\ba}$, where $\tilde\phi^{\ba}_{ij}$ and $\tilde\phi^{\ba}_{0,ij}$ are the full-error-prone and error-prone-only analogues of the gold-standard contrast. Stacking these into the bivariate response and fitting an intercept-only multivariate regression on a constant returns the bivariate intercept $(\hat\tau^{\ba}_1,\;\hat\zeta^{\ba})^{T}$ with cluster-robust covariance, computed as in \eqref{eq:vhat},
\[
\widehat{\bm\Sigma} = \begin{pmatrix}\widehat\Omega & \widehat\Gamma\\ \widehat\Gamma & \widehat V_\zeta\end{pmatrix}.
\]
The joint estimator's point form and variance are
\begin{equation}\label{eq:cv-est}
\hat\xi^{\ba} = \hat\tau^{\ba}_1 + \hat b\,\hat\zeta^{\ba},\quad \hat b = -\widehat\Gamma/\widehat V_\zeta, \quad \widehat\var\left(\hat\xi^{\ba}\right) = \widehat\Omega - \widehat\Gamma^{2}/\widehat V_\zeta = \widehat\var\left(\hat\tau^{\ba}_1\right)(1-\hat\rho^{2}),
\end{equation}
where $\hat\rho=\widehat\Gamma/\sqrt{\widehat\Omega\widehat V_\zeta}$.  Each influence function also carries an outcome-regression prediction $\hat\mu(\ba,\bX_{ij})$ that is evaluated on every observation, but the reweighted residual part of $\hat\phi^{\ba}_{ij}$ is nonzero only on $\{Z_{ij}=1\}$, that of $\tilde\phi^{\ba}_{0,ij}$ only on $\{Z_{ij}=0\}$, and that of $\tilde\phi^{\ba}_{ij}$ on the full error-prone sample $\{Z_{ij}=0\}\cup\{S_{ij}=1\}$.  The gold-standard residual overlaps the control variate only through $\tilde\phi^{\ba}_{ij}$ on the validation subsample $\{S_{ij}=1\}$, since $\tilde\phi^{\ba}_{0,ij}$ is confined to $\{Z_{ij}=0\}$; an observation therefore contributes to the cross-cluster sum that defines $\widehat\Gamma$ only when it lies in $\{S_{ij}=1\}$.  This is the formal sense in which the control variate covariance is identified by the validation overlap.  Each component of \eqref{eq:cv-est} is doubly robust for its own target, and the linear combination preserves the consistency of $\hat\tau^{\ba}_1$ for $\tau^{\ba}$ because $\hat\zeta^{\ba}$ targets zero in expectation under Assumption~4 (exchangeable error-prone sampling).

\section{Variation: Mismeasured Covariate and Exposure}
\label{sec:supp-variations-cov}

The main-text algorithm targets misclassified exposures with all other variables observed exactly.  The same calibration, AIPW, and control variate construction extends to settings in which a baseline covariate is also subject to measurement error.  The gold-standard estimator $\hat\tau^{\ba}_1$ is unchanged, the control variate $\hat\zeta^{\ba}$ retains its mean-zero property, and the joint moment regression of Section~\ref{sec:supp-cv} provides $\hat\xi^{\ba}$ with the same closed-form variance reduction $\widehat\var\left(\hat\xi^{\ba}\right) = \widehat\Omega - \widehat\Gamma^{2}/\widehat V_\zeta$.  The modifications affect only the variables used to build the error-prone and validation nuisance fits, and the exchangeable-sampling assumption that anchors the mean-zero property.

Let $\bX_{ij}$ denote the true baseline covariate vector and $\bX^*_{ij}$ an error-prone surrogate observed on the error-prone-only and validation subsamples.  Gold-standard-only observations record $(\bX_{ij},\bA_{ij})$, error-prone-only observations record $(\bX^*_{ij},\bA^*_{ij})$, and validation observations record both pairs.  Throughout this section, $Z_{ij}=1$ marks the observations where the exposure and the covariate are both measured without error, the intersection of their gold-standard sets, and $S_{ij}=1$ the validation subset where the true pair $(\bX_{ij},\bA_{ij})$ and the surrogate pair $(\bX^*_{ij},\bA^*_{ij})$ are both recorded.  Assumption~4 of the main text is replaced by its surrogate-covariate analogue: the conditional mean outcome given the error-prone exposure and surrogate covariate agrees across the error-prone-only ($Z_{ij}=0$) and validation ($S_{ij}=1$) subsamples,
\[
\E\!\left[Y_{ij} \mid \bA^*_{ij}=\ba, \bX^*_{ij}, Z_{ij}=0\right] = \E\!\left[Y_{ij} \mid \bA^*_{ij}=\ba, \bX^*_{ij}, S_{ij}=1\right] \quad \text{for all } \ba \in \{0,1\}^2,
\]
so that the error-prone outcome regression in $(\bA^*_{ij},\bX^*_{ij})$ shares a common population limit across the error-prone-only and validation subsamples.

Explicitly, the construction proceeds in four steps:
\begin{enumerate}
\item \textit{Gold-standard estimator.} On the gold-standard subsample $\{Z_{ij}=1\}$, construct the calibration weights $\hat\gamma^{\ba}_1(\bX_{ij})$, the outcome regression $\hat\mu_1(\ba,\bX_{ij})$, the pseudo-outcome $\hat\psi^{\ba}_1$, and the estimator $\hat\tau^{\ba}_1$ from the true covariate $\bX_{ij}$ and true exposure $\bA_{ij}$, exactly as in Sections~\ref{sec:supp-calibration} and~\ref{sec:supp-aipw}.  This step is unchanged from the misclassified-exposure algorithm and targets $\tau^{\ba}$.
\item \textit{Full error-prone estimator.} On the full error-prone sample $\{(1-Z_{ij})+S_{ij}=1\}$, solve for calibration weights that balance the surrogate covariate margin,
\[
\sum_{i,j} \ind\{(1-Z_{ij})+S_{ij}=1,\,\bA^*_{ij}=\ba\}\,\tilde\gamma^{\ba}(\bX^*_{ij})\,\bc(\bX^*_{ij}) \;=\; \sum_{i,j} \ind\{(1-Z_{ij})+S_{ij}=1\}\,q_{ij}\,\bc(\bX^*_{ij}),
\]
where the target accumulates over every observation on which $\bX^*$ is recorded, namely the error-prone-only and validation subsamples.  Fit the outcome regression $\tilde\mu(\ba,\bX^*_{ij}) = \hat\E[Y_{ij}\mid \bA^*_{ij}=\ba,\bX^*_{ij},(1-Z_{ij})+S_{ij}=1]$ using $\bX^*_{ij}$ throughout, including on the validation observations where $\bX_{ij}$ is also recorded, then form $\tilde\psi^{\ba}$ and $\tilde\tau^{\ba}$ as in Section~\ref{sec:supp-aipw} with $(\bX^*_{ij},\bA^*_{ij})$ replacing $(\bX_{ij},\bA_{ij})$.
\item \textit{Error-prone-only estimator.} On the error-prone-only subsample $\{Z_{ij}=0\}$, solve for calibration weights from the same surrogate covariate,
\[
\sum_{i,j} \ind(Z_{ij}=0,\bA^*_{ij}=\ba)\,\tilde\gamma^{\ba}_0(\bX^*_{ij})\,\bc(\bX^*_{ij}) \;=\; \sum_{i,j} \ind\{(1-Z_{ij})+S_{ij}=1\}\,q_{ij}\,\bc(\bX^*_{ij}),
\]
so that both error-prone estimators calibrate to the same target population, on which $\bX^*$ is recorded.  Fit $\tilde\mu_0(\ba,\bX^*_{ij}) = \hat\E[Y_{ij}\mid \bA^*_{ij}=\ba,\bX^*_{ij},Z_{ij}=0]$, and form $\tilde\psi^{\ba}_0$ and $\tilde\tau^{\ba}_0$ in the same way.
\item \textit{Control variate and joint regression.} Set $\hat\zeta^{\ba} = \tilde\tau^{\ba} - \tilde\tau^{\ba}_0$ and run the intercept-only joint moment regression of Section~\ref{sec:supp-cv}, which returns $\hat\xi^{\ba}$ and its cluster-robust variance $\widehat\var\left(\hat\xi^{\ba}\right) = \widehat\Omega - \widehat\Gamma^{2}/\widehat V_\zeta$.
\end{enumerate}

Both estimators are built from the same surrogate $\bX^*$, and the full-sample estimator $\tilde\tau^{\ba}$ uses $\bX^*$ even on the validation observations where $\bX_{ij}$ is also recorded. Thus, under the exchangeable-sampling condition, they share a common population limit and $\hat\zeta^{\ba} = \tilde\tau^{\ba} - \tilde\tau^{\ba}_0$ retains its mean-zero property. No separate marginalization step is required to return the surrogate-based estimators to the true-covariate scale; conditioning the error-prone nuisances on $\bX^*$ already averages over $\bX$ within each surrogate stratum, whereas replacing $\bX^*$ with $\bX$ would change the population limit and reintroduce bias into the control variate. Consistency for $\tau^{\ba}$ therefore relies on the gold-standard estimator of Step~1 alone, with the error-prone estimators contributing only variance reduction.  That reduction depends on the within-validation correlation between $\hat\phi^{\ba}_{ij}$ and $\tilde\phi^{\ba}_{ij}-\tilde\phi^{\ba}_{0,ij}$, which reflects the agreement between $\bX$ and $\bX^*$ as well as between $\bA$ and $\bA^*$.

This variation validates the exposure and the covariate together, so $Z_{ij}$ and $S_{ij}$ each refer to a single subset. Relaxing the design to one in which the two are validated on separate subsets, so that their gold-standard sets differ, is a direction for future work.

\section{Nuisance Estimation in Practice}
\label{sec:supp-nuisance}

The calibration weights are obtained from \eqref{eq:dual-cal} via BFGS as described above.  The outcome regressions $\hat\mu_1$, $\tilde\mu_0$, and $\tilde\mu$ are estimated separately on each subsample by a SuperLearner ensemble\cite{vanderLaan2007} with library $\{\texttt{SL.mean}, \texttt{SL.glmnet}, \texttt{SL.earth}\}$: a sample mean, $\ell_1$-penalized (lasso) regression, and degree-two multivariate adaptive regression splines.\cite{friedman1991} Implementing flexible machine learning within doubly robust causal estimation generally requires sample splitting or cross-fitting to break the dependence between the nuisance fits and the data used to solve the estimating equations.\cite{chernozhukov2018, naimi2023}  We avoid cross-fitting by restricting the library to learners whose fits lie in a bounded-variation, Donsker function class (a constant, a linear predictor with bounded $\ell_1$ norm, and piecewise-linear splines), since a convex combination of finitely many Donsker classes is itself Donsker.\cite{vandervaart1996} The SuperLearner fit therefore lies in a Donsker class, and its empirical-process remainder is controlled without sample splitting; the same property underlies data-adaptive estimators confined to a bounded-variation class, such as the highly adaptive lasso.\cite{benkeser2016}  Calibration covariates and outcome-regression covariates coincide: $X_0$ and $X_1$ in the simulation, and the adjustment covariates (age, sex, height, weight, and F508del genotype) in the application.  Misspecification in the simulation is induced in the data-generating mechanism by replacing $\bX$ with the transformed covariates $\mathbf{U}=(U_0,U_1)$ in the propensity (\texttt{ps-mis}) or in the outcome surface (\texttt{out-mis}).  The analysis covariates remain $\bX$ in all scenarios.  Estimates are winsorized at the $1$st and $99$th percentiles before computing bias and RMSE.  Coverage and CI lengths use the unwinsorized confidence intervals.

\section{Efficiency of the Validation-Only Control Variate}
\label{sec:supp-val-efficiency}

The control variate in Section~3.2 of the main text is built to keep every nuisance fit on a large sample.  It pairs the error-prone-only estimator $\tilde{\tau}^{\ba}_0$ with the full error-prone estimator $\tilde{\tau}^{\ba}$, both estimated on the abundant error-prone data rather than on the sparse validation subsample.  That is a deliberate choice, but not the only one available.  As noted in Section~3.2, a natural alternative replaces $\tilde{\tau}^{\ba}$ with the validation-only estimator $\tilde{\tau}^{\ba}_{\val}$, forming the mean-zero contrast $\hat{\zeta}^{\ba} = \tilde{\tau}^{\ba}_{\val} - \tilde{\tau}^{\ba}_0$; this is the form used to build the univariate control variate,\cite{barnatchez2025flexible} and we write the resulting estimator $\hat{\xi}^{\ba}_{\val}$.  By concentrating the validation signal in a single estimator rather than diluting it across the full error-prone sample, this alternative should sharpen the control variate, at the cost of a more demanding nuisance-estimation problem.  To see how that tradeoff resolves, we repeat the simulation of Section~4 with $\hat{\xi}^{\ba}_{\val}$ in place of $\hat{\xi}^{\ba}$, holding the data-generating process, the competing estimators, and every other implementation choice fixed.

\begin{figure}[H]
\centering
\includegraphics[width=\linewidth]{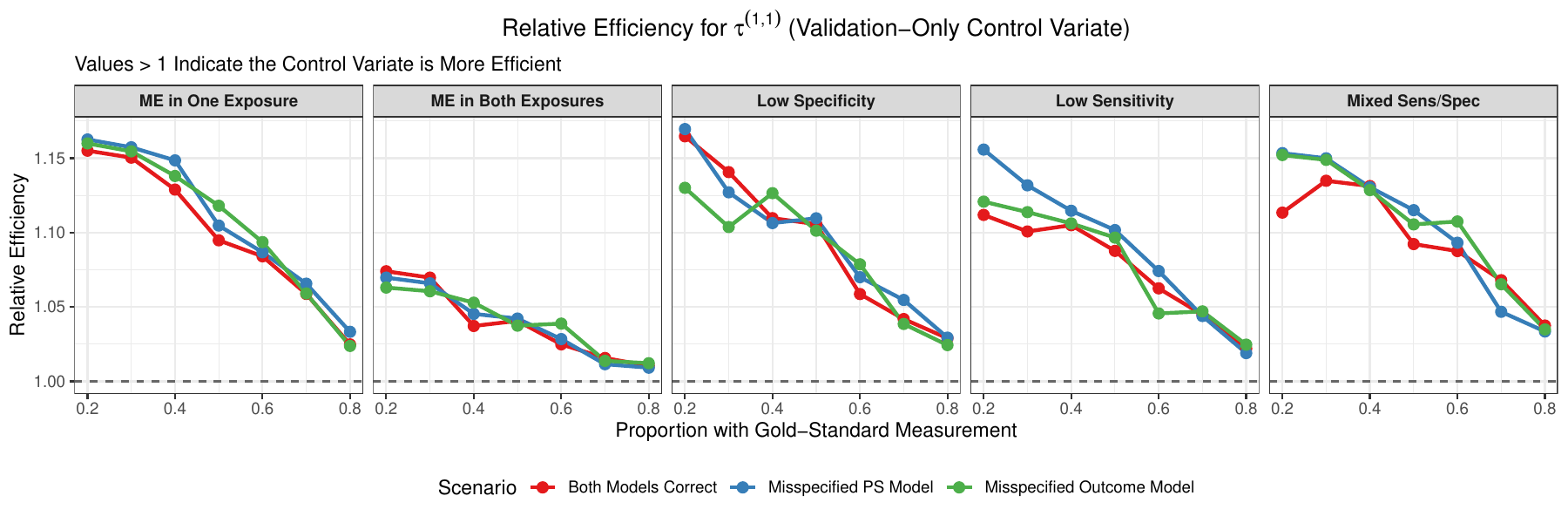}
\caption{Relative efficiency of the validation-only control variate $\hat{\xi}^{(1,1)}_{\val}$ versus the gold-standard estimator for the joint effect $\tau^{(1,1)}$, computed as $\widehat{\var}\left(\hat{\tau}^{(1,1)}_{\gs}\right)/\widehat{\var}\left(\hat{\xi}^{(1,1)}_{\val}\right)$.  Values above one indicate variance reduction.  Panels correspond to the five measurement-error configurations; the three lines correspond to the model-specification scenarios (Correct, PS Misspec., OR Misspec.).}
\label{fig:supp-efficiency-val}
\end{figure}

The alternative delivers on its promise.  For the joint effect $\tau^{(1,1)}$, the relative efficiency of $\hat{\xi}^{\ba}_{\val}$ reaches $1.17$, an efficiency gain of roughly $17\%$ and more than triple the $5\%$ attained by the full-error-prone construction; the marginal effects $\tau^{(0,1)}$ and $\tau^{(1,0)}$ improve by up to $16\%$ and $13\%$, and even the interaction $\Delta$, the hardest contrast to sharpen, gains as much as $10\%$ (Figure~\ref{fig:supp-efficiency-val}).  The gains keep the same shape seen in the main text, largest where gold-standard observations are scarce and fading as $\eta$ grows, since that is where the error-prone sample carries the most additional information.  The improvement also costs nothing in validity.  The estimator $\hat{\xi}^{\ba}_{\val}$ stays unbiased across every configuration, and its coverage tracks that of the gold-standard and full-error-prone estimators, sharing the same mild undercoverage at small $\eta$ and no more.

The mechanism is the one anticipated in Section~4.  Placing the entire validation signal in a single estimator raises the correlation $\rho$ between the control variate and the gold-standard estimator, and because the adjusted variance scales with $1 - \rho^2$, even a modest gain in $\rho$ produces a visible reduction in variance.  This advantage carries a condition.  Since $\tilde{\tau}^{\ba}_{\val}$ fits its calibration weights and outcome regression on the validation observations alone, it requires a validation sample large enough to support flexible, data-adaptive estimation.  That condition is met comfortably in these simulations, where the validation subsample still numbers in the hundreds even at $\eta = 0.2$, yet it fails in our application, where the validation observations, split across the four exposure combinations, leave too few in each cell for the SuperLearner ensemble to fit.  We therefore retain the full-error-prone construction for the main analysis and report the validation-only results here as a picture of the efficiency that a larger validation sample would make available.

\section{Complete Simulation Results}
\label{sec:supp-simresults}

Tables~\ref{tab:sim-bias-01}--\ref{tab:sim-var-int} report the simulation results from Section~4 of the main text.  We present one table for each combination of performance metric (bias, RMSE, $95\%$ CI coverage, empirical variance) and causal contrast ($\tau^{(0,1)}$, $\tau^{(1,0)}$, $\tau^{(1,1)}$, $\Delta$), for sixteen tables in total.  Within each table, the rows are the five measurement-error configurations crossed with the seven gold-standard proportions $\eta\in\{0.2,\dots,0.8\}$, and the columns are the four estimators (Naive, EP, Gold, CV) reported separately under each of the three model-specification scenarios (Correct, PS Misspec., OR Misspec.).  The five measurement-error configurations are labelled descriptively: ``ME in One Exposure'' for $(0.7,0.7,0.9,0.9)$; ``ME in Both Exposures'' for $(0.7,0.7,0.7,0.7)$; ``Low Specificity'' for $(0.9,0.7,0.9,0.7)$; ``Low Sensitivity'' for $(0.7,0.9,0.7,0.9)$; and ``Mixed Sens/Spec'' for $(0.7,0.9,0.9,0.7)$, in which $A_0$ has elevated false negatives and $A_1$ has elevated false positives.  The Oracle benchmark is omitted from the tables for compactness since its bias never exceeds $0.05$ in magnitude, its $95\%$ CI coverage falls in $[0.93, 0.96]$, and its variance is bounded above by $0.08$ across all configurations.  Tables~\ref{tab:sim-eff-01}--\ref{tab:sim-eff-int} report the relative efficiency $\widehat\var\left(\hat\tau^{\ba}_1\right)/\widehat\var\left(\hat\xi^{\ba}\right)$ for each parameter.

Some patterns recur.  The Naive and EP estimators carry substantial bias that grows with the severity of misclassification of the exposures entering each contrast, and persists across all proportions.  Bias is largest in magnitude under ``ME in One Exposure'' for the joint effect $\tau^{(1,1)}$ and the first-exposure effect $\tau^{(1,0)}$, both of which involve the heavily misclassified first exposure, and largest in magnitude under ``ME in Both Exposures'' for the second-exposure effect $\tau^{(0,1)}$ and the interaction $\Delta$; EP coverage falls to roughly $10\%$ in the worst cases.  The Gold and CV estimators are nearly unbiased across all measurement-error configurations and all model-specification scenarios, consistent with double robustness. However, both undercover at small $\eta$, most notably for the joint and marginal effects, with coverage as low as roughly $92\%$ at $\eta=0.2$ and moving towards the nominal level by $\eta=0.8$ (between roughly $94$ and $95\%$ across the four causal contrasts).  The CV estimator nearly always improves on the Gold estimator's variance, with maximum gains of approximately $4\%$ for $\tau^{(0,1)}$, $4\%$ for $\tau^{(1,0)}$, $5\%$ for $\tau^{(1,1)}$, and $3\%$ for $\Delta$.

\begin{sidewaystable}[p]
\centering
\footnotesize
\setlength{\tabcolsep}{5pt}
\renewcommand{\arraystretch}{0.95}
\begin{tabular}{llrrrrrrrrrrrr}
\toprule
 &  & \multicolumn{4}{c}{Correct} & \multicolumn{4}{c}{PS Misspec.} & \multicolumn{4}{c}{OR Misspec.} \\
\cmidrule(lr){3-6} \cmidrule(lr){7-10} \cmidrule(lr){11-14}
ME configuration & $\eta$ & Naive & EP & Gold & CV & Naive & EP & Gold & CV & Naive & EP & Gold & CV \\
\midrule
ME in One Exposure & $0.2$ & -0.18 & -0.23 & -0.00 & -0.00 & -0.24 & -0.28 & -0.01 & -0.01 & -0.24 & -0.29 & -0.02 & -0.01 \\
 & $0.3$ & -0.15 & -0.24 & -0.00 & 0.00 & -0.21 & -0.28 & -0.00 & 0.00 & -0.21 & -0.29 & -0.01 & -0.00 \\
 & $0.4$ & -0.13 & -0.24 & -0.00 & 0.00 & -0.19 & -0.28 & -0.00 & 0.00 & -0.18 & -0.29 & -0.00 & 0.00 \\
 & $0.5$ & -0.10 & -0.25 & 0.00 & 0.00 & -0.16 & -0.28 & 0.00 & 0.00 & -0.15 & -0.29 & -0.00 & 0.00 \\
 & $0.6$ & -0.08 & -0.25 & 0.01 & 0.01 & -0.13 & -0.27 & 0.00 & 0.01 & -0.12 & -0.30 & 0.00 & 0.01 \\
 & $0.7$ & -0.06 & -0.26 & -0.00 & -0.00 & -0.10 & -0.28 & 0.00 & 0.00 & -0.10 & -0.31 & -0.00 & -0.00 \\
 & $0.8$ & -0.04 & -0.27 & -0.00 & -0.00 & -0.07 & -0.26 & -0.00 & 0.00 & -0.06 & -0.30 & 0.00 & 0.01 \\
\midrule
ME in Both Exposures & $0.2$ & -0.41 & -0.51 & 0.01 & 0.00 & -0.45 & -0.53 & -0.00 & -0.00 & -0.45 & -0.53 & -0.01 & -0.01 \\
 & $0.3$ & -0.35 & -0.51 & -0.00 & -0.00 & -0.40 & -0.52 & 0.00 & 0.00 & -0.39 & -0.53 & -0.01 & -0.01 \\
 & $0.4$ & -0.30 & -0.51 & 0.01 & 0.01 & -0.35 & -0.52 & -0.01 & -0.01 & -0.34 & -0.53 & -0.01 & -0.00 \\
 & $0.5$ & -0.25 & -0.52 & -0.00 & -0.00 & -0.30 & -0.52 & -0.00 & -0.00 & -0.30 & -0.53 & -0.01 & -0.01 \\
 & $0.6$ & -0.19 & -0.50 & 0.00 & 0.00 & -0.24 & -0.51 & 0.00 & 0.00 & -0.23 & -0.54 & 0.00 & 0.00 \\
 & $0.7$ & -0.15 & -0.52 & -0.00 & -0.00 & -0.19 & -0.52 & -0.00 & -0.00 & -0.18 & -0.53 & -0.00 & -0.00 \\
 & $0.8$ & -0.10 & -0.51 & -0.00 & -0.00 & -0.13 & -0.52 & -0.00 & -0.00 & -0.12 & -0.53 & -0.00 & -0.00 \\
\midrule
Low Specificity & $0.2$ & -0.23 & -0.30 & -0.00 & -0.00 & -0.28 & -0.33 & -0.00 & -0.00 & -0.26 & -0.33 & -0.02 & -0.01 \\
 & $0.3$ & -0.21 & -0.31 & -0.01 & -0.01 & -0.25 & -0.33 & 0.00 & 0.00 & -0.22 & -0.33 & -0.01 & -0.00 \\
 & $0.4$ & -0.17 & -0.31 & -0.00 & -0.00 & -0.22 & -0.33 & 0.00 & 0.00 & -0.20 & -0.34 & -0.00 & 0.00 \\
 & $0.5$ & -0.14 & -0.31 & 0.00 & 0.00 & -0.19 & -0.33 & -0.01 & -0.01 & -0.16 & -0.34 & -0.00 & 0.00 \\
 & $0.6$ & -0.11 & -0.32 & 0.00 & 0.00 & -0.15 & -0.33 & -0.00 & -0.00 & -0.13 & -0.34 & -0.01 & -0.00 \\
 & $0.7$ & -0.09 & -0.34 & -0.00 & -0.00 & -0.12 & -0.32 & -0.00 & -0.00 & -0.09 & -0.34 & -0.00 & 0.00 \\
 & $0.8$ & -0.05 & -0.33 & 0.00 & 0.00 & -0.09 & -0.33 & -0.01 & -0.01 & -0.06 & -0.36 & -0.00 & -0.00 \\
\midrule
Low Sensitivity & $0.2$ & -0.30 & -0.37 & 0.01 & 0.01 & -0.32 & -0.38 & 0.01 & 0.01 & -0.34 & -0.39 & -0.02 & -0.01 \\
 & $0.3$ & -0.26 & -0.36 & 0.01 & 0.02 & -0.28 & -0.38 & -0.00 & -0.00 & -0.31 & -0.39 & -0.01 & -0.00 \\
 & $0.4$ & -0.22 & -0.36 & 0.00 & 0.01 & -0.24 & -0.37 & -0.01 & -0.00 & -0.27 & -0.39 & -0.00 & 0.00 \\
 & $0.5$ & -0.19 & -0.36 & -0.00 & 0.00 & -0.20 & -0.37 & -0.00 & 0.00 & -0.23 & -0.39 & -0.01 & -0.00 \\
 & $0.6$ & -0.15 & -0.36 & 0.01 & 0.01 & -0.17 & -0.37 & -0.00 & -0.00 & -0.19 & -0.39 & 0.00 & 0.01 \\
 & $0.7$ & -0.11 & -0.35 & 0.00 & 0.01 & -0.12 & -0.36 & -0.00 & 0.00 & -0.15 & -0.38 & -0.00 & 0.00 \\
 & $0.8$ & -0.07 & -0.35 & 0.00 & 0.01 & -0.08 & -0.36 & -0.00 & 0.00 & -0.11 & -0.40 & -0.01 & -0.00 \\
\midrule
Mixed Sens/Spec & $0.2$ & -0.27 & -0.33 & 0.01 & 0.00 & -0.33 & -0.37 & -0.00 & 0.00 & -0.31 & -0.37 & -0.03 & -0.01 \\
 & $0.3$ & -0.24 & -0.33 & -0.01 & -0.01 & -0.31 & -0.37 & -0.01 & -0.00 & -0.28 & -0.37 & -0.01 & -0.00 \\
 & $0.4$ & -0.20 & -0.33 & 0.01 & 0.01 & -0.28 & -0.37 & 0.00 & 0.01 & -0.25 & -0.38 & -0.01 & -0.00 \\
 & $0.5$ & -0.17 & -0.34 & 0.00 & 0.00 & -0.25 & -0.36 & -0.00 & 0.00 & -0.21 & -0.38 & -0.00 & 0.00 \\
 & $0.6$ & -0.14 & -0.35 & 0.00 & 0.00 & -0.21 & -0.36 & -0.00 & 0.00 & -0.17 & -0.38 & -0.00 & 0.00 \\
 & $0.7$ & -0.12 & -0.37 & -0.01 & -0.01 & -0.17 & -0.37 & 0.00 & 0.00 & -0.13 & -0.38 & 0.00 & 0.00 \\
 & $0.8$ & -0.08 & -0.37 & -0.00 & -0.00 & -0.12 & -0.36 & 0.00 & 0.00 & -0.09 & -0.39 & -0.00 & -0.00 \\
\bottomrule
\end{tabular}
\caption{Bias for $\tau^{(0,1)}$ across measurement-error configurations (rows) and gold-standard proportions $\eta$ (rows within each configuration), separately for each model-specification scenario (column groups: Correct, PS Misspec., OR Misspec.).  Estimators within each scenario are Naive, EP (error-prone), Gold, and CV (control variate); the Oracle benchmark is omitted (see Section~\ref{sec:supp-simresults}).}
\label{tab:sim-bias-01}
\end{sidewaystable}

\begin{sidewaystable}[p]
\centering
\footnotesize
\setlength{\tabcolsep}{5pt}
\renewcommand{\arraystretch}{0.95}
\begin{tabular}{llrrrrrrrrrrrr}
\toprule
 &  & \multicolumn{4}{c}{Correct} & \multicolumn{4}{c}{PS Misspec.} & \multicolumn{4}{c}{OR Misspec.} \\
\cmidrule(lr){3-6} \cmidrule(lr){7-10} \cmidrule(lr){11-14}
ME configuration & $\eta$ & Naive & EP & Gold & CV & Naive & EP & Gold & CV & Naive & EP & Gold & CV \\
\midrule
ME in One Exposure & $0.2$ & 0.46 & 0.45 & 0.00 & 0.01 & 0.36 & 0.38 & 0.00 & 0.00 & 0.09 & 0.14 & 0.04 & 0.04 \\
 & $0.3$ & 0.43 & 0.41 & 0.01 & 0.01 & 0.31 & 0.34 & -0.00 & -0.00 & 0.07 & 0.14 & 0.03 & 0.03 \\
 & $0.4$ & 0.39 & 0.38 & 0.00 & 0.01 & 0.26 & 0.31 & -0.00 & -0.00 & 0.05 & 0.14 & 0.01 & 0.01 \\
 & $0.5$ & 0.35 & 0.34 & -0.00 & -0.00 & 0.22 & 0.29 & -0.01 & -0.01 & 0.04 & 0.13 & 0.00 & 0.00 \\
 & $0.6$ & 0.32 & 0.32 & 0.00 & 0.00 & 0.18 & 0.26 & 0.00 & 0.00 & 0.03 & 0.14 & -0.02 & -0.01 \\
 & $0.7$ & 0.26 & 0.27 & 0.00 & -0.00 & 0.13 & 0.22 & -0.00 & -0.01 & 0.01 & 0.13 & -0.02 & -0.02 \\
 & $0.8$ & 0.21 & 0.25 & 0.00 & -0.00 & 0.09 & 0.21 & -0.00 & -0.00 & -0.01 & 0.14 & -0.03 & -0.03 \\
\midrule
ME in Both Exposures & $0.2$ & 0.42 & 0.38 & 0.01 & 0.01 & 0.31 & 0.34 & 0.00 & 0.00 & 0.04 & 0.09 & 0.03 & 0.03 \\
 & $0.3$ & 0.41 & 0.36 & 0.00 & 0.00 & 0.27 & 0.30 & 0.00 & 0.00 & 0.04 & 0.09 & 0.02 & 0.02 \\
 & $0.4$ & 0.39 & 0.33 & -0.00 & -0.00 & 0.23 & 0.27 & -0.00 & -0.00 & 0.03 & 0.09 & 0.01 & 0.01 \\
 & $0.5$ & 0.36 & 0.29 & -0.00 & -0.00 & 0.19 & 0.24 & -0.00 & -0.01 & 0.02 & 0.09 & -0.00 & -0.00 \\
 & $0.6$ & 0.33 & 0.27 & 0.01 & 0.00 & 0.16 & 0.21 & 0.00 & 0.00 & 0.02 & 0.09 & -0.00 & -0.00 \\
 & $0.7$ & 0.29 & 0.24 & -0.00 & -0.00 & 0.11 & 0.18 & -0.00 & -0.00 & 0.00 & 0.09 & -0.03 & -0.03 \\
 & $0.8$ & 0.24 & 0.21 & -0.00 & -0.00 & 0.08 & 0.14 & -0.00 & -0.00 & -0.00 & 0.11 & -0.03 & -0.03 \\
\midrule
Low Specificity & $0.2$ & 0.33 & 0.31 & 0.01 & 0.01 & 0.19 & 0.20 & -0.00 & -0.00 & 0.04 & 0.06 & 0.03 & 0.03 \\
 & $0.3$ & 0.31 & 0.28 & -0.01 & -0.02 & 0.17 & 0.19 & -0.01 & -0.00 & 0.03 & 0.06 & 0.00 & 0.01 \\
 & $0.4$ & 0.29 & 0.27 & -0.00 & -0.01 & 0.14 & 0.17 & -0.00 & 0.00 & 0.02 & 0.05 & 0.01 & 0.02 \\
 & $0.5$ & 0.27 & 0.25 & 0.01 & 0.00 & 0.11 & 0.14 & -0.00 & -0.00 & 0.02 & 0.06 & 0.00 & 0.00 \\
 & $0.6$ & 0.24 & 0.22 & -0.01 & -0.01 & 0.09 & 0.13 & 0.00 & 0.00 & 0.01 & 0.06 & -0.01 & -0.01 \\
 & $0.7$ & 0.21 & 0.19 & -0.00 & -0.01 & 0.07 & 0.12 & -0.00 & -0.00 & -0.00 & 0.05 & -0.03 & -0.03 \\
 & $0.8$ & 0.16 & 0.19 & -0.01 & -0.01 & 0.04 & 0.09 & -0.00 & -0.00 & -0.01 & 0.05 & -0.02 & -0.02 \\
\midrule
Low Sensitivity & $0.2$ & 0.23 & 0.21 & 0.01 & 0.02 & 0.17 & 0.18 & 0.01 & 0.01 & -0.03 & -0.00 & 0.04 & 0.03 \\
 & $0.3$ & 0.21 & 0.19 & 0.01 & 0.02 & 0.15 & 0.15 & -0.00 & -0.01 & -0.04 & -0.00 & 0.02 & 0.01 \\
 & $0.4$ & 0.20 & 0.18 & -0.00 & 0.00 & 0.13 & 0.13 & -0.01 & -0.01 & -0.04 & -0.00 & 0.01 & 0.01 \\
 & $0.5$ & 0.19 & 0.16 & 0.01 & 0.01 & 0.11 & 0.12 & 0.00 & -0.00 & -0.05 & 0.00 & -0.01 & -0.01 \\
 & $0.6$ & 0.17 & 0.15 & 0.01 & 0.01 & 0.09 & 0.10 & -0.00 & -0.00 & -0.05 & -0.00 & -0.01 & -0.01 \\
 & $0.7$ & 0.14 & 0.13 & 0.00 & 0.00 & 0.07 & 0.07 & 0.00 & 0.00 & -0.05 & -0.01 & -0.02 & -0.02 \\
 & $0.8$ & 0.10 & 0.10 & -0.00 & -0.00 & 0.05 & 0.07 & 0.01 & 0.00 & -0.06 & -0.01 & -0.04 & -0.04 \\
\midrule
Mixed Sens/Spec & $0.2$ & 0.27 & 0.27 & 0.00 & 0.01 & 0.20 & 0.21 & -0.00 & -0.00 & 0.01 & 0.04 & 0.04 & 0.03 \\
 & $0.3$ & 0.25 & 0.25 & -0.00 & 0.00 & 0.17 & 0.20 & -0.00 & -0.00 & -0.01 & 0.04 & 0.02 & 0.01 \\
 & $0.4$ & 0.23 & 0.24 & 0.02 & 0.02 & 0.15 & 0.17 & 0.00 & 0.00 & -0.01 & 0.05 & 0.02 & 0.02 \\
 & $0.5$ & 0.20 & 0.22 & -0.00 & -0.00 & 0.13 & 0.16 & -0.00 & -0.00 & -0.03 & 0.04 & -0.01 & -0.01 \\
 & $0.6$ & 0.16 & 0.20 & -0.00 & -0.00 & 0.10 & 0.13 & 0.00 & -0.00 & -0.03 & 0.04 & -0.02 & -0.02 \\
 & $0.7$ & 0.13 & 0.17 & -0.00 & -0.00 & 0.08 & 0.11 & 0.00 & 0.00 & -0.03 & 0.04 & -0.02 & -0.02 \\
 & $0.8$ & 0.09 & 0.15 & 0.01 & 0.01 & 0.05 & 0.10 & 0.00 & 0.00 & -0.04 & 0.04 & -0.03 & -0.03 \\
\bottomrule
\end{tabular}
\caption{Bias for $\tau^{(1,0)}$ across measurement-error configurations (rows) and gold-standard proportions $\eta$ (rows within each configuration), separately for each model-specification scenario (column groups: Correct, PS Misspec., OR Misspec.).  Estimators within each scenario are Naive, EP (error-prone), Gold, and CV (control variate); the Oracle benchmark is omitted (see Section~\ref{sec:supp-simresults}).}
\label{tab:sim-bias-10}
\end{sidewaystable}

\begin{sidewaystable}[p]
\centering
\footnotesize
\setlength{\tabcolsep}{5pt}
\renewcommand{\arraystretch}{0.95}
\begin{tabular}{llrrrrrrrrrrrr}
\toprule
 &  & \multicolumn{4}{c}{Correct} & \multicolumn{4}{c}{PS Misspec.} & \multicolumn{4}{c}{OR Misspec.} \\
\cmidrule(lr){3-6} \cmidrule(lr){7-10} \cmidrule(lr){11-14}
ME configuration & $\eta$ & Naive & EP & Gold & CV & Naive & EP & Gold & CV & Naive & EP & Gold & CV \\
\midrule
ME in One Exposure & $0.2$ & 0.50 & 0.53 & 0.01 & 0.00 & 0.40 & 0.44 & -0.00 & 0.01 & 0.12 & 0.18 & 0.04 & 0.05 \\
 & $0.3$ & 0.45 & 0.49 & -0.00 & -0.01 & 0.35 & 0.41 & 0.00 & 0.01 & 0.09 & 0.17 & 0.02 & 0.03 \\
 & $0.4$ & 0.40 & 0.45 & 0.00 & -0.00 & 0.30 & 0.38 & -0.01 & -0.00 & 0.07 & 0.17 & 0.01 & 0.01 \\
 & $0.5$ & 0.34 & 0.42 & -0.00 & -0.01 & 0.25 & 0.34 & -0.00 & 0.00 & 0.04 & 0.16 & -0.00 & -0.00 \\
 & $0.6$ & 0.28 & 0.37 & 0.00 & -0.00 & 0.20 & 0.32 & 0.01 & 0.01 & 0.02 & 0.15 & -0.01 & -0.01 \\
 & $0.7$ & 0.22 & 0.34 & -0.00 & -0.00 & 0.15 & 0.28 & -0.00 & -0.00 & 0.00 & 0.15 & -0.02 & -0.02 \\
 & $0.8$ & 0.15 & 0.30 & 0.00 & -0.00 & 0.10 & 0.26 & -0.00 & -0.00 & -0.01 & 0.16 & -0.03 & -0.03 \\
\midrule
ME in Both Exposures & $0.2$ & 0.29 & 0.29 & 0.01 & 0.01 & 0.23 & 0.23 & -0.00 & -0.00 & -0.05 & -0.02 & 0.04 & 0.04 \\
 & $0.3$ & 0.26 & 0.26 & 0.00 & -0.00 & 0.20 & 0.20 & -0.00 & -0.00 & -0.06 & -0.02 & 0.01 & 0.01 \\
 & $0.4$ & 0.23 & 0.23 & 0.00 & -0.00 & 0.17 & 0.17 & -0.01 & -0.01 & -0.07 & -0.03 & 0.00 & 0.00 \\
 & $0.5$ & 0.19 & 0.19 & -0.00 & -0.01 & 0.15 & 0.14 & 0.00 & 0.00 & -0.07 & -0.02 & -0.00 & -0.00 \\
 & $0.6$ & 0.16 & 0.17 & -0.00 & -0.00 & 0.12 & 0.11 & -0.00 & 0.00 & -0.07 & -0.02 & -0.01 & -0.01 \\
 & $0.7$ & 0.12 & 0.14 & -0.00 & -0.00 & 0.09 & 0.09 & -0.01 & -0.00 & -0.07 & -0.01 & -0.03 & -0.02 \\
 & $0.8$ & 0.09 & 0.11 & -0.00 & -0.00 & 0.06 & 0.05 & -0.00 & -0.00 & -0.07 & -0.02 & -0.03 & -0.03 \\
\midrule
Low Specificity & $0.2$ & 0.30 & 0.30 & -0.00 & -0.02 & 0.18 & 0.18 & 0.00 & 0.02 & 0.02 & 0.02 & 0.04 & 0.04 \\
 & $0.3$ & 0.26 & 0.26 & -0.02 & -0.03 & 0.16 & 0.16 & 0.00 & 0.01 & 0.02 & 0.02 & 0.01 & 0.02 \\
 & $0.4$ & 0.24 & 0.25 & -0.00 & -0.01 & 0.14 & 0.14 & 0.00 & 0.01 & 0.01 & 0.01 & 0.01 & 0.01 \\
 & $0.5$ & 0.22 & 0.22 & 0.00 & -0.00 & 0.11 & 0.12 & -0.00 & -0.00 & 0.00 & 0.01 & 0.00 & 0.00 \\
 & $0.6$ & 0.18 & 0.20 & -0.00 & -0.01 & 0.10 & 0.11 & -0.00 & -0.00 & -0.01 & 0.01 & -0.02 & -0.02 \\
 & $0.7$ & 0.15 & 0.16 & 0.00 & -0.01 & 0.08 & 0.09 & 0.00 & 0.00 & -0.01 & 0.01 & -0.03 & -0.03 \\
 & $0.8$ & 0.11 & 0.15 & 0.00 & -0.00 & 0.05 & 0.07 & -0.00 & -0.00 & -0.02 & 0.00 & -0.03 & -0.03 \\
\midrule
Low Sensitivity & $0.2$ & 0.10 & 0.12 & 0.00 & 0.01 & 0.11 & 0.11 & 0.00 & 0.00 & -0.12 & -0.10 & 0.03 & 0.02 \\
 & $0.3$ & 0.08 & 0.10 & 0.00 & 0.01 & 0.10 & 0.09 & -0.01 & -0.01 & -0.13 & -0.10 & 0.02 & 0.01 \\
 & $0.4$ & 0.07 & 0.08 & 0.00 & 0.00 & 0.09 & 0.07 & -0.00 & -0.00 & -0.13 & -0.11 & 0.01 & 0.00 \\
 & $0.5$ & 0.05 & 0.07 & -0.00 & 0.00 & 0.08 & 0.05 & -0.00 & -0.00 & -0.13 & -0.10 & -0.01 & -0.01 \\
 & $0.6$ & 0.04 & 0.05 & 0.00 & 0.01 & 0.06 & 0.03 & 0.00 & -0.00 & -0.12 & -0.10 & -0.01 & -0.01 \\
 & $0.7$ & 0.02 & 0.04 & -0.00 & -0.00 & 0.05 & 0.01 & 0.00 & 0.00 & -0.11 & -0.10 & -0.02 & -0.02 \\
 & $0.8$ & 0.02 & 0.01 & 0.00 & 0.00 & 0.03 & -0.00 & 0.00 & 0.00 & -0.10 & -0.11 & -0.03 & -0.03 \\
\midrule
Mixed Sens/Spec & $0.2$ & 0.19 & 0.19 & 0.00 & 0.00 & 0.16 & 0.15 & 0.00 & 0.00 & -0.04 & -0.03 & 0.03 & 0.02 \\
 & $0.3$ & 0.17 & 0.17 & -0.00 & -0.01 & 0.14 & 0.13 & -0.00 & -0.00 & -0.05 & -0.03 & 0.02 & 0.01 \\
 & $0.4$ & 0.15 & 0.16 & 0.01 & 0.01 & 0.12 & 0.11 & 0.00 & 0.00 & -0.06 & -0.04 & 0.01 & 0.00 \\
 & $0.5$ & 0.12 & 0.13 & -0.00 & -0.00 & 0.10 & 0.10 & 0.00 & 0.00 & -0.06 & -0.04 & -0.00 & -0.01 \\
 & $0.6$ & 0.10 & 0.12 & 0.00 & 0.00 & 0.08 & 0.08 & -0.00 & -0.00 & -0.06 & -0.04 & -0.02 & -0.02 \\
 & $0.7$ & 0.07 & 0.08 & -0.01 & -0.01 & 0.06 & 0.05 & 0.00 & 0.00 & -0.06 & -0.05 & -0.02 & -0.02 \\
 & $0.8$ & 0.05 & 0.06 & -0.01 & -0.01 & 0.04 & 0.03 & -0.00 & -0.00 & -0.05 & -0.06 & -0.03 & -0.03 \\
\bottomrule
\end{tabular}
\caption{Bias for $\tau^{(1,1)}$ across measurement-error configurations (rows) and gold-standard proportions $\eta$ (rows within each configuration), separately for each model-specification scenario (column groups: Correct, PS Misspec., OR Misspec.).  Estimators within each scenario are Naive, EP (error-prone), Gold, and CV (control variate); the Oracle benchmark is omitted (see Section~\ref{sec:supp-simresults}).}
\label{tab:sim-bias-11}
\end{sidewaystable}

\begin{sidewaystable}[p]
\centering
\footnotesize
\setlength{\tabcolsep}{5pt}
\renewcommand{\arraystretch}{0.95}
\begin{tabular}{llrrrrrrrrrrrr}
\toprule
 &  & \multicolumn{4}{c}{Correct} & \multicolumn{4}{c}{PS Misspec.} & \multicolumn{4}{c}{OR Misspec.} \\
\cmidrule(lr){3-6} \cmidrule(lr){7-10} \cmidrule(lr){11-14}
ME configuration & $\eta$ & Naive & EP & Gold & CV & Naive & EP & Gold & CV & Naive & EP & Gold & CV \\
\midrule
ME in One Exposure & $0.2$ & 0.22 & 0.31 & 0.00 & 0.00 & 0.28 & 0.34 & 0.01 & 0.01 & 0.27 & 0.33 & 0.02 & 0.02 \\
 & $0.3$ & 0.17 & 0.31 & -0.01 & -0.02 & 0.25 & 0.34 & 0.01 & 0.01 & 0.23 & 0.32 & 0.01 & 0.01 \\
 & $0.4$ & 0.13 & 0.31 & 0.00 & -0.00 & 0.22 & 0.34 & -0.00 & -0.00 & 0.20 & 0.32 & 0.01 & 0.00 \\
 & $0.5$ & 0.09 & 0.32 & -0.00 & -0.01 & 0.19 & 0.33 & 0.00 & 0.00 & 0.16 & 0.32 & -0.00 & -0.00 \\
 & $0.6$ & 0.04 & 0.30 & -0.01 & -0.01 & 0.15 & 0.33 & -0.00 & -0.00 & 0.12 & 0.31 & -0.00 & -0.00 \\
 & $0.7$ & 0.01 & 0.33 & 0.00 & 0.00 & 0.12 & 0.35 & 0.00 & 0.00 & 0.09 & 0.33 & 0.00 & -0.00 \\
 & $0.8$ & -0.02 & 0.32 & 0.00 & 0.00 & 0.08 & 0.31 & 0.00 & -0.00 & 0.05 & 0.31 & -0.00 & -0.00 \\
\midrule
ME in Both Exposures & $0.2$ & 0.28 & 0.41 & -0.00 & 0.00 & 0.37 & 0.42 & 0.00 & 0.00 & 0.35 & 0.42 & 0.02 & 0.02 \\
 & $0.3$ & 0.21 & 0.41 & 0.01 & 0.01 & 0.34 & 0.43 & -0.00 & -0.00 & 0.29 & 0.41 & 0.01 & 0.01 \\
 & $0.4$ & 0.14 & 0.41 & -0.00 & -0.01 & 0.30 & 0.42 & 0.01 & 0.01 & 0.25 & 0.41 & 0.00 & 0.00 \\
 & $0.5$ & 0.09 & 0.42 & 0.00 & 0.00 & 0.26 & 0.42 & 0.01 & 0.01 & 0.20 & 0.42 & 0.01 & 0.01 \\
 & $0.6$ & 0.02 & 0.40 & -0.01 & -0.01 & 0.21 & 0.41 & -0.00 & -0.00 & 0.15 & 0.43 & -0.01 & -0.01 \\
 & $0.7$ & -0.02 & 0.42 & 0.00 & 0.00 & 0.17 & 0.43 & 0.00 & 0.00 & 0.11 & 0.43 & 0.01 & 0.01 \\
 & $0.8$ & -0.05 & 0.42 & 0.00 & 0.00 & 0.11 & 0.43 & -0.00 & -0.00 & 0.06 & 0.40 & -0.00 & -0.00 \\
\midrule
Low Specificity & $0.2$ & 0.20 & 0.28 & -0.01 & -0.02 & 0.27 & 0.31 & 0.01 & 0.01 & 0.24 & 0.29 & 0.03 & 0.02 \\
 & $0.3$ & 0.17 & 0.29 & 0.00 & -0.00 & 0.24 & 0.30 & 0.01 & 0.01 & 0.21 & 0.29 & 0.02 & 0.01 \\
 & $0.4$ & 0.13 & 0.28 & 0.01 & 0.00 & 0.22 & 0.31 & 0.00 & 0.00 & 0.18 & 0.30 & 0.00 & -0.01 \\
 & $0.5$ & 0.09 & 0.29 & -0.01 & -0.01 & 0.19 & 0.31 & 0.01 & 0.01 & 0.15 & 0.30 & -0.00 & -0.00 \\
 & $0.6$ & 0.05 & 0.29 & 0.00 & 0.00 & 0.16 & 0.30 & -0.01 & -0.00 & 0.11 & 0.29 & 0.00 & -0.00 \\
 & $0.7$ & 0.03 & 0.31 & 0.01 & 0.01 & 0.13 & 0.30 & 0.01 & 0.01 & 0.08 & 0.29 & -0.00 & -0.00 \\
 & $0.8$ & -0.00 & 0.30 & 0.00 & 0.00 & 0.09 & 0.32 & 0.01 & 0.01 & 0.05 & 0.30 & -0.00 & -0.00 \\
\midrule
Low Sensitivity & $0.2$ & 0.18 & 0.27 & -0.02 & -0.02 & 0.26 & 0.31 & -0.01 & -0.01 & 0.25 & 0.30 & 0.01 & 0.00 \\
 & $0.3$ & 0.13 & 0.27 & -0.03 & -0.03 & 0.24 & 0.32 & 0.00 & 0.00 & 0.21 & 0.29 & 0.01 & -0.00 \\
 & $0.4$ & 0.09 & 0.27 & 0.00 & -0.00 & 0.20 & 0.31 & 0.01 & 0.01 & 0.18 & 0.29 & 0.00 & -0.00 \\
 & $0.5$ & 0.05 & 0.27 & -0.01 & -0.01 & 0.16 & 0.30 & -0.00 & -0.00 & 0.15 & 0.28 & 0.01 & 0.00 \\
 & $0.6$ & 0.02 & 0.26 & -0.01 & -0.01 & 0.14 & 0.30 & 0.01 & 0.00 & 0.12 & 0.29 & -0.00 & -0.01 \\
 & $0.7$ & -0.01 & 0.26 & -0.01 & -0.01 & 0.10 & 0.29 & 0.00 & 0.00 & 0.09 & 0.29 & -0.00 & -0.00 \\
 & $0.8$ & -0.02 & 0.27 & 0.00 & 0.00 & 0.06 & 0.28 & -0.00 & -0.00 & 0.07 & 0.31 & 0.01 & 0.01 \\
\midrule
Mixed Sens/Spec & $0.2$ & 0.19 & 0.25 & -0.01 & -0.01 & 0.29 & 0.31 & 0.00 & 0.00 & 0.27 & 0.29 & 0.02 & 0.01 \\
 & $0.3$ & 0.16 & 0.25 & 0.01 & 0.01 & 0.27 & 0.30 & 0.01 & 0.00 & 0.24 & 0.30 & 0.01 & 0.00 \\
 & $0.4$ & 0.12 & 0.25 & -0.02 & -0.02 & 0.25 & 0.31 & -0.00 & -0.00 & 0.21 & 0.29 & 0.00 & -0.01 \\
 & $0.5$ & 0.10 & 0.26 & 0.00 & -0.00 & 0.22 & 0.29 & 0.00 & 0.00 & 0.18 & 0.30 & 0.01 & 0.01 \\
 & $0.6$ & 0.08 & 0.27 & 0.00 & 0.00 & 0.19 & 0.31 & 0.00 & 0.00 & 0.15 & 0.29 & 0.01 & 0.00 \\
 & $0.7$ & 0.06 & 0.27 & 0.00 & 0.00 & 0.16 & 0.31 & -0.00 & -0.00 & 0.11 & 0.29 & -0.00 & -0.00 \\
 & $0.8$ & 0.03 & 0.28 & -0.01 & -0.01 & 0.11 & 0.29 & -0.00 & -0.00 & 0.08 & 0.29 & 0.01 & 0.00 \\
\bottomrule
\end{tabular}
\caption{Bias for $\Delta$ (interaction) across measurement-error configurations (rows) and gold-standard proportions $\eta$ (rows within each configuration), separately for each model-specification scenario (column groups: Correct, PS Misspec., OR Misspec.).  Estimators within each scenario are Naive, EP (error-prone), Gold, and CV (control variate); the Oracle benchmark is omitted (see Section~\ref{sec:supp-simresults}).}
\label{tab:sim-bias-int}
\end{sidewaystable}

\begin{sidewaystable}[p]
\centering
\footnotesize
\setlength{\tabcolsep}{5pt}
\renewcommand{\arraystretch}{0.95}
\begin{tabular}{llrrrrrrrrrrrr}
\toprule
 &  & \multicolumn{4}{c}{Correct} & \multicolumn{4}{c}{PS Misspec.} & \multicolumn{4}{c}{OR Misspec.} \\
\cmidrule(lr){3-6} \cmidrule(lr){7-10} \cmidrule(lr){11-14}
ME configuration & $\eta$ & Naive & EP & Gold & CV & Naive & EP & Gold & CV & Naive & EP & Gold & CV \\
\midrule
ME in One Exposure & $0.2$ & 0.23 & 0.29 & 0.55 & 0.54 & 0.28 & 0.33 & 0.56 & 0.55 & 0.28 & 0.33 & 0.55 & 0.54 \\
 & $0.3$ & 0.21 & 0.30 & 0.40 & 0.40 & 0.26 & 0.33 & 0.40 & 0.40 & 0.26 & 0.34 & 0.41 & 0.40 \\
 & $0.4$ & 0.20 & 0.32 & 0.31 & 0.31 & 0.24 & 0.34 & 0.32 & 0.31 & 0.24 & 0.36 & 0.32 & 0.32 \\
 & $0.5$ & 0.18 & 0.34 & 0.27 & 0.26 & 0.22 & 0.37 & 0.26 & 0.26 & 0.21 & 0.38 & 0.27 & 0.27 \\
 & $0.6$ & 0.17 & 0.38 & 0.23 & 0.23 & 0.20 & 0.39 & 0.23 & 0.22 & 0.20 & 0.41 & 0.24 & 0.23 \\
 & $0.7$ & 0.17 & 0.44 & 0.20 & 0.20 & 0.18 & 0.45 & 0.20 & 0.19 & 0.19 & 0.46 & 0.20 & 0.20 \\
 & $0.8$ & 0.16 & 0.55 & 0.18 & 0.18 & 0.17 & 0.54 & 0.18 & 0.17 & 0.17 & 0.57 & 0.19 & 0.19 \\
\midrule
ME in Both Exposures & $0.2$ & 0.43 & 0.53 & 0.55 & 0.55 & 0.47 & 0.55 & 0.57 & 0.57 & 0.47 & 0.56 & 0.55 & 0.55 \\
 & $0.3$ & 0.38 & 0.54 & 0.41 & 0.41 & 0.43 & 0.55 & 0.40 & 0.40 & 0.42 & 0.56 & 0.41 & 0.41 \\
 & $0.4$ & 0.33 & 0.55 & 0.32 & 0.32 & 0.38 & 0.56 & 0.32 & 0.32 & 0.37 & 0.57 & 0.32 & 0.32 \\
 & $0.5$ & 0.29 & 0.57 & 0.26 & 0.26 & 0.34 & 0.57 & 0.27 & 0.27 & 0.33 & 0.58 & 0.27 & 0.27 \\
 & $0.6$ & 0.25 & 0.57 & 0.23 & 0.23 & 0.28 & 0.58 & 0.22 & 0.22 & 0.28 & 0.60 & 0.23 & 0.23 \\
 & $0.7$ & 0.21 & 0.62 & 0.20 & 0.20 & 0.25 & 0.62 & 0.20 & 0.20 & 0.24 & 0.64 & 0.21 & 0.21 \\
 & $0.8$ & 0.19 & 0.69 & 0.19 & 0.19 & 0.20 & 0.70 & 0.18 & 0.18 & 0.20 & 0.71 & 0.19 & 0.19 \\
\midrule
Low Specificity & $0.2$ & 0.29 & 0.35 & 0.55 & 0.54 & 0.32 & 0.38 & 0.55 & 0.55 & 0.31 & 0.38 & 0.55 & 0.54 \\
 & $0.3$ & 0.27 & 0.37 & 0.41 & 0.40 & 0.30 & 0.39 & 0.41 & 0.41 & 0.28 & 0.39 & 0.41 & 0.40 \\
 & $0.4$ & 0.24 & 0.38 & 0.32 & 0.31 & 0.27 & 0.40 & 0.32 & 0.32 & 0.26 & 0.41 & 0.32 & 0.32 \\
 & $0.5$ & 0.22 & 0.41 & 0.27 & 0.27 & 0.25 & 0.42 & 0.26 & 0.26 & 0.23 & 0.43 & 0.26 & 0.26 \\
 & $0.6$ & 0.20 & 0.45 & 0.23 & 0.23 & 0.22 & 0.46 & 0.22 & 0.22 & 0.21 & 0.46 & 0.23 & 0.23 \\
 & $0.7$ & 0.18 & 0.52 & 0.20 & 0.20 & 0.20 & 0.51 & 0.20 & 0.20 & 0.19 & 0.51 & 0.21 & 0.20 \\
 & $0.8$ & 0.17 & 0.62 & 0.19 & 0.19 & 0.18 & 0.64 & 0.18 & 0.18 & 0.18 & 0.63 & 0.19 & 0.19 \\
\midrule
Low Sensitivity & $0.2$ & 0.33 & 0.40 & 0.55 & 0.55 & 0.35 & 0.41 & 0.56 & 0.55 & 0.37 & 0.42 & 0.56 & 0.55 \\
 & $0.3$ & 0.29 & 0.40 & 0.40 & 0.39 & 0.32 & 0.41 & 0.41 & 0.40 & 0.34 & 0.43 & 0.40 & 0.40 \\
 & $0.4$ & 0.27 & 0.41 & 0.32 & 0.32 & 0.28 & 0.41 & 0.32 & 0.31 & 0.31 & 0.44 & 0.32 & 0.32 \\
 & $0.5$ & 0.24 & 0.42 & 0.26 & 0.26 & 0.25 & 0.42 & 0.26 & 0.26 & 0.27 & 0.44 & 0.26 & 0.26 \\
 & $0.6$ & 0.21 & 0.45 & 0.23 & 0.23 & 0.22 & 0.45 & 0.22 & 0.22 & 0.24 & 0.47 & 0.23 & 0.22 \\
 & $0.7$ & 0.19 & 0.49 & 0.20 & 0.20 & 0.19 & 0.49 & 0.20 & 0.19 & 0.22 & 0.51 & 0.21 & 0.20 \\
 & $0.8$ & 0.17 & 0.57 & 0.18 & 0.18 & 0.17 & 0.56 & 0.18 & 0.17 & 0.19 & 0.61 & 0.19 & 0.18 \\
\midrule
Mixed Sens/Spec & $0.2$ & 0.31 & 0.37 & 0.55 & 0.54 & 0.36 & 0.41 & 0.57 & 0.56 & 0.34 & 0.40 & 0.55 & 0.55 \\
 & $0.3$ & 0.28 & 0.37 & 0.40 & 0.40 & 0.34 & 0.40 & 0.42 & 0.41 & 0.32 & 0.41 & 0.41 & 0.40 \\
 & $0.4$ & 0.25 & 0.38 & 0.32 & 0.32 & 0.31 & 0.41 & 0.33 & 0.32 & 0.29 & 0.42 & 0.33 & 0.32 \\
 & $0.5$ & 0.23 & 0.40 & 0.27 & 0.27 & 0.28 & 0.42 & 0.26 & 0.26 & 0.26 & 0.44 & 0.27 & 0.27 \\
 & $0.6$ & 0.21 & 0.43 & 0.23 & 0.23 & 0.26 & 0.45 & 0.23 & 0.22 & 0.23 & 0.46 & 0.23 & 0.23 \\
 & $0.7$ & 0.19 & 0.48 & 0.20 & 0.20 & 0.22 & 0.49 & 0.20 & 0.19 & 0.20 & 0.50 & 0.21 & 0.20 \\
 & $0.8$ & 0.17 & 0.55 & 0.18 & 0.18 & 0.19 & 0.57 & 0.18 & 0.17 & 0.18 & 0.58 & 0.19 & 0.19 \\
\bottomrule
\end{tabular}
\caption{Root mean squared error (RMSE) for $\tau^{(0,1)}$ across measurement-error configurations (rows) and gold-standard proportions $\eta$ (rows within each configuration), separately for each model-specification scenario (column groups: Correct, PS Misspec., OR Misspec.).  Estimators within each scenario are Naive, EP (error-prone), Gold, and CV (control variate); the Oracle benchmark is omitted (see Section~\ref{sec:supp-simresults}).}
\label{tab:sim-rmse-01}
\end{sidewaystable}

\begin{sidewaystable}[p]
\centering
\footnotesize
\setlength{\tabcolsep}{5pt}
\renewcommand{\arraystretch}{0.95}
\begin{tabular}{llrrrrrrrrrrrr}
\toprule
 &  & \multicolumn{4}{c}{Correct} & \multicolumn{4}{c}{PS Misspec.} & \multicolumn{4}{c}{OR Misspec.} \\
\cmidrule(lr){3-6} \cmidrule(lr){7-10} \cmidrule(lr){11-14}
ME configuration & $\eta$ & Naive & EP & Gold & CV & Naive & EP & Gold & CV & Naive & EP & Gold & CV \\
\midrule
ME in One Exposure & $0.2$ & 0.49 & 0.48 & 0.78 & 0.77 & 0.38 & 0.42 & 0.47 & 0.47 & 0.19 & 0.23 & 0.77 & 0.77 \\
 & $0.3$ & 0.46 & 0.46 & 0.56 & 0.56 & 0.34 & 0.39 & 0.35 & 0.35 & 0.18 & 0.24 & 0.58 & 0.57 \\
 & $0.4$ & 0.43 & 0.43 & 0.45 & 0.45 & 0.30 & 0.38 & 0.28 & 0.28 & 0.18 & 0.25 & 0.45 & 0.45 \\
 & $0.5$ & 0.39 & 0.42 & 0.37 & 0.37 & 0.27 & 0.38 & 0.24 & 0.24 & 0.18 & 0.27 & 0.38 & 0.38 \\
 & $0.6$ & 0.37 & 0.43 & 0.32 & 0.32 & 0.23 & 0.39 & 0.21 & 0.21 & 0.18 & 0.32 & 0.32 & 0.32 \\
 & $0.7$ & 0.32 & 0.44 & 0.28 & 0.28 & 0.20 & 0.43 & 0.18 & 0.18 & 0.18 & 0.37 & 0.28 & 0.28 \\
 & $0.8$ & 0.28 & 0.52 & 0.25 & 0.25 & 0.18 & 0.55 & 0.17 & 0.17 & 0.19 & 0.48 & 0.26 & 0.26 \\
\midrule
ME in Both Exposures & $0.2$ & 0.45 & 0.42 & 0.77 & 0.77 & 0.34 & 0.38 & 0.48 & 0.48 & 0.16 & 0.19 & 0.76 & 0.76 \\
 & $0.3$ & 0.44 & 0.40 & 0.56 & 0.56 & 0.31 & 0.35 & 0.35 & 0.35 & 0.16 & 0.21 & 0.58 & 0.58 \\
 & $0.4$ & 0.42 & 0.39 & 0.46 & 0.45 & 0.28 & 0.34 & 0.28 & 0.28 & 0.16 & 0.23 & 0.46 & 0.46 \\
 & $0.5$ & 0.39 & 0.37 & 0.37 & 0.37 & 0.24 & 0.34 & 0.24 & 0.24 & 0.16 & 0.26 & 0.37 & 0.37 \\
 & $0.6$ & 0.38 & 0.39 & 0.32 & 0.32 & 0.22 & 0.36 & 0.20 & 0.20 & 0.17 & 0.29 & 0.33 & 0.33 \\
 & $0.7$ & 0.34 & 0.43 & 0.28 & 0.28 & 0.19 & 0.41 & 0.19 & 0.19 & 0.17 & 0.36 & 0.28 & 0.28 \\
 & $0.8$ & 0.30 & 0.51 & 0.25 & 0.25 & 0.17 & 0.51 & 0.17 & 0.17 & 0.18 & 0.48 & 0.25 & 0.25 \\
\midrule
Low Specificity & $0.2$ & 0.38 & 0.37 & 0.77 & 0.75 & 0.25 & 0.28 & 0.48 & 0.47 & 0.19 & 0.21 & 0.79 & 0.78 \\
 & $0.3$ & 0.36 & 0.36 & 0.57 & 0.57 & 0.24 & 0.29 & 0.35 & 0.35 & 0.19 & 0.23 & 0.57 & 0.57 \\
 & $0.4$ & 0.35 & 0.36 & 0.45 & 0.44 & 0.21 & 0.29 & 0.28 & 0.28 & 0.19 & 0.25 & 0.45 & 0.44 \\
 & $0.5$ & 0.33 & 0.37 & 0.37 & 0.37 & 0.19 & 0.32 & 0.24 & 0.23 & 0.19 & 0.28 & 0.37 & 0.37 \\
 & $0.6$ & 0.30 & 0.39 & 0.32 & 0.31 & 0.18 & 0.37 & 0.21 & 0.20 & 0.19 & 0.33 & 0.32 & 0.32 \\
 & $0.7$ & 0.28 & 0.44 & 0.28 & 0.27 & 0.17 & 0.44 & 0.19 & 0.19 & 0.19 & 0.41 & 0.28 & 0.28 \\
 & $0.8$ & 0.26 & 0.56 & 0.24 & 0.25 & 0.16 & 0.58 & 0.17 & 0.17 & 0.20 & 0.53 & 0.25 & 0.25 \\
\midrule
Low Sensitivity & $0.2$ & 0.28 & 0.28 & 0.77 & 0.76 & 0.22 & 0.24 & 0.47 & 0.47 & 0.16 & 0.17 & 0.78 & 0.76 \\
 & $0.3$ & 0.27 & 0.27 & 0.58 & 0.57 & 0.20 & 0.23 & 0.35 & 0.35 & 0.17 & 0.19 & 0.57 & 0.56 \\
 & $0.4$ & 0.27 & 0.28 & 0.45 & 0.44 & 0.19 & 0.24 & 0.29 & 0.28 & 0.18 & 0.21 & 0.45 & 0.44 \\
 & $0.5$ & 0.26 & 0.29 & 0.37 & 0.37 & 0.18 & 0.26 & 0.24 & 0.23 & 0.18 & 0.25 & 0.38 & 0.37 \\
 & $0.6$ & 0.25 & 0.33 & 0.32 & 0.31 & 0.17 & 0.29 & 0.20 & 0.20 & 0.19 & 0.29 & 0.32 & 0.32 \\
 & $0.7$ & 0.24 & 0.38 & 0.28 & 0.28 & 0.16 & 0.36 & 0.19 & 0.18 & 0.19 & 0.36 & 0.28 & 0.28 \\
 & $0.8$ & 0.23 & 0.49 & 0.25 & 0.25 & 0.16 & 0.46 & 0.17 & 0.17 & 0.20 & 0.47 & 0.25 & 0.25 \\
\midrule
Mixed Sens/Spec & $0.2$ & 0.34 & 0.35 & 0.76 & 0.75 & 0.26 & 0.29 & 0.48 & 0.47 & 0.20 & 0.23 & 0.78 & 0.77 \\
 & $0.3$ & 0.32 & 0.35 & 0.57 & 0.57 & 0.24 & 0.30 & 0.35 & 0.35 & 0.20 & 0.25 & 0.57 & 0.56 \\
 & $0.4$ & 0.31 & 0.37 & 0.46 & 0.45 & 0.22 & 0.30 & 0.28 & 0.28 & 0.21 & 0.28 & 0.45 & 0.44 \\
 & $0.5$ & 0.29 & 0.38 & 0.37 & 0.37 & 0.20 & 0.33 & 0.24 & 0.24 & 0.21 & 0.31 & 0.37 & 0.37 \\
 & $0.6$ & 0.26 & 0.41 & 0.32 & 0.32 & 0.19 & 0.37 & 0.21 & 0.20 & 0.21 & 0.36 & 0.32 & 0.31 \\
 & $0.7$ & 0.25 & 0.49 & 0.28 & 0.28 & 0.17 & 0.44 & 0.19 & 0.18 & 0.21 & 0.46 & 0.28 & 0.28 \\
 & $0.8$ & 0.23 & 0.62 & 0.25 & 0.25 & 0.16 & 0.60 & 0.17 & 0.17 & 0.21 & 0.61 & 0.25 & 0.25 \\
\bottomrule
\end{tabular}
\caption{Root mean squared error (RMSE) for $\tau^{(1,0)}$ across measurement-error configurations (rows) and gold-standard proportions $\eta$ (rows within each configuration), separately for each model-specification scenario (column groups: Correct, PS Misspec., OR Misspec.).  Estimators within each scenario are Naive, EP (error-prone), Gold, and CV (control variate); the Oracle benchmark is omitted (see Section~\ref{sec:supp-simresults}).}
\label{tab:sim-rmse-10}
\end{sidewaystable}

\begin{sidewaystable}[p]
\centering
\footnotesize
\setlength{\tabcolsep}{5pt}
\renewcommand{\arraystretch}{0.95}
\begin{tabular}{llrrrrrrrrrrrr}
\toprule
 &  & \multicolumn{4}{c}{Correct} & \multicolumn{4}{c}{PS Misspec.} & \multicolumn{4}{c}{OR Misspec.} \\
\cmidrule(lr){3-6} \cmidrule(lr){7-10} \cmidrule(lr){11-14}
ME configuration & $\eta$ & Naive & EP & Gold & CV & Naive & EP & Gold & CV & Naive & EP & Gold & CV \\
\midrule
ME in One Exposure & $0.2$ & 0.53 & 0.55 & 0.58 & 0.57 & 0.43 & 0.47 & 0.52 & 0.51 & 0.19 & 0.25 & 0.59 & 0.58 \\
 & $0.3$ & 0.48 & 0.52 & 0.43 & 0.42 & 0.38 & 0.45 & 0.38 & 0.37 & 0.18 & 0.25 & 0.42 & 0.42 \\
 & $0.4$ & 0.42 & 0.49 & 0.33 & 0.32 & 0.34 & 0.43 & 0.30 & 0.29 & 0.17 & 0.27 & 0.33 & 0.33 \\
 & $0.5$ & 0.37 & 0.48 & 0.28 & 0.27 & 0.29 & 0.42 & 0.25 & 0.24 & 0.16 & 0.29 & 0.27 & 0.27 \\
 & $0.6$ & 0.32 & 0.47 & 0.24 & 0.23 & 0.25 & 0.42 & 0.22 & 0.21 & 0.16 & 0.33 & 0.24 & 0.23 \\
 & $0.7$ & 0.27 & 0.49 & 0.21 & 0.20 & 0.21 & 0.46 & 0.19 & 0.19 & 0.15 & 0.39 & 0.20 & 0.20 \\
 & $0.8$ & 0.22 & 0.57 & 0.18 & 0.18 & 0.18 & 0.54 & 0.17 & 0.17 & 0.16 & 0.50 & 0.19 & 0.19 \\
\midrule
ME in Both Exposures & $0.2$ & 0.33 & 0.33 & 0.58 & 0.57 & 0.27 & 0.28 & 0.52 & 0.52 & 0.16 & 0.17 & 0.57 & 0.57 \\
 & $0.3$ & 0.30 & 0.32 & 0.43 & 0.42 & 0.25 & 0.27 & 0.38 & 0.38 & 0.16 & 0.18 & 0.43 & 0.43 \\
 & $0.4$ & 0.27 & 0.31 & 0.34 & 0.34 & 0.23 & 0.27 & 0.30 & 0.30 & 0.16 & 0.21 & 0.34 & 0.33 \\
 & $0.5$ & 0.25 & 0.31 & 0.27 & 0.27 & 0.21 & 0.28 & 0.26 & 0.25 & 0.17 & 0.24 & 0.28 & 0.28 \\
 & $0.6$ & 0.22 & 0.33 & 0.23 & 0.23 & 0.19 & 0.30 & 0.21 & 0.21 & 0.16 & 0.28 & 0.24 & 0.23 \\
 & $0.7$ & 0.20 & 0.39 & 0.21 & 0.21 & 0.17 & 0.37 & 0.19 & 0.19 & 0.17 & 0.35 & 0.21 & 0.21 \\
 & $0.8$ & 0.18 & 0.50 & 0.18 & 0.18 & 0.16 & 0.48 & 0.17 & 0.17 & 0.17 & 0.48 & 0.19 & 0.19 \\
\midrule
Low Specificity & $0.2$ & 0.34 & 0.35 & 0.59 & 0.58 & 0.24 & 0.26 & 0.52 & 0.51 & 0.16 & 0.18 & 0.59 & 0.58 \\
 & $0.3$ & 0.31 & 0.33 & 0.42 & 0.42 & 0.22 & 0.25 & 0.38 & 0.37 & 0.16 & 0.20 & 0.42 & 0.42 \\
 & $0.4$ & 0.29 & 0.33 & 0.33 & 0.33 & 0.21 & 0.26 & 0.30 & 0.30 & 0.16 & 0.22 & 0.33 & 0.33 \\
 & $0.5$ & 0.27 & 0.33 & 0.27 & 0.27 & 0.19 & 0.28 & 0.26 & 0.25 & 0.16 & 0.25 & 0.28 & 0.27 \\
 & $0.6$ & 0.24 & 0.35 & 0.23 & 0.23 & 0.18 & 0.32 & 0.22 & 0.22 & 0.16 & 0.30 & 0.23 & 0.23 \\
 & $0.7$ & 0.22 & 0.40 & 0.20 & 0.20 & 0.17 & 0.39 & 0.19 & 0.19 & 0.16 & 0.37 & 0.21 & 0.20 \\
 & $0.8$ & 0.19 & 0.51 & 0.18 & 0.18 & 0.16 & 0.53 & 0.18 & 0.18 & 0.16 & 0.49 & 0.19 & 0.18 \\
\midrule
Low Sensitivity & $0.2$ & 0.19 & 0.22 & 0.58 & 0.57 & 0.19 & 0.21 & 0.52 & 0.51 & 0.20 & 0.21 & 0.58 & 0.57 \\
 & $0.3$ & 0.18 & 0.23 & 0.43 & 0.42 & 0.18 & 0.22 & 0.38 & 0.37 & 0.20 & 0.22 & 0.42 & 0.41 \\
 & $0.4$ & 0.17 & 0.24 & 0.33 & 0.33 & 0.18 & 0.23 & 0.30 & 0.30 & 0.21 & 0.25 & 0.33 & 0.33 \\
 & $0.5$ & 0.17 & 0.28 & 0.28 & 0.27 & 0.17 & 0.27 & 0.25 & 0.24 & 0.20 & 0.29 & 0.28 & 0.27 \\
 & $0.6$ & 0.16 & 0.32 & 0.23 & 0.23 & 0.16 & 0.30 & 0.22 & 0.21 & 0.20 & 0.33 & 0.24 & 0.23 \\
 & $0.7$ & 0.16 & 0.40 & 0.20 & 0.20 & 0.16 & 0.38 & 0.19 & 0.19 & 0.19 & 0.41 & 0.21 & 0.21 \\
 & $0.8$ & 0.16 & 0.54 & 0.19 & 0.19 & 0.16 & 0.51 & 0.18 & 0.17 & 0.19 & 0.56 & 0.19 & 0.19 \\
\midrule
Mixed Sens/Spec & $0.2$ & 0.25 & 0.27 & 0.59 & 0.57 & 0.22 & 0.23 & 0.52 & 0.50 & 0.16 & 0.18 & 0.58 & 0.57 \\
 & $0.3$ & 0.23 & 0.26 & 0.42 & 0.41 & 0.20 & 0.23 & 0.39 & 0.38 & 0.17 & 0.20 & 0.43 & 0.42 \\
 & $0.4$ & 0.22 & 0.27 & 0.33 & 0.33 & 0.20 & 0.24 & 0.30 & 0.30 & 0.17 & 0.22 & 0.33 & 0.33 \\
 & $0.5$ & 0.20 & 0.28 & 0.28 & 0.27 & 0.19 & 0.26 & 0.25 & 0.25 & 0.17 & 0.25 & 0.27 & 0.27 \\
 & $0.6$ & 0.19 & 0.31 & 0.23 & 0.23 & 0.17 & 0.31 & 0.22 & 0.21 & 0.17 & 0.30 & 0.23 & 0.23 \\
 & $0.7$ & 0.17 & 0.37 & 0.21 & 0.20 & 0.17 & 0.37 & 0.19 & 0.19 & 0.17 & 0.37 & 0.21 & 0.20 \\
 & $0.8$ & 0.16 & 0.49 & 0.18 & 0.18 & 0.16 & 0.50 & 0.18 & 0.17 & 0.17 & 0.49 & 0.19 & 0.19 \\
\bottomrule
\end{tabular}
\caption{Root mean squared error (RMSE) for $\tau^{(1,1)}$ across measurement-error configurations (rows) and gold-standard proportions $\eta$ (rows within each configuration), separately for each model-specification scenario (column groups: Correct, PS Misspec., OR Misspec.).  Estimators within each scenario are Naive, EP (error-prone), Gold, and CV (control variate); the Oracle benchmark is omitted (see Section~\ref{sec:supp-simresults}).}
\label{tab:sim-rmse-11}
\end{sidewaystable}

\begin{sidewaystable}[p]
\centering
\footnotesize
\setlength{\tabcolsep}{5pt}
\renewcommand{\arraystretch}{0.95}
\begin{tabular}{llrrrrrrrrrrrr}
\toprule
 &  & \multicolumn{4}{c}{Correct} & \multicolumn{4}{c}{PS Misspec.} & \multicolumn{4}{c}{OR Misspec.} \\
\cmidrule(lr){3-6} \cmidrule(lr){7-10} \cmidrule(lr){11-14}
ME configuration & $\eta$ & Naive & EP & Gold & CV & Naive & EP & Gold & CV & Naive & EP & Gold & CV \\
\midrule
ME in One Exposure & $0.2$ & 0.31 & 0.40 & 0.91 & 0.91 & 0.35 & 0.41 & 0.73 & 0.73 & 0.35 & 0.41 & 0.92 & 0.91 \\
 & $0.3$ & 0.28 & 0.41 & 0.67 & 0.67 & 0.33 & 0.43 & 0.53 & 0.53 & 0.32 & 0.42 & 0.67 & 0.67 \\
 & $0.4$ & 0.26 & 0.43 & 0.54 & 0.53 & 0.30 & 0.45 & 0.42 & 0.42 & 0.30 & 0.44 & 0.53 & 0.53 \\
 & $0.5$ & 0.25 & 0.47 & 0.44 & 0.44 & 0.28 & 0.48 & 0.35 & 0.35 & 0.28 & 0.47 & 0.45 & 0.45 \\
 & $0.6$ & 0.25 & 0.51 & 0.39 & 0.39 & 0.26 & 0.52 & 0.30 & 0.30 & 0.26 & 0.52 & 0.39 & 0.39 \\
 & $0.7$ & 0.24 & 0.62 & 0.34 & 0.34 & 0.24 & 0.61 & 0.27 & 0.27 & 0.25 & 0.60 & 0.34 & 0.34 \\
 & $0.8$ & 0.25 & 0.76 & 0.31 & 0.31 & 0.22 & 0.75 & 0.24 & 0.24 & 0.26 & 0.76 & 0.31 & 0.31 \\
\midrule
ME in Both Exposures & $0.2$ & 0.35 & 0.47 & 0.90 & 0.90 & 0.42 & 0.48 & 0.75 & 0.75 & 0.41 & 0.48 & 0.90 & 0.90 \\
 & $0.3$ & 0.30 & 0.49 & 0.67 & 0.67 & 0.40 & 0.50 & 0.53 & 0.53 & 0.36 & 0.49 & 0.68 & 0.68 \\
 & $0.4$ & 0.26 & 0.51 & 0.54 & 0.54 & 0.36 & 0.51 & 0.43 & 0.42 & 0.33 & 0.51 & 0.54 & 0.54 \\
 & $0.5$ & 0.24 & 0.54 & 0.45 & 0.45 & 0.33 & 0.54 & 0.35 & 0.35 & 0.30 & 0.54 & 0.44 & 0.44 \\
 & $0.6$ & 0.23 & 0.57 & 0.39 & 0.39 & 0.29 & 0.57 & 0.30 & 0.30 & 0.27 & 0.59 & 0.40 & 0.40 \\
 & $0.7$ & 0.24 & 0.66 & 0.34 & 0.34 & 0.27 & 0.66 & 0.27 & 0.27 & 0.26 & 0.67 & 0.34 & 0.34 \\
 & $0.8$ & 0.25 & 0.79 & 0.31 & 0.31 & 0.24 & 0.79 & 0.24 & 0.24 & 0.25 & 0.80 & 0.31 & 0.31 \\
\midrule
Low Specificity & $0.2$ & 0.30 & 0.38 & 0.91 & 0.90 & 0.35 & 0.40 & 0.73 & 0.72 & 0.33 & 0.39 & 0.92 & 0.91 \\
 & $0.3$ & 0.29 & 0.40 & 0.68 & 0.67 & 0.32 & 0.41 & 0.54 & 0.54 & 0.31 & 0.40 & 0.68 & 0.67 \\
 & $0.4$ & 0.26 & 0.42 & 0.54 & 0.53 & 0.30 & 0.43 & 0.43 & 0.43 & 0.29 & 0.43 & 0.53 & 0.53 \\
 & $0.5$ & 0.25 & 0.45 & 0.44 & 0.44 & 0.29 & 0.47 & 0.35 & 0.35 & 0.28 & 0.46 & 0.45 & 0.44 \\
 & $0.6$ & 0.25 & 0.52 & 0.39 & 0.38 & 0.26 & 0.52 & 0.30 & 0.30 & 0.27 & 0.51 & 0.39 & 0.39 \\
 & $0.7$ & 0.25 & 0.61 & 0.34 & 0.34 & 0.24 & 0.60 & 0.27 & 0.27 & 0.26 & 0.61 & 0.34 & 0.34 \\
 & $0.8$ & 0.25 & 0.77 & 0.31 & 0.31 & 0.22 & 0.79 & 0.24 & 0.24 & 0.26 & 0.77 & 0.31 & 0.31 \\
\midrule
Low Sensitivity & $0.2$ & 0.29 & 0.38 & 0.91 & 0.91 & 0.33 & 0.40 & 0.73 & 0.72 & 0.33 & 0.39 & 0.91 & 0.90 \\
 & $0.3$ & 0.27 & 0.40 & 0.67 & 0.66 & 0.32 & 0.42 & 0.54 & 0.54 & 0.31 & 0.41 & 0.67 & 0.67 \\
 & $0.4$ & 0.26 & 0.42 & 0.54 & 0.53 & 0.29 & 0.43 & 0.43 & 0.43 & 0.30 & 0.44 & 0.53 & 0.53 \\
 & $0.5$ & 0.25 & 0.46 & 0.45 & 0.44 & 0.27 & 0.46 & 0.35 & 0.35 & 0.28 & 0.47 & 0.45 & 0.45 \\
 & $0.6$ & 0.24 & 0.52 & 0.39 & 0.38 & 0.25 & 0.51 & 0.30 & 0.29 & 0.27 & 0.54 & 0.39 & 0.39 \\
 & $0.7$ & 0.25 & 0.62 & 0.34 & 0.34 & 0.23 & 0.60 & 0.26 & 0.26 & 0.26 & 0.64 & 0.34 & 0.34 \\
 & $0.8$ & 0.26 & 0.81 & 0.31 & 0.31 & 0.22 & 0.76 & 0.24 & 0.24 & 0.26 & 0.83 & 0.31 & 0.31 \\
\midrule
Mixed Sens/Spec & $0.2$ & 0.31 & 0.37 & 0.89 & 0.88 & 0.36 & 0.40 & 0.74 & 0.74 & 0.36 & 0.40 & 0.92 & 0.91 \\
 & $0.3$ & 0.30 & 0.40 & 0.68 & 0.68 & 0.34 & 0.40 & 0.55 & 0.55 & 0.34 & 0.42 & 0.68 & 0.67 \\
 & $0.4$ & 0.28 & 0.43 & 0.54 & 0.54 & 0.33 & 0.44 & 0.42 & 0.42 & 0.32 & 0.44 & 0.54 & 0.53 \\
 & $0.5$ & 0.28 & 0.47 & 0.45 & 0.45 & 0.30 & 0.46 & 0.35 & 0.35 & 0.31 & 0.48 & 0.45 & 0.45 \\
 & $0.6$ & 0.27 & 0.53 & 0.39 & 0.38 & 0.29 & 0.53 & 0.30 & 0.30 & 0.29 & 0.55 & 0.38 & 0.38 \\
 & $0.7$ & 0.27 & 0.64 & 0.34 & 0.34 & 0.26 & 0.62 & 0.27 & 0.26 & 0.28 & 0.66 & 0.35 & 0.34 \\
 & $0.8$ & 0.26 & 0.81 & 0.31 & 0.31 & 0.24 & 0.79 & 0.24 & 0.24 & 0.28 & 0.84 & 0.31 & 0.31 \\
\bottomrule
\end{tabular}
\caption{Root mean squared error (RMSE) for $\Delta$ (interaction) across measurement-error configurations (rows) and gold-standard proportions $\eta$ (rows within each configuration), separately for each model-specification scenario (column groups: Correct, PS Misspec., OR Misspec.).  Estimators within each scenario are Naive, EP (error-prone), Gold, and CV (control variate); the Oracle benchmark is omitted (see Section~\ref{sec:supp-simresults}).}
\label{tab:sim-rmse-int}
\end{sidewaystable}

\begin{sidewaystable}[p]
\centering
\footnotesize
\setlength{\tabcolsep}{5pt}
\renewcommand{\arraystretch}{0.95}
\begin{tabular}{llrrrrrrrrrrrr}
\toprule
 &  & \multicolumn{4}{c}{Correct} & \multicolumn{4}{c}{PS Misspec.} & \multicolumn{4}{c}{OR Misspec.} \\
\cmidrule(lr){3-6} \cmidrule(lr){7-10} \cmidrule(lr){11-14}
ME configuration & $\eta$ & Naive & EP & Gold & CV & Naive & EP & Gold & CV & Naive & EP & Gold & CV \\
\midrule
ME in One Exposure & $0.2$ & 0.78 & 0.71 & 0.93 & 0.93 & 0.65 & 0.62 & 0.94 & 0.93 & 0.65 & 0.59 & 0.93 & 0.93 \\
 & $0.3$ & 0.83 & 0.74 & 0.94 & 0.94 & 0.72 & 0.68 & 0.95 & 0.94 & 0.71 & 0.64 & 0.94 & 0.94 \\
 & $0.4$ & 0.86 & 0.78 & 0.95 & 0.95 & 0.77 & 0.73 & 0.95 & 0.95 & 0.78 & 0.70 & 0.94 & 0.94 \\
 & $0.5$ & 0.90 & 0.81 & 0.94 & 0.94 & 0.81 & 0.78 & 0.95 & 0.94 & 0.85 & 0.76 & 0.94 & 0.94 \\
 & $0.6$ & 0.92 & 0.85 & 0.94 & 0.94 & 0.86 & 0.84 & 0.95 & 0.95 & 0.88 & 0.82 & 0.93 & 0.93 \\
 & $0.7$ & 0.93 & 0.87 & 0.94 & 0.94 & 0.90 & 0.86 & 0.95 & 0.95 & 0.91 & 0.85 & 0.95 & 0.95 \\
 & $0.8$ & 0.95 & 0.90 & 0.95 & 0.95 & 0.93 & 0.90 & 0.95 & 0.95 & 0.93 & 0.89 & 0.95 & 0.94 \\
\midrule
ME in Both Exposures & $0.2$ & 0.21 & 0.13 & 0.93 & 0.93 & 0.14 & 0.12 & 0.94 & 0.93 & 0.14 & 0.10 & 0.94 & 0.94 \\
 & $0.3$ & 0.34 & 0.20 & 0.94 & 0.94 & 0.24 & 0.19 & 0.94 & 0.94 & 0.24 & 0.17 & 0.94 & 0.94 \\
 & $0.4$ & 0.48 & 0.28 & 0.94 & 0.94 & 0.35 & 0.27 & 0.95 & 0.94 & 0.37 & 0.26 & 0.94 & 0.94 \\
 & $0.5$ & 0.61 & 0.40 & 0.95 & 0.95 & 0.49 & 0.40 & 0.94 & 0.94 & 0.50 & 0.38 & 0.94 & 0.94 \\
 & $0.6$ & 0.76 & 0.55 & 0.94 & 0.94 & 0.64 & 0.54 & 0.95 & 0.95 & 0.67 & 0.52 & 0.94 & 0.94 \\
 & $0.7$ & 0.84 & 0.67 & 0.95 & 0.95 & 0.76 & 0.67 & 0.95 & 0.95 & 0.78 & 0.65 & 0.95 & 0.95 \\
 & $0.8$ & 0.91 & 0.78 & 0.95 & 0.95 & 0.86 & 0.78 & 0.95 & 0.95 & 0.88 & 0.77 & 0.95 & 0.95 \\
\midrule
Low Specificity & $0.2$ & 0.72 & 0.66 & 0.93 & 0.93 & 0.60 & 0.58 & 0.94 & 0.94 & 0.66 & 0.60 & 0.94 & 0.94 \\
 & $0.3$ & 0.76 & 0.68 & 0.94 & 0.93 & 0.67 & 0.65 & 0.94 & 0.94 & 0.74 & 0.65 & 0.94 & 0.94 \\
 & $0.4$ & 0.82 & 0.73 & 0.95 & 0.94 & 0.73 & 0.71 & 0.94 & 0.94 & 0.78 & 0.69 & 0.94 & 0.94 \\
 & $0.5$ & 0.86 & 0.77 & 0.94 & 0.94 & 0.78 & 0.76 & 0.95 & 0.95 & 0.84 & 0.74 & 0.95 & 0.95 \\
 & $0.6$ & 0.90 & 0.82 & 0.94 & 0.94 & 0.83 & 0.81 & 0.94 & 0.94 & 0.87 & 0.80 & 0.94 & 0.94 \\
 & $0.7$ & 0.92 & 0.85 & 0.95 & 0.95 & 0.88 & 0.86 & 0.94 & 0.94 & 0.91 & 0.85 & 0.94 & 0.94 \\
 & $0.8$ & 0.94 & 0.88 & 0.95 & 0.95 & 0.91 & 0.88 & 0.95 & 0.95 & 0.92 & 0.88 & 0.94 & 0.95 \\
\midrule
Low Sensitivity & $0.2$ & 0.42 & 0.34 & 0.93 & 0.93 & 0.40 & 0.33 & 0.94 & 0.94 & 0.30 & 0.27 & 0.93 & 0.93 \\
 & $0.3$ & 0.55 & 0.42 & 0.94 & 0.94 & 0.50 & 0.40 & 0.95 & 0.94 & 0.41 & 0.35 & 0.94 & 0.94 \\
 & $0.4$ & 0.64 & 0.52 & 0.94 & 0.94 & 0.63 & 0.51 & 0.95 & 0.95 & 0.53 & 0.46 & 0.94 & 0.94 \\
 & $0.5$ & 0.75 & 0.62 & 0.94 & 0.94 & 0.73 & 0.61 & 0.95 & 0.95 & 0.65 & 0.59 & 0.95 & 0.94 \\
 & $0.6$ & 0.84 & 0.71 & 0.95 & 0.95 & 0.80 & 0.69 & 0.94 & 0.94 & 0.75 & 0.68 & 0.95 & 0.95 \\
 & $0.7$ & 0.89 & 0.80 & 0.94 & 0.94 & 0.87 & 0.79 & 0.95 & 0.95 & 0.82 & 0.77 & 0.95 & 0.94 \\
 & $0.8$ & 0.92 & 0.86 & 0.95 & 0.95 & 0.91 & 0.85 & 0.95 & 0.95 & 0.89 & 0.83 & 0.95 & 0.95 \\
\midrule
Mixed Sens/Spec & $0.2$ & 0.52 & 0.45 & 0.93 & 0.93 & 0.35 & 0.34 & 0.94 & 0.94 & 0.40 & 0.35 & 0.94 & 0.93 \\
 & $0.3$ & 0.62 & 0.50 & 0.94 & 0.94 & 0.43 & 0.43 & 0.94 & 0.94 & 0.50 & 0.42 & 0.94 & 0.94 \\
 & $0.4$ & 0.72 & 0.58 & 0.94 & 0.94 & 0.51 & 0.52 & 0.94 & 0.94 & 0.60 & 0.49 & 0.94 & 0.94 \\
 & $0.5$ & 0.78 & 0.65 & 0.94 & 0.95 & 0.60 & 0.62 & 0.94 & 0.94 & 0.70 & 0.58 & 0.94 & 0.94 \\
 & $0.6$ & 0.84 & 0.72 & 0.95 & 0.94 & 0.70 & 0.72 & 0.94 & 0.94 & 0.80 & 0.68 & 0.95 & 0.95 \\
 & $0.7$ & 0.88 & 0.78 & 0.95 & 0.95 & 0.78 & 0.79 & 0.95 & 0.95 & 0.86 & 0.76 & 0.94 & 0.94 \\
 & $0.8$ & 0.92 & 0.85 & 0.95 & 0.95 & 0.86 & 0.85 & 0.95 & 0.95 & 0.90 & 0.83 & 0.95 & 0.94 \\
\bottomrule
\end{tabular}
\caption{Empirical $95\%$ CI coverage for $\tau^{(0,1)}$ across measurement-error configurations (rows) and gold-standard proportions $\eta$ (rows within each configuration), separately for each model-specification scenario (column groups: Correct, PS Misspec., OR Misspec.).  Estimators within each scenario are Naive, EP (error-prone), Gold, and CV (control variate); the Oracle benchmark is omitted (see Section~\ref{sec:supp-simresults}).}
\label{tab:sim-cp-01}
\end{sidewaystable}

\begin{sidewaystable}[p]
\centering
\footnotesize
\setlength{\tabcolsep}{5pt}
\renewcommand{\arraystretch}{0.95}
\begin{tabular}{llrrrrrrrrrrrr}
\toprule
 &  & \multicolumn{4}{c}{Correct} & \multicolumn{4}{c}{PS Misspec.} & \multicolumn{4}{c}{OR Misspec.} \\
\cmidrule(lr){3-6} \cmidrule(lr){7-10} \cmidrule(lr){11-14}
ME configuration & $\eta$ & Naive & EP & Gold & CV & Naive & EP & Gold & CV & Naive & EP & Gold & CV \\
\midrule
ME in One Exposure & $0.2$ & 0.23 & 0.31 & 0.92 & 0.93 & 0.34 & 0.41 & 0.94 & 0.93 & 0.91 & 0.88 & 0.93 & 0.93 \\
 & $0.3$ & 0.30 & 0.44 & 0.94 & 0.94 & 0.46 & 0.55 & 0.94 & 0.94 & 0.93 & 0.89 & 0.94 & 0.94 \\
 & $0.4$ & 0.38 & 0.57 & 0.94 & 0.94 & 0.57 & 0.69 & 0.95 & 0.94 & 0.94 & 0.90 & 0.94 & 0.94 \\
 & $0.5$ & 0.48 & 0.70 & 0.94 & 0.94 & 0.68 & 0.77 & 0.95 & 0.95 & 0.94 & 0.92 & 0.94 & 0.94 \\
 & $0.6$ & 0.57 & 0.79 & 0.95 & 0.95 & 0.77 & 0.86 & 0.95 & 0.94 & 0.94 & 0.92 & 0.94 & 0.94 \\
 & $0.7$ & 0.71 & 0.87 & 0.95 & 0.95 & 0.85 & 0.90 & 0.94 & 0.94 & 0.96 & 0.93 & 0.94 & 0.94 \\
 & $0.8$ & 0.82 & 0.90 & 0.94 & 0.94 & 0.91 & 0.92 & 0.95 & 0.95 & 0.95 & 0.93 & 0.94 & 0.94 \\
\midrule
ME in Both Exposures & $0.2$ & 0.25 & 0.40 & 0.93 & 0.93 & 0.45 & 0.49 & 0.94 & 0.94 & 0.94 & 0.92 & 0.94 & 0.94 \\
 & $0.3$ & 0.29 & 0.53 & 0.94 & 0.94 & 0.57 & 0.63 & 0.94 & 0.94 & 0.94 & 0.92 & 0.94 & 0.94 \\
 & $0.4$ & 0.36 & 0.65 & 0.94 & 0.94 & 0.67 & 0.75 & 0.94 & 0.94 & 0.94 & 0.93 & 0.94 & 0.94 \\
 & $0.5$ & 0.45 & 0.77 & 0.95 & 0.94 & 0.75 & 0.83 & 0.94 & 0.94 & 0.95 & 0.93 & 0.95 & 0.94 \\
 & $0.6$ & 0.52 & 0.84 & 0.94 & 0.94 & 0.81 & 0.88 & 0.95 & 0.95 & 0.95 & 0.94 & 0.94 & 0.94 \\
 & $0.7$ & 0.64 & 0.89 & 0.94 & 0.94 & 0.88 & 0.92 & 0.94 & 0.94 & 0.95 & 0.93 & 0.94 & 0.94 \\
 & $0.8$ & 0.76 & 0.92 & 0.94 & 0.94 & 0.92 & 0.93 & 0.95 & 0.95 & 0.95 & 0.94 & 0.94 & 0.94 \\
\midrule
Low Specificity & $0.2$ & 0.59 & 0.67 & 0.94 & 0.94 & 0.79 & 0.81 & 0.94 & 0.94 & 0.95 & 0.94 & 0.93 & 0.93 \\
 & $0.3$ & 0.64 & 0.75 & 0.93 & 0.93 & 0.81 & 0.85 & 0.94 & 0.94 & 0.94 & 0.94 & 0.94 & 0.94 \\
 & $0.4$ & 0.67 & 0.81 & 0.94 & 0.94 & 0.85 & 0.90 & 0.94 & 0.95 & 0.94 & 0.95 & 0.94 & 0.94 \\
 & $0.5$ & 0.71 & 0.86 & 0.94 & 0.94 & 0.89 & 0.92 & 0.95 & 0.95 & 0.95 & 0.95 & 0.95 & 0.95 \\
 & $0.6$ & 0.76 & 0.89 & 0.94 & 0.94 & 0.91 & 0.92 & 0.95 & 0.94 & 0.95 & 0.95 & 0.94 & 0.94 \\
 & $0.7$ & 0.82 & 0.92 & 0.94 & 0.94 & 0.92 & 0.93 & 0.94 & 0.94 & 0.95 & 0.94 & 0.94 & 0.94 \\
 & $0.8$ & 0.87 & 0.93 & 0.95 & 0.95 & 0.93 & 0.94 & 0.94 & 0.94 & 0.95 & 0.94 & 0.94 & 0.94 \\
\midrule
Low Sensitivity & $0.2$ & 0.73 & 0.78 & 0.93 & 0.93 & 0.77 & 0.80 & 0.94 & 0.94 & 0.94 & 0.95 & 0.94 & 0.93 \\
 & $0.3$ & 0.75 & 0.83 & 0.94 & 0.93 & 0.82 & 0.87 & 0.94 & 0.94 & 0.94 & 0.94 & 0.94 & 0.94 \\
 & $0.4$ & 0.79 & 0.86 & 0.94 & 0.94 & 0.85 & 0.90 & 0.94 & 0.94 & 0.94 & 0.95 & 0.94 & 0.94 \\
 & $0.5$ & 0.82 & 0.89 & 0.94 & 0.94 & 0.88 & 0.92 & 0.95 & 0.94 & 0.94 & 0.95 & 0.94 & 0.94 \\
 & $0.6$ & 0.84 & 0.92 & 0.94 & 0.94 & 0.91 & 0.93 & 0.95 & 0.95 & 0.93 & 0.94 & 0.94 & 0.94 \\
 & $0.7$ & 0.89 & 0.93 & 0.95 & 0.94 & 0.93 & 0.94 & 0.94 & 0.94 & 0.93 & 0.94 & 0.94 & 0.94 \\
 & $0.8$ & 0.91 & 0.94 & 0.95 & 0.94 & 0.93 & 0.95 & 0.94 & 0.94 & 0.93 & 0.94 & 0.94 & 0.94 \\
\midrule
Mixed Sens/Spec & $0.2$ & 0.75 & 0.78 & 0.94 & 0.93 & 0.76 & 0.80 & 0.94 & 0.93 & 0.95 & 0.94 & 0.93 & 0.93 \\
 & $0.3$ & 0.78 & 0.82 & 0.94 & 0.93 & 0.80 & 0.84 & 0.94 & 0.94 & 0.95 & 0.95 & 0.94 & 0.94 \\
 & $0.4$ & 0.80 & 0.85 & 0.94 & 0.93 & 0.84 & 0.89 & 0.95 & 0.95 & 0.94 & 0.94 & 0.94 & 0.94 \\
 & $0.5$ & 0.85 & 0.88 & 0.95 & 0.95 & 0.87 & 0.90 & 0.95 & 0.94 & 0.94 & 0.95 & 0.95 & 0.95 \\
 & $0.6$ & 0.89 & 0.91 & 0.95 & 0.95 & 0.89 & 0.92 & 0.95 & 0.94 & 0.94 & 0.95 & 0.95 & 0.95 \\
 & $0.7$ & 0.91 & 0.93 & 0.95 & 0.94 & 0.92 & 0.94 & 0.94 & 0.94 & 0.94 & 0.94 & 0.95 & 0.95 \\
 & $0.8$ & 0.93 & 0.94 & 0.95 & 0.95 & 0.94 & 0.94 & 0.95 & 0.95 & 0.94 & 0.95 & 0.94 & 0.94 \\
\bottomrule
\end{tabular}
\caption{Empirical $95\%$ CI coverage for $\tau^{(1,0)}$ across measurement-error configurations (rows) and gold-standard proportions $\eta$ (rows within each configuration), separately for each model-specification scenario (column groups: Correct, PS Misspec., OR Misspec.).  Estimators within each scenario are Naive, EP (error-prone), Gold, and CV (control variate); the Oracle benchmark is omitted (see Section~\ref{sec:supp-simresults}).}
\label{tab:sim-cp-10}
\end{sidewaystable}

\begin{sidewaystable}[p]
\centering
\footnotesize
\setlength{\tabcolsep}{5pt}
\renewcommand{\arraystretch}{0.95}
\begin{tabular}{llrrrrrrrrrrrr}
\toprule
 &  & \multicolumn{4}{c}{Correct} & \multicolumn{4}{c}{PS Misspec.} & \multicolumn{4}{c}{OR Misspec.} \\
\cmidrule(lr){3-6} \cmidrule(lr){7-10} \cmidrule(lr){11-14}
ME configuration & $\eta$ & Naive & EP & Gold & CV & Naive & EP & Gold & CV & Naive & EP & Gold & CV \\
\midrule
ME in One Exposure & $0.2$ & 0.11 & 0.14 & 0.93 & 0.93 & 0.26 & 0.28 & 0.93 & 0.93 & 0.87 & 0.82 & 0.93 & 0.93 \\
 & $0.3$ & 0.18 & 0.27 & 0.94 & 0.94 & 0.37 & 0.42 & 0.94 & 0.94 & 0.91 & 0.85 & 0.94 & 0.94 \\
 & $0.4$ & 0.29 & 0.44 & 0.94 & 0.94 & 0.49 & 0.57 & 0.95 & 0.95 & 0.92 & 0.87 & 0.94 & 0.94 \\
 & $0.5$ & 0.43 & 0.60 & 0.94 & 0.94 & 0.63 & 0.70 & 0.95 & 0.95 & 0.94 & 0.90 & 0.95 & 0.95 \\
 & $0.6$ & 0.57 & 0.73 & 0.94 & 0.94 & 0.74 & 0.80 & 0.95 & 0.95 & 0.94 & 0.91 & 0.94 & 0.94 \\
 & $0.7$ & 0.72 & 0.83 & 0.94 & 0.94 & 0.83 & 0.86 & 0.95 & 0.95 & 0.96 & 0.92 & 0.95 & 0.95 \\
 & $0.8$ & 0.84 & 0.89 & 0.95 & 0.95 & 0.90 & 0.91 & 0.95 & 0.95 & 0.94 & 0.93 & 0.95 & 0.94 \\
\midrule
ME in Both Exposures & $0.2$ & 0.51 & 0.60 & 0.93 & 0.93 & 0.66 & 0.73 & 0.93 & 0.93 & 0.93 & 0.94 & 0.94 & 0.94 \\
 & $0.3$ & 0.59 & 0.72 & 0.94 & 0.94 & 0.73 & 0.79 & 0.94 & 0.94 & 0.93 & 0.95 & 0.94 & 0.94 \\
 & $0.4$ & 0.69 & 0.80 & 0.94 & 0.94 & 0.78 & 0.86 & 0.94 & 0.95 & 0.92 & 0.95 & 0.94 & 0.94 \\
 & $0.5$ & 0.76 & 0.87 & 0.95 & 0.95 & 0.83 & 0.90 & 0.94 & 0.94 & 0.92 & 0.95 & 0.94 & 0.94 \\
 & $0.6$ & 0.82 & 0.91 & 0.95 & 0.95 & 0.87 & 0.93 & 0.95 & 0.95 & 0.93 & 0.95 & 0.94 & 0.94 \\
 & $0.7$ & 0.87 & 0.92 & 0.94 & 0.94 & 0.91 & 0.93 & 0.95 & 0.95 & 0.93 & 0.95 & 0.95 & 0.95 \\
 & $0.8$ & 0.91 & 0.93 & 0.95 & 0.95 & 0.93 & 0.94 & 0.95 & 0.95 & 0.93 & 0.94 & 0.94 & 0.94 \\
\midrule
Low Specificity & $0.2$ & 0.56 & 0.65 & 0.93 & 0.92 & 0.78 & 0.83 & 0.93 & 0.93 & 0.95 & 0.95 & 0.93 & 0.93 \\
 & $0.3$ & 0.64 & 0.74 & 0.94 & 0.94 & 0.82 & 0.87 & 0.94 & 0.94 & 0.95 & 0.95 & 0.94 & 0.94 \\
 & $0.4$ & 0.68 & 0.80 & 0.95 & 0.94 & 0.85 & 0.90 & 0.94 & 0.94 & 0.95 & 0.95 & 0.94 & 0.95 \\
 & $0.5$ & 0.73 & 0.86 & 0.95 & 0.95 & 0.88 & 0.93 & 0.94 & 0.94 & 0.95 & 0.95 & 0.95 & 0.95 \\
 & $0.6$ & 0.80 & 0.89 & 0.94 & 0.94 & 0.90 & 0.93 & 0.94 & 0.94 & 0.95 & 0.94 & 0.95 & 0.95 \\
 & $0.7$ & 0.85 & 0.92 & 0.95 & 0.95 & 0.92 & 0.94 & 0.95 & 0.95 & 0.95 & 0.95 & 0.95 & 0.95 \\
 & $0.8$ & 0.89 & 0.93 & 0.95 & 0.95 & 0.93 & 0.93 & 0.95 & 0.94 & 0.95 & 0.94 & 0.94 & 0.94 \\
\midrule
Low Sensitivity & $0.2$ & 0.90 & 0.90 & 0.93 & 0.93 & 0.88 & 0.90 & 0.94 & 0.93 & 0.88 & 0.92 & 0.93 & 0.93 \\
 & $0.3$ & 0.92 & 0.92 & 0.94 & 0.94 & 0.90 & 0.93 & 0.94 & 0.94 & 0.87 & 0.92 & 0.95 & 0.95 \\
 & $0.4$ & 0.93 & 0.93 & 0.94 & 0.95 & 0.91 & 0.94 & 0.95 & 0.94 & 0.86 & 0.92 & 0.95 & 0.94 \\
 & $0.5$ & 0.94 & 0.93 & 0.94 & 0.94 & 0.92 & 0.94 & 0.95 & 0.95 & 0.87 & 0.93 & 0.94 & 0.94 \\
 & $0.6$ & 0.94 & 0.94 & 0.95 & 0.95 & 0.93 & 0.95 & 0.95 & 0.95 & 0.87 & 0.94 & 0.94 & 0.94 \\
 & $0.7$ & 0.94 & 0.94 & 0.95 & 0.95 & 0.94 & 0.95 & 0.95 & 0.95 & 0.89 & 0.94 & 0.95 & 0.95 \\
 & $0.8$ & 0.94 & 0.94 & 0.94 & 0.94 & 0.94 & 0.94 & 0.95 & 0.94 & 0.90 & 0.93 & 0.94 & 0.94 \\
\midrule
Mixed Sens/Spec & $0.2$ & 0.78 & 0.81 & 0.92 & 0.92 & 0.82 & 0.86 & 0.94 & 0.94 & 0.94 & 0.94 & 0.94 & 0.94 \\
 & $0.3$ & 0.82 & 0.85 & 0.94 & 0.94 & 0.85 & 0.89 & 0.94 & 0.94 & 0.94 & 0.94 & 0.94 & 0.94 \\
 & $0.4$ & 0.85 & 0.88 & 0.94 & 0.94 & 0.87 & 0.92 & 0.94 & 0.94 & 0.93 & 0.94 & 0.95 & 0.95 \\
 & $0.5$ & 0.89 & 0.92 & 0.94 & 0.94 & 0.90 & 0.93 & 0.95 & 0.95 & 0.93 & 0.94 & 0.95 & 0.95 \\
 & $0.6$ & 0.90 & 0.93 & 0.95 & 0.95 & 0.91 & 0.93 & 0.94 & 0.94 & 0.93 & 0.94 & 0.95 & 0.94 \\
 & $0.7$ & 0.93 & 0.94 & 0.95 & 0.95 & 0.93 & 0.94 & 0.95 & 0.94 & 0.93 & 0.94 & 0.95 & 0.95 \\
 & $0.8$ & 0.95 & 0.94 & 0.95 & 0.95 & 0.94 & 0.94 & 0.95 & 0.94 & 0.94 & 0.94 & 0.95 & 0.95 \\
\bottomrule
\end{tabular}
\caption{Empirical $95\%$ CI coverage for $\tau^{(1,1)}$ across measurement-error configurations (rows) and gold-standard proportions $\eta$ (rows within each configuration), separately for each model-specification scenario (column groups: Correct, PS Misspec., OR Misspec.).  Estimators within each scenario are Naive, EP (error-prone), Gold, and CV (control variate); the Oracle benchmark is omitted (see Section~\ref{sec:supp-simresults}).}
\label{tab:sim-cp-11}
\end{sidewaystable}

\begin{sidewaystable}[p]
\centering
\footnotesize
\setlength{\tabcolsep}{5pt}
\renewcommand{\arraystretch}{0.95}
\begin{tabular}{llrrrrrrrrrrrr}
\toprule
 &  & \multicolumn{4}{c}{Correct} & \multicolumn{4}{c}{PS Misspec.} & \multicolumn{4}{c}{OR Misspec.} \\
\cmidrule(lr){3-6} \cmidrule(lr){7-10} \cmidrule(lr){11-14}
ME configuration & $\eta$ & Naive & EP & Gold & CV & Naive & EP & Gold & CV & Naive & EP & Gold & CV \\
\midrule
ME in One Exposure & $0.2$ & 0.83 & 0.75 & 0.94 & 0.94 & 0.73 & 0.71 & 0.95 & 0.94 & 0.76 & 0.72 & 0.94 & 0.94 \\
 & $0.3$ & 0.88 & 0.78 & 0.95 & 0.94 & 0.78 & 0.74 & 0.95 & 0.95 & 0.81 & 0.78 & 0.94 & 0.94 \\
 & $0.4$ & 0.91 & 0.81 & 0.95 & 0.95 & 0.82 & 0.79 & 0.95 & 0.95 & 0.85 & 0.81 & 0.95 & 0.95 \\
 & $0.5$ & 0.93 & 0.84 & 0.95 & 0.95 & 0.86 & 0.83 & 0.95 & 0.94 & 0.89 & 0.85 & 0.94 & 0.94 \\
 & $0.6$ & 0.94 & 0.89 & 0.94 & 0.94 & 0.89 & 0.86 & 0.94 & 0.94 & 0.92 & 0.88 & 0.95 & 0.95 \\
 & $0.7$ & 0.95 & 0.90 & 0.94 & 0.94 & 0.91 & 0.89 & 0.95 & 0.94 & 0.93 & 0.90 & 0.94 & 0.94 \\
 & $0.8$ & 0.95 & 0.92 & 0.95 & 0.95 & 0.94 & 0.92 & 0.95 & 0.95 & 0.94 & 0.93 & 0.94 & 0.94 \\
\midrule
ME in Both Exposures & $0.2$ & 0.75 & 0.60 & 0.94 & 0.94 & 0.58 & 0.57 & 0.94 & 0.94 & 0.62 & 0.57 & 0.94 & 0.94 \\
 & $0.3$ & 0.84 & 0.66 & 0.94 & 0.94 & 0.63 & 0.62 & 0.95 & 0.95 & 0.73 & 0.65 & 0.94 & 0.94 \\
 & $0.4$ & 0.90 & 0.71 & 0.95 & 0.94 & 0.71 & 0.70 & 0.95 & 0.95 & 0.80 & 0.71 & 0.94 & 0.94 \\
 & $0.5$ & 0.93 & 0.76 & 0.94 & 0.94 & 0.77 & 0.75 & 0.95 & 0.95 & 0.86 & 0.77 & 0.95 & 0.95 \\
 & $0.6$ & 0.95 & 0.83 & 0.94 & 0.94 & 0.83 & 0.82 & 0.95 & 0.95 & 0.90 & 0.82 & 0.94 & 0.94 \\
 & $0.7$ & 0.95 & 0.87 & 0.95 & 0.95 & 0.86 & 0.86 & 0.95 & 0.94 & 0.92 & 0.86 & 0.95 & 0.94 \\
 & $0.8$ & 0.94 & 0.90 & 0.95 & 0.94 & 0.92 & 0.90 & 0.95 & 0.94 & 0.95 & 0.91 & 0.95 & 0.95 \\
\midrule
Low Specificity & $0.2$ & 0.87 & 0.81 & 0.94 & 0.94 & 0.75 & 0.75 & 0.95 & 0.95 & 0.81 & 0.80 & 0.94 & 0.94 \\
 & $0.3$ & 0.89 & 0.81 & 0.94 & 0.94 & 0.79 & 0.79 & 0.94 & 0.94 & 0.85 & 0.82 & 0.94 & 0.94 \\
 & $0.4$ & 0.92 & 0.85 & 0.95 & 0.95 & 0.82 & 0.83 & 0.94 & 0.94 & 0.88 & 0.84 & 0.95 & 0.95 \\
 & $0.5$ & 0.94 & 0.87 & 0.95 & 0.95 & 0.85 & 0.86 & 0.95 & 0.95 & 0.90 & 0.87 & 0.95 & 0.95 \\
 & $0.6$ & 0.95 & 0.89 & 0.95 & 0.95 & 0.88 & 0.88 & 0.95 & 0.95 & 0.93 & 0.90 & 0.94 & 0.94 \\
 & $0.7$ & 0.95 & 0.91 & 0.94 & 0.94 & 0.91 & 0.91 & 0.95 & 0.95 & 0.94 & 0.90 & 0.95 & 0.95 \\
 & $0.8$ & 0.95 & 0.92 & 0.95 & 0.95 & 0.92 & 0.92 & 0.95 & 0.95 & 0.94 & 0.92 & 0.95 & 0.95 \\
\midrule
Low Sensitivity & $0.2$ & 0.88 & 0.82 & 0.93 & 0.93 & 0.78 & 0.75 & 0.95 & 0.95 & 0.82 & 0.79 & 0.94 & 0.94 \\
 & $0.3$ & 0.91 & 0.85 & 0.94 & 0.94 & 0.80 & 0.78 & 0.94 & 0.94 & 0.85 & 0.83 & 0.95 & 0.95 \\
 & $0.4$ & 0.93 & 0.87 & 0.94 & 0.94 & 0.84 & 0.83 & 0.95 & 0.95 & 0.88 & 0.84 & 0.95 & 0.95 \\
 & $0.5$ & 0.94 & 0.89 & 0.95 & 0.95 & 0.88 & 0.85 & 0.94 & 0.94 & 0.91 & 0.88 & 0.95 & 0.95 \\
 & $0.6$ & 0.95 & 0.91 & 0.95 & 0.95 & 0.90 & 0.89 & 0.95 & 0.95 & 0.92 & 0.89 & 0.95 & 0.94 \\
 & $0.7$ & 0.95 & 0.92 & 0.95 & 0.95 & 0.93 & 0.91 & 0.95 & 0.95 & 0.93 & 0.91 & 0.95 & 0.94 \\
 & $0.8$ & 0.95 & 0.94 & 0.95 & 0.95 & 0.94 & 0.92 & 0.95 & 0.95 & 0.93 & 0.93 & 0.94 & 0.94 \\
\midrule
Mixed Sens/Spec & $0.2$ & 0.87 & 0.85 & 0.94 & 0.94 & 0.72 & 0.75 & 0.94 & 0.94 & 0.81 & 0.81 & 0.94 & 0.94 \\
 & $0.3$ & 0.89 & 0.86 & 0.94 & 0.94 & 0.75 & 0.80 & 0.94 & 0.94 & 0.83 & 0.83 & 0.94 & 0.94 \\
 & $0.4$ & 0.92 & 0.88 & 0.95 & 0.95 & 0.78 & 0.82 & 0.94 & 0.94 & 0.86 & 0.86 & 0.94 & 0.95 \\
 & $0.5$ & 0.93 & 0.89 & 0.95 & 0.95 & 0.82 & 0.86 & 0.94 & 0.94 & 0.89 & 0.88 & 0.95 & 0.95 \\
 & $0.6$ & 0.94 & 0.90 & 0.94 & 0.94 & 0.84 & 0.88 & 0.95 & 0.95 & 0.91 & 0.90 & 0.95 & 0.95 \\
 & $0.7$ & 0.94 & 0.91 & 0.95 & 0.95 & 0.88 & 0.90 & 0.95 & 0.95 & 0.93 & 0.91 & 0.94 & 0.94 \\
 & $0.8$ & 0.95 & 0.93 & 0.96 & 0.95 & 0.92 & 0.92 & 0.95 & 0.95 & 0.93 & 0.92 & 0.95 & 0.95 \\
\bottomrule
\end{tabular}
\caption{Empirical $95\%$ CI coverage for $\Delta$ (interaction) across measurement-error configurations (rows) and gold-standard proportions $\eta$ (rows within each configuration), separately for each model-specification scenario (column groups: Correct, PS Misspec., OR Misspec.).  Estimators within each scenario are Naive, EP (error-prone), Gold, and CV (control variate); the Oracle benchmark is omitted (see Section~\ref{sec:supp-simresults}).}
\label{tab:sim-cp-int}
\end{sidewaystable}

\begin{sidewaystable}[p]
\centering
\footnotesize
\setlength{\tabcolsep}{5pt}
\renewcommand{\arraystretch}{0.95}
\begin{tabular}{llrrrrrrrrrrrr}
\toprule
 &  & \multicolumn{4}{c}{Correct} & \multicolumn{4}{c}{PS Misspec.} & \multicolumn{4}{c}{OR Misspec.} \\
\cmidrule(lr){3-6} \cmidrule(lr){7-10} \cmidrule(lr){11-14}
ME configuration & $\eta$ & Naive & EP & Gold & CV & Naive & EP & Gold & CV & Naive & EP & Gold & CV \\
\midrule
ME in One Exposure & $0.2$ & 2.28 & 2.86 & 29.84 & 28.70 & 2.25 & 2.81 & 31.15 & 30.46 & 2.24 & 2.79 & 30.20 & 29.07 \\
 & $0.3$ & 2.25 & 3.41 & 16.21 & 15.61 & 2.26 & 3.31 & 16.28 & 15.77 & 2.33 & 3.34 & 16.68 & 16.08 \\
 & $0.4$ & 2.28 & 4.02 & 9.86 & 9.54 & 2.20 & 4.15 & 10.20 & 9.77 & 2.37 & 4.28 & 10.43 & 10.05 \\
 & $0.5$ & 2.34 & 5.55 & 7.07 & 6.83 & 2.27 & 5.58 & 6.90 & 6.71 & 2.26 & 5.63 & 7.20 & 7.04 \\
 & $0.6$ & 2.43 & 8.02 & 5.26 & 5.14 & 2.31 & 7.84 & 5.17 & 4.98 & 2.53 & 8.06 & 5.58 & 5.46 \\
 & $0.7$ & 2.58 & 12.48 & 4.19 & 4.12 & 2.33 & 12.11 & 3.85 & 3.78 & 2.47 & 12.15 & 4.07 & 4.00 \\
 & $0.8$ & 2.55 & 22.44 & 3.34 & 3.32 & 2.26 & 22.14 & 3.11 & 3.06 & 2.61 & 23.26 & 3.48 & 3.42 \\
\midrule
ME in Both Exposures & $0.2$ & 2.02 & 2.53 & 30.24 & 29.91 & 2.15 & 2.64 & 32.38 & 32.03 & 2.09 & 2.63 & 30.30 & 29.74 \\
 & $0.3$ & 2.13 & 3.18 & 16.63 & 16.53 & 2.13 & 3.21 & 16.38 & 16.18 & 2.16 & 3.21 & 16.58 & 16.51 \\
 & $0.4$ & 2.13 & 3.93 & 10.44 & 10.34 & 2.16 & 4.02 & 10.40 & 10.26 & 2.13 & 4.08 & 10.34 & 10.29 \\
 & $0.5$ & 2.20 & 5.21 & 6.98 & 6.98 & 2.16 & 5.42 & 7.37 & 7.30 & 2.23 & 5.32 & 7.24 & 7.20 \\
 & $0.6$ & 2.32 & 7.53 & 5.42 & 5.39 & 2.22 & 7.85 & 4.98 & 4.99 & 2.38 & 7.33 & 5.51 & 5.48 \\
 & $0.7$ & 2.40 & 12.18 & 4.14 & 4.12 & 2.24 & 11.80 & 3.86 & 3.83 & 2.46 & 12.35 & 4.23 & 4.23 \\
 & $0.8$ & 2.52 & 21.56 & 3.45 & 3.45 & 2.26 & 21.95 & 3.10 & 3.08 & 2.51 & 22.65 & 3.47 & 3.47 \\
\midrule
Low Specificity & $0.2$ & 2.76 & 3.57 & 30.55 & 29.70 & 2.70 & 3.53 & 30.55 & 29.75 & 2.69 & 3.52 & 30.31 & 29.30 \\
 & $0.3$ & 2.73 & 4.29 & 16.50 & 16.10 & 2.67 & 4.38 & 16.95 & 16.71 & 2.71 & 4.12 & 16.67 & 16.28 \\
 & $0.4$ & 2.70 & 5.25 & 9.99 & 9.72 & 2.54 & 5.25 & 10.45 & 10.22 & 2.74 & 5.14 & 10.24 & 10.08 \\
 & $0.5$ & 2.70 & 6.71 & 7.21 & 7.07 & 2.48 & 7.03 & 6.96 & 6.82 & 2.67 & 7.03 & 7.00 & 6.80 \\
 & $0.6$ & 2.63 & 9.85 & 5.33 & 5.23 & 2.52 & 10.09 & 5.01 & 4.91 & 2.77 & 9.55 & 5.44 & 5.33 \\
 & $0.7$ & 2.67 & 15.42 & 4.15 & 4.04 & 2.44 & 15.98 & 4.03 & 3.98 & 2.70 & 14.95 & 4.22 & 4.14 \\
 & $0.8$ & 2.71 & 27.48 & 3.48 & 3.42 & 2.35 & 29.88 & 3.16 & 3.12 & 2.78 & 27.22 & 3.57 & 3.54 \\
\midrule
Low Sensitivity & $0.2$ & 1.83 & 2.30 & 30.41 & 29.74 & 1.99 & 2.38 & 31.88 & 30.65 & 1.84 & 2.32 & 30.85 & 30.10 \\
 & $0.3$ & 1.89 & 2.71 & 15.81 & 15.46 & 2.07 & 2.87 & 16.68 & 16.14 & 1.89 & 2.66 & 16.30 & 15.85 \\
 & $0.4$ & 2.02 & 3.48 & 10.45 & 10.20 & 2.09 & 3.57 & 10.13 & 9.89 & 2.01 & 3.72 & 10.40 & 10.11 \\
 & $0.5$ & 2.05 & 4.74 & 6.94 & 6.80 & 2.07 & 4.51 & 6.88 & 6.74 & 1.98 & 4.73 & 7.02 & 6.88 \\
 & $0.6$ & 2.18 & 6.85 & 5.11 & 5.07 & 2.14 & 6.63 & 5.05 & 4.94 & 2.15 & 6.97 & 5.13 & 5.05 \\
 & $0.7$ & 2.30 & 10.99 & 4.06 & 4.03 & 2.19 & 10.41 & 3.82 & 3.75 & 2.38 & 11.39 & 4.25 & 4.18 \\
 & $0.8$ & 2.40 & 19.63 & 3.33 & 3.31 & 2.27 & 18.22 & 3.09 & 3.04 & 2.49 & 20.96 & 3.46 & 3.41 \\
\midrule
Mixed Sens/Spec & $0.2$ & 2.12 & 2.49 & 29.76 & 28.89 & 2.04 & 2.46 & 32.07 & 31.49 & 1.95 & 2.36 & 30.52 & 29.70 \\
 & $0.3$ & 2.01 & 2.85 & 16.21 & 15.78 & 2.00 & 2.89 & 17.57 & 17.10 & 2.01 & 2.82 & 16.84 & 16.39 \\
 & $0.4$ & 2.09 & 3.52 & 10.35 & 10.04 & 2.03 & 3.69 & 10.59 & 10.33 & 2.06 & 3.49 & 10.57 & 10.31 \\
 & $0.5$ & 2.15 & 4.54 & 7.29 & 7.11 & 2.09 & 4.81 & 6.98 & 6.79 & 2.15 & 4.71 & 7.15 & 7.03 \\
 & $0.6$ & 2.27 & 6.35 & 5.32 & 5.20 & 2.15 & 6.88 & 5.16 & 5.06 & 2.26 & 6.50 & 5.26 & 5.18 \\
 & $0.7$ & 2.41 & 9.98 & 4.10 & 4.03 & 2.09 & 10.77 & 3.84 & 3.76 & 2.39 & 10.58 & 4.23 & 4.15 \\
 & $0.8$ & 2.41 & 17.05 & 3.38 & 3.34 & 2.15 & 19.33 & 3.10 & 3.04 & 2.45 & 18.14 & 3.46 & 3.44 \\
\bottomrule
\end{tabular}
\caption{Empirical variance ($\times 100$) for $\tau^{(0,1)}$ across measurement-error configurations (rows) and gold-standard proportions $\eta$ (rows within each configuration), separately for each model-specification scenario (column groups: Correct, PS Misspec., OR Misspec.).  Estimators within each scenario are Naive, EP (error-prone), Gold, and CV (control variate); the Oracle benchmark is omitted (see Section~\ref{sec:supp-simresults}).}
\label{tab:sim-var-01}
\end{sidewaystable}

\begin{sidewaystable}[p]
\centering
\footnotesize
\setlength{\tabcolsep}{5pt}
\renewcommand{\arraystretch}{0.95}
\begin{tabular}{llrrrrrrrrrrrr}
\toprule
 &  & \multicolumn{4}{c}{Correct} & \multicolumn{4}{c}{PS Misspec.} & \multicolumn{4}{c}{OR Misspec.} \\
\cmidrule(lr){3-6} \cmidrule(lr){7-10} \cmidrule(lr){11-14}
ME configuration & $\eta$ & Naive & EP & Gold & CV & Naive & EP & Gold & CV & Naive & EP & Gold & CV \\
\midrule
ME in One Exposure & $0.2$ & 2.79 & 3.18 & 61.29 & 59.58 & 2.15 & 2.91 & 22.35 & 22.01 & 2.80 & 3.31 & 59.84 & 58.50 \\
 & $0.3$ & 2.85 & 3.63 & 31.63 & 31.23 & 2.15 & 3.45 & 12.38 & 12.19 & 2.74 & 3.69 & 33.03 & 32.77 \\
 & $0.4$ & 2.94 & 4.38 & 20.20 & 19.99 & 2.28 & 4.50 & 7.85 & 7.82 & 2.85 & 4.45 & 20.25 & 19.83 \\
 & $0.5$ & 3.04 & 5.81 & 13.93 & 13.80 & 2.25 & 6.24 & 5.66 & 5.60 & 2.97 & 5.67 & 14.29 & 14.12 \\
 & $0.6$ & 3.32 & 8.02 & 10.13 & 10.02 & 2.18 & 8.59 & 4.31 & 4.27 & 3.09 & 8.05 & 10.40 & 10.33 \\
 & $0.7$ & 3.29 & 11.71 & 7.64 & 7.59 & 2.22 & 13.84 & 3.40 & 3.38 & 3.19 & 11.68 & 7.90 & 7.85 \\
 & $0.8$ & 3.74 & 21.02 & 6.30 & 6.30 & 2.21 & 25.50 & 2.86 & 2.84 & 3.57 & 21.33 & 6.47 & 6.48 \\
\midrule
ME in Both Exposures & $0.2$ & 2.45 & 2.89 & 59.57 & 59.09 & 2.20 & 2.82 & 23.18 & 22.98 & 2.38 & 2.88 & 57.33 & 57.14 \\
 & $0.3$ & 2.48 & 3.48 & 31.68 & 31.47 & 2.24 & 3.48 & 12.29 & 12.16 & 2.46 & 3.42 & 33.35 & 33.21 \\
 & $0.4$ & 2.68 & 4.52 & 20.75 & 20.62 & 2.21 & 4.34 & 8.08 & 8.01 & 2.62 & 4.35 & 20.77 & 20.71 \\
 & $0.5$ & 2.81 & 5.60 & 13.77 & 13.72 & 2.26 & 5.99 & 5.87 & 5.81 & 2.67 & 5.72 & 13.73 & 13.75 \\
 & $0.6$ & 2.93 & 7.90 & 10.39 & 10.34 & 2.25 & 8.15 & 4.15 & 4.14 & 2.80 & 7.87 & 10.64 & 10.57 \\
 & $0.7$ & 3.19 & 12.54 & 7.82 & 7.77 & 2.23 & 13.13 & 3.44 & 3.44 & 3.01 & 12.43 & 7.96 & 7.98 \\
 & $0.8$ & 3.55 & 21.31 & 6.27 & 6.27 & 2.20 & 24.03 & 2.84 & 2.80 & 3.28 & 21.70 & 6.24 & 6.25 \\
\midrule
Low Specificity & $0.2$ & 3.50 & 4.24 & 58.89 & 56.67 & 2.71 & 3.76 & 22.72 & 21.99 & 3.42 & 4.17 & 62.44 & 60.84 \\
 & $0.3$ & 3.52 & 4.82 & 32.83 & 32.02 & 2.72 & 4.67 & 12.43 & 12.05 & 3.52 & 4.91 & 32.78 & 32.03 \\
 & $0.4$ & 3.47 & 5.67 & 20.20 & 19.75 & 2.56 & 5.69 & 7.89 & 7.64 & 3.47 & 5.82 & 20.09 & 19.78 \\
 & $0.5$ & 3.58 & 7.15 & 13.99 & 13.98 & 2.56 & 8.07 & 5.57 & 5.47 & 3.52 & 7.39 & 13.79 & 13.60 \\
 & $0.6$ & 3.56 & 10.44 & 9.98 & 9.84 & 2.47 & 11.77 & 4.23 & 4.17 & 3.48 & 10.46 & 10.25 & 10.13 \\
 & $0.7$ & 3.83 & 15.33 & 7.62 & 7.54 & 2.49 & 17.87 & 3.54 & 3.48 & 3.67 & 16.42 & 7.81 & 7.72 \\
 & $0.8$ & 3.91 & 27.55 & 5.99 & 6.00 & 2.37 & 33.07 & 2.88 & 2.84 & 3.92 & 28.05 & 6.44 & 6.39 \\
\midrule
Low Sensitivity & $0.2$ & 2.72 & 3.11 & 59.79 & 58.03 & 1.95 & 2.52 & 22.41 & 21.96 & 2.61 & 3.00 & 59.97 & 58.15 \\
 & $0.3$ & 2.84 & 3.66 & 33.08 & 32.42 & 1.99 & 3.00 & 12.50 & 12.14 & 2.74 & 3.71 & 32.04 & 31.35 \\
 & $0.4$ & 3.00 & 4.66 & 19.94 & 19.59 & 2.08 & 3.78 & 8.27 & 8.03 & 2.87 & 4.59 & 19.86 & 19.54 \\
 & $0.5$ & 3.05 & 5.81 & 13.85 & 13.61 & 2.07 & 5.16 & 5.64 & 5.48 & 3.06 & 6.08 & 14.20 & 13.88 \\
 & $0.6$ & 3.29 & 8.38 & 9.97 & 9.78 & 2.00 & 7.64 & 4.17 & 4.04 & 3.34 & 8.66 & 10.23 & 10.13 \\
 & $0.7$ & 3.56 & 12.60 & 7.66 & 7.62 & 2.07 & 12.07 & 3.46 & 3.36 & 3.46 & 13.04 & 7.96 & 7.88 \\
 & $0.8$ & 4.00 & 22.79 & 6.14 & 6.11 & 2.23 & 20.62 & 2.95 & 2.90 & 3.74 & 22.08 & 6.28 & 6.23 \\
\midrule
Mixed Sens/Spec & $0.2$ & 4.23 & 5.01 & 57.79 & 55.83 & 2.68 & 3.83 & 22.76 & 22.21 & 4.10 & 4.94 & 61.21 & 58.83 \\
 & $0.3$ & 4.41 & 6.17 & 32.63 & 32.09 & 2.66 & 4.84 & 12.50 & 12.10 & 4.18 & 5.99 & 32.46 & 31.46 \\
 & $0.4$ & 4.46 & 7.70 & 20.76 & 20.34 & 2.56 & 6.04 & 7.88 & 7.73 & 4.29 & 7.60 & 20.24 & 19.77 \\
 & $0.5$ & 4.36 & 9.75 & 13.96 & 13.66 & 2.46 & 8.11 & 5.70 & 5.55 & 4.23 & 9.43 & 13.86 & 13.68 \\
 & $0.6$ & 4.39 & 13.00 & 10.10 & 9.97 & 2.50 & 12.18 & 4.30 & 4.18 & 4.18 & 12.92 & 10.08 & 9.84 \\
 & $0.7$ & 4.53 & 20.88 & 7.77 & 7.67 & 2.37 & 18.21 & 3.45 & 3.38 & 4.45 & 21.06 & 7.79 & 7.74 \\
 & $0.8$ & 4.37 & 35.99 & 6.26 & 6.21 & 2.29 & 34.75 & 2.82 & 2.81 & 4.35 & 36.51 & 6.40 & 6.37 \\
\bottomrule
\end{tabular}
\caption{Empirical variance ($\times 100$) for $\tau^{(1,0)}$ across measurement-error configurations (rows) and gold-standard proportions $\eta$ (rows within each configuration), separately for each model-specification scenario (column groups: Correct, PS Misspec., OR Misspec.).  Estimators within each scenario are Naive, EP (error-prone), Gold, and CV (control variate); the Oracle benchmark is omitted (see Section~\ref{sec:supp-simresults}).}
\label{tab:sim-var-10}
\end{sidewaystable}

\begin{sidewaystable}[p]
\centering
\footnotesize
\setlength{\tabcolsep}{5pt}
\renewcommand{\arraystretch}{0.95}
\begin{tabular}{llrrrrrrrrrrrr}
\toprule
 &  & \multicolumn{4}{c}{Correct} & \multicolumn{4}{c}{PS Misspec.} & \multicolumn{4}{c}{OR Misspec.} \\
\cmidrule(lr){3-6} \cmidrule(lr){7-10} \cmidrule(lr){11-14}
ME configuration & $\eta$ & Naive & EP & Gold & CV & Naive & EP & Gold & CV & Naive & EP & Gold & CV \\
\midrule
ME in One Exposure & $0.2$ & 2.42 & 2.94 & 33.93 & 32.61 & 2.28 & 2.90 & 27.47 & 26.26 & 2.34 & 2.98 & 34.41 & 33.11 \\
 & $0.3$ & 2.38 & 3.53 & 18.12 & 17.37 & 2.30 & 3.53 & 14.55 & 13.99 & 2.30 & 3.41 & 17.87 & 17.38 \\
 & $0.4$ & 2.37 & 4.30 & 10.71 & 10.34 & 2.27 & 4.33 & 9.00 & 8.63 & 2.28 & 4.30 & 11.21 & 10.74 \\
 & $0.5$ & 2.31 & 5.71 & 7.69 & 7.48 & 2.32 & 5.76 & 6.16 & 5.92 & 2.29 & 5.71 & 7.34 & 7.10 \\
 & $0.6$ & 2.44 & 8.27 & 5.62 & 5.50 & 2.21 & 7.95 & 4.73 & 4.57 & 2.44 & 8.44 & 5.58 & 5.45 \\
 & $0.7$ & 2.50 & 12.78 & 4.33 & 4.20 & 2.34 & 13.03 & 3.63 & 3.54 & 2.28 & 12.78 & 4.12 & 4.02 \\
 & $0.8$ & 2.36 & 23.60 & 3.36 & 3.30 & 2.30 & 22.87 & 3.02 & 2.98 & 2.49 & 22.85 & 3.49 & 3.44 \\
\midrule
ME in Both Exposures & $0.2$ & 2.20 & 2.78 & 33.37 & 32.96 & 2.21 & 2.78 & 27.36 & 26.76 & 2.18 & 2.79 & 32.87 & 32.38 \\
 & $0.3$ & 2.13 & 3.25 & 18.27 & 17.97 & 2.15 & 3.37 & 14.60 & 14.39 & 2.17 & 3.22 & 18.55 & 18.20 \\
 & $0.4$ & 2.29 & 4.22 & 11.32 & 11.23 & 2.24 & 4.31 & 8.97 & 8.81 & 2.19 & 4.22 & 11.24 & 11.14 \\
 & $0.5$ & 2.32 & 5.62 & 7.55 & 7.55 & 2.35 & 5.86 & 6.56 & 6.44 & 2.28 & 5.70 & 7.74 & 7.60 \\
 & $0.6$ & 2.29 & 8.10 & 5.42 & 5.37 & 2.29 & 7.70 & 4.58 & 4.51 & 2.28 & 7.83 & 5.54 & 5.46 \\
 & $0.7$ & 2.45 & 12.96 & 4.32 & 4.26 & 2.21 & 12.83 & 3.60 & 3.56 & 2.35 & 12.55 & 4.26 & 4.23 \\
 & $0.8$ & 2.43 & 23.91 & 3.40 & 3.39 & 2.24 & 22.91 & 2.99 & 2.99 & 2.41 & 23.31 & 3.47 & 3.46 \\
\midrule
Low Specificity & $0.2$ & 2.68 & 3.43 & 34.88 & 33.28 & 2.45 & 3.19 & 27.15 & 26.04 & 2.60 & 3.28 & 34.45 & 32.97 \\
 & $0.3$ & 2.60 & 3.91 & 17.99 & 17.44 & 2.35 & 3.82 & 14.70 & 14.03 & 2.54 & 3.79 & 18.02 & 17.28 \\
 & $0.4$ & 2.55 & 4.67 & 10.93 & 10.59 & 2.43 & 4.88 & 9.20 & 8.87 & 2.52 & 4.75 & 11.10 & 10.60 \\
 & $0.5$ & 2.53 & 6.09 & 7.43 & 7.22 & 2.27 & 6.32 & 6.50 & 6.27 & 2.58 & 6.20 & 7.73 & 7.48 \\
 & $0.6$ & 2.59 & 8.76 & 5.45 & 5.28 & 2.29 & 9.06 & 4.74 & 4.63 & 2.56 & 8.93 & 5.23 & 5.12 \\
 & $0.7$ & 2.48 & 13.66 & 4.19 & 4.04 & 2.32 & 14.76 & 3.72 & 3.64 & 2.51 & 13.45 & 4.21 & 4.09 \\
 & $0.8$ & 2.51 & 23.38 & 3.40 & 3.32 & 2.41 & 27.29 & 3.19 & 3.10 & 2.46 & 23.57 & 3.42 & 3.34 \\
\midrule
Low Sensitivity & $0.2$ & 2.56 & 3.46 & 34.21 & 32.80 & 2.39 & 3.16 & 27.27 & 26.12 & 2.49 & 3.32 & 33.78 & 32.25 \\
 & $0.3$ & 2.45 & 4.07 & 18.24 & 17.55 & 2.39 & 3.78 & 14.33 & 13.86 & 2.43 & 3.98 & 17.54 & 17.02 \\
 & $0.4$ & 2.51 & 5.22 & 11.04 & 10.65 & 2.33 & 4.59 & 9.08 & 8.83 & 2.45 & 5.20 & 10.98 & 10.68 \\
 & $0.5$ & 2.54 & 7.44 & 7.57 & 7.34 & 2.34 & 6.73 & 6.04 & 5.86 & 2.45 & 7.22 & 7.62 & 7.41 \\
 & $0.6$ & 2.52 & 9.98 & 5.41 & 5.29 & 2.29 & 9.07 & 4.64 & 4.52 & 2.47 & 9.99 & 5.62 & 5.49 \\
 & $0.7$ & 2.45 & 15.70 & 4.19 & 4.09 & 2.35 & 14.19 & 3.77 & 3.67 & 2.42 & 16.09 & 4.27 & 4.17 \\
 & $0.8$ & 2.63 & 29.45 & 3.58 & 3.52 & 2.36 & 25.71 & 3.09 & 3.06 & 2.50 & 30.15 & 3.49 & 3.43 \\
\midrule
Mixed Sens/Spec & $0.2$ & 2.62 & 3.31 & 34.46 & 32.98 & 2.37 & 2.99 & 26.67 & 25.35 & 2.53 & 3.20 & 33.69 & 32.13 \\
 & $0.3$ & 2.53 & 3.79 & 17.44 & 16.67 & 2.32 & 3.57 & 14.87 & 14.28 & 2.51 & 3.75 & 18.39 & 17.82 \\
 & $0.4$ & 2.57 & 4.66 & 11.05 & 10.66 & 2.34 & 4.47 & 9.17 & 8.80 & 2.53 & 4.87 & 11.07 & 10.66 \\
 & $0.5$ & 2.52 & 5.93 & 7.63 & 7.43 & 2.39 & 5.93 & 6.33 & 6.11 & 2.48 & 6.27 & 7.43 & 7.23 \\
 & $0.6$ & 2.52 & 8.50 & 5.47 & 5.32 & 2.34 & 8.80 & 4.72 & 4.61 & 2.49 & 9.07 & 5.48 & 5.33 \\
 & $0.7$ & 2.46 & 13.40 & 4.21 & 4.11 & 2.36 & 13.53 & 3.71 & 3.63 & 2.53 & 13.65 & 4.18 & 4.12 \\
 & $0.8$ & 2.43 & 23.99 & 3.37 & 3.28 & 2.32 & 24.96 & 3.09 & 3.04 & 2.49 & 24.05 & 3.46 & 3.39 \\
\bottomrule
\end{tabular}
\caption{Empirical variance ($\times 100$) for $\tau^{(1,1)}$ across measurement-error configurations (rows) and gold-standard proportions $\eta$ (rows within each configuration), separately for each model-specification scenario (column groups: Correct, PS Misspec., OR Misspec.).  Estimators within each scenario are Naive, EP (error-prone), Gold, and CV (control variate); the Oracle benchmark is omitted (see Section~\ref{sec:supp-simresults}).}
\label{tab:sim-var-11}
\end{sidewaystable}

\begin{sidewaystable}[p]
\centering
\footnotesize
\setlength{\tabcolsep}{5pt}
\renewcommand{\arraystretch}{0.95}
\begin{tabular}{llrrrrrrrrrrrr}
\toprule
 &  & \multicolumn{4}{c}{Correct} & \multicolumn{4}{c}{PS Misspec.} & \multicolumn{4}{c}{OR Misspec.} \\
\cmidrule(lr){3-6} \cmidrule(lr){7-10} \cmidrule(lr){11-14}
ME configuration & $\eta$ & Naive & EP & Gold & CV & Naive & EP & Gold & CV & Naive & EP & Gold & CV \\
\midrule
ME in One Exposure & $0.2$ & 4.79 & 5.85 & 83.14 & 82.08 & 4.28 & 5.59 & 53.52 & 52.75 & 4.83 & 6.01 & 83.72 & 82.21 \\
 & $0.3$ & 4.97 & 6.99 & 44.79 & 44.21 & 4.39 & 6.58 & 28.33 & 28.07 & 4.99 & 7.24 & 45.52 & 45.36 \\
 & $0.4$ & 5.26 & 8.97 & 28.89 & 28.62 & 4.31 & 8.35 & 17.88 & 17.76 & 5.21 & 9.14 & 28.52 & 28.19 \\
 & $0.5$ & 5.34 & 11.90 & 19.48 & 19.32 & 4.34 & 11.46 & 12.32 & 12.23 & 5.21 & 12.11 & 20.65 & 20.42 \\
 & $0.6$ & 5.91 & 17.10 & 15.29 & 15.18 & 4.40 & 16.34 & 9.22 & 9.14 & 5.46 & 17.34 & 15.55 & 15.41 \\
 & $0.7$ & 5.91 & 27.02 & 11.39 & 11.36 & 4.30 & 24.98 & 7.09 & 7.03 & 5.59 & 25.54 & 11.68 & 11.60 \\
 & $0.8$ & 6.22 & 47.21 & 9.60 & 9.59 & 4.30 & 46.22 & 5.72 & 5.69 & 6.22 & 48.09 & 9.84 & 9.83 \\
\midrule
ME in Both Exposures & $0.2$ & 4.39 & 5.42 & 81.44 & 81.34 & 4.29 & 5.40 & 55.83 & 55.76 & 4.48 & 5.57 & 81.44 & 81.43 \\
 & $0.3$ & 4.78 & 6.90 & 44.39 & 44.30 & 4.45 & 6.83 & 28.63 & 28.55 & 4.65 & 6.80 & 45.98 & 46.03 \\
 & $0.4$ & 4.97 & 8.81 & 28.92 & 28.91 & 4.18 & 8.20 & 18.15 & 18.06 & 4.76 & 8.78 & 29.65 & 29.67 \\
 & $0.5$ & 5.14 & 11.43 & 20.03 & 19.94 & 4.16 & 11.34 & 12.48 & 12.45 & 4.91 & 11.57 & 19.39 & 19.41 \\
 & $0.6$ & 5.42 & 16.25 & 15.53 & 15.52 & 4.23 & 15.76 & 8.96 & 8.95 & 5.28 & 16.59 & 15.72 & 15.75 \\
 & $0.7$ & 5.70 & 26.30 & 11.29 & 11.24 & 4.28 & 24.73 & 7.13 & 7.11 & 5.51 & 25.99 & 11.71 & 11.71 \\
 & $0.8$ & 6.15 & 45.65 & 9.54 & 9.57 & 4.29 & 44.71 & 5.74 & 5.74 & 5.68 & 47.35 & 9.58 & 9.59 \\
\midrule
Low Specificity & $0.2$ & 5.31 & 6.37 & 83.09 & 81.80 & 4.51 & 6.06 & 52.83 & 51.79 & 5.10 & 6.31 & 83.84 & 81.96 \\
 & $0.3$ & 5.34 & 7.70 & 45.98 & 45.21 & 4.60 & 7.42 & 29.31 & 28.94 & 5.39 & 7.58 & 46.37 & 45.48 \\
 & $0.4$ & 5.45 & 9.53 & 28.71 & 28.36 & 4.34 & 8.78 & 18.53 & 18.15 & 5.42 & 9.45 & 28.40 & 28.22 \\
 & $0.5$ & 5.68 & 12.11 & 19.67 & 19.60 & 4.43 & 12.38 & 12.10 & 12.00 & 5.57 & 12.71 & 19.82 & 19.68 \\
 & $0.6$ & 5.78 & 18.27 & 14.89 & 14.74 & 4.45 & 18.13 & 9.00 & 8.89 & 5.75 & 17.62 & 15.10 & 15.02 \\
 & $0.7$ & 6.24 & 27.29 & 11.59 & 11.48 & 4.34 & 27.75 & 7.08 & 7.03 & 6.09 & 29.07 & 11.50 & 11.42 \\
 & $0.8$ & 6.40 & 50.57 & 9.37 & 9.35 & 4.16 & 51.70 & 5.59 & 5.55 & 6.47 & 50.10 & 9.68 & 9.67 \\
\midrule
Low Sensitivity & $0.2$ & 5.26 & 6.75 & 82.99 & 82.34 & 4.42 & 5.99 & 53.30 & 52.35 & 5.13 & 6.72 & 83.00 & 81.65 \\
 & $0.3$ & 5.40 & 8.20 & 44.22 & 43.77 & 4.54 & 7.23 & 29.11 & 28.73 & 5.23 & 7.90 & 44.82 & 44.31 \\
 & $0.4$ & 5.70 & 10.56 & 28.65 & 28.43 & 4.58 & 9.13 & 18.34 & 18.07 & 5.49 & 10.87 & 27.95 & 27.67 \\
 & $0.5$ & 5.76 & 14.29 & 19.87 & 19.62 & 4.42 & 12.35 & 12.53 & 12.34 & 5.45 & 14.07 & 20.56 & 20.35 \\
 & $0.6$ & 5.95 & 20.08 & 14.85 & 14.63 & 4.31 & 17.38 & 8.83 & 8.69 & 5.97 & 20.97 & 14.93 & 14.88 \\
 & $0.7$ & 6.19 & 32.08 & 11.51 & 11.43 & 4.20 & 27.84 & 6.85 & 6.75 & 6.15 & 32.86 & 11.61 & 11.51 \\
 & $0.8$ & 6.52 & 57.87 & 9.45 & 9.42 & 4.33 & 49.19 & 5.70 & 5.66 & 6.44 & 59.62 & 9.63 & 9.54 \\
\midrule
Mixed Sens/Spec & $0.2$ & 6.16 & 7.41 & 79.27 & 78.13 & 4.67 & 6.31 & 55.39 & 54.61 & 6.00 & 7.39 & 84.05 & 81.92 \\
 & $0.3$ & 6.49 & 9.26 & 46.68 & 46.32 & 4.51 & 7.32 & 30.33 & 29.88 & 6.12 & 9.12 & 46.07 & 45.07 \\
 & $0.4$ & 6.44 & 11.72 & 29.17 & 28.72 & 4.43 & 9.76 & 17.96 & 17.71 & 6.32 & 11.29 & 28.67 & 28.07 \\
 & $0.5$ & 6.58 & 15.17 & 20.47 & 20.14 & 4.26 & 12.92 & 12.53 & 12.23 & 6.19 & 14.62 & 20.33 & 20.19 \\
 & $0.6$ & 6.74 & 21.19 & 14.94 & 14.81 & 4.47 & 18.55 & 9.22 & 9.09 & 6.39 & 21.10 & 14.52 & 14.41 \\
 & $0.7$ & 7.03 & 33.71 & 11.57 & 11.50 & 4.22 & 28.94 & 7.07 & 7.00 & 6.87 & 34.71 & 11.96 & 11.86 \\
 & $0.8$ & 6.64 & 58.47 & 9.39 & 9.33 & 4.24 & 53.76 & 5.60 & 5.55 & 6.91 & 61.75 & 9.74 & 9.74 \\
\bottomrule
\end{tabular}
\caption{Empirical variance ($\times 100$) for $\Delta$ (interaction) across measurement-error configurations (rows) and gold-standard proportions $\eta$ (rows within each configuration), separately for each model-specification scenario (column groups: Correct, PS Misspec., OR Misspec.).  Estimators within each scenario are Naive, EP (error-prone), Gold, and CV (control variate); the Oracle benchmark is omitted (see Section~\ref{sec:supp-simresults}).}
\label{tab:sim-var-int}
\end{sidewaystable}

\begin{table}[H]
\centering
\small
\setlength{\tabcolsep}{6pt}
\begin{tabular}{llccccccc}
\toprule
Scenario & ME configuration & $\eta=0.2$ & $\eta=0.3$ & $\eta=0.4$ & $\eta=0.5$ & $\eta=0.6$ & $\eta=0.7$ & $\eta=0.8$ \\
\midrule
Correct & ME in One Exposure & 1.040 & 1.038 & 1.033 & 1.035 & 1.024 & 1.018 & 1.007 \\
 & ME in Both Exposures & 1.011 & 1.006 & 1.010 & 1.001 & 1.004 & 1.004 & 1.002 \\
 & Low Specificity & 1.029 & 1.025 & 1.028 & 1.020 & 1.019 & 1.028 & 1.017 \\
 & Low Sensitivity & 1.022 & 1.022 & 1.024 & 1.020 & 1.009 & 1.008 & 1.004 \\
 & Mixed Sens/Spec & 1.030 & 1.028 & 1.031 & 1.024 & 1.023 & 1.016 & 1.012 \\
\midrule
PS Misspec. & ME in One Exposure & 1.023 & 1.033 & 1.044 & 1.029 & 1.037 & 1.019 & 1.019 \\
 & ME in Both Exposures & 1.011 & 1.012 & 1.013 & 1.009 & 0.999 & 1.007 & 1.005 \\
 & Low Specificity & 1.027 & 1.015 & 1.022 & 1.020 & 1.020 & 1.013 & 1.014 \\
 & Low Sensitivity & 1.040 & 1.033 & 1.023 & 1.021 & 1.022 & 1.018 & 1.015 \\
 & Mixed Sens/Spec & 1.018 & 1.027 & 1.025 & 1.027 & 1.019 & 1.023 & 1.021 \\
\midrule
OR Misspec. & ME in One Exposure & 1.039 & 1.037 & 1.038 & 1.023 & 1.021 & 1.016 & 1.017 \\
 & ME in Both Exposures & 1.019 & 1.004 & 1.005 & 1.005 & 1.006 & 1.001 & 1.000 \\
 & Low Specificity & 1.035 & 1.023 & 1.016 & 1.030 & 1.021 & 1.019 & 1.007 \\
 & Low Sensitivity & 1.025 & 1.028 & 1.029 & 1.020 & 1.016 & 1.015 & 1.013 \\
 & Mixed Sens/Spec & 1.028 & 1.028 & 1.026 & 1.017 & 1.015 & 1.019 & 1.005 \\
\bottomrule
\end{tabular}
\caption{Relative efficiency of the control variate estimator versus the gold-standard estimator, $\widehat{\mathrm{Var}}\left(\hat{\tau}_{\gs}^{(0,1)}\right)/\widehat{\mathrm{Var}}\left(\hat{\xi}^{(0,1)}\right)$, for the parameter $\tau^{(0,1)}$.  Values exceeding $1$ indicate efficiency gains from the control variate adjustment.}
\label{tab:sim-eff-01}
\end{table}

\begin{table}[H]
\centering
\small
\setlength{\tabcolsep}{6pt}
\begin{tabular}{llccccccc}
\toprule
Scenario & ME configuration & $\eta=0.2$ & $\eta=0.3$ & $\eta=0.4$ & $\eta=0.5$ & $\eta=0.6$ & $\eta=0.7$ & $\eta=0.8$ \\
\midrule
Correct & ME in One Exposure & 1.029 & 1.013 & 1.011 & 1.010 & 1.010 & 1.007 & 1.000 \\
 & ME in Both Exposures & 1.008 & 1.007 & 1.006 & 1.004 & 1.004 & 1.006 & 1.001 \\
 & Low Specificity & 1.039 & 1.025 & 1.023 & 1.001 & 1.014 & 1.010 & 0.998 \\
 & Low Sensitivity & 1.030 & 1.021 & 1.018 & 1.018 & 1.020 & 1.005 & 1.004 \\
 & Mixed Sens/Spec & 1.035 & 1.017 & 1.020 & 1.022 & 1.013 & 1.014 & 1.008 \\
\midrule
PS Misspec. & ME in One Exposure & 1.016 & 1.016 & 1.003 & 1.012 & 1.011 & 1.007 & 1.007 \\
 & ME in Both Exposures & 1.009 & 1.011 & 1.009 & 1.010 & 1.002 & 1.002 & 1.012 \\
 & Low Specificity & 1.033 & 1.031 & 1.032 & 1.020 & 1.015 & 1.017 & 1.013 \\
 & Low Sensitivity & 1.020 & 1.030 & 1.029 & 1.030 & 1.031 & 1.029 & 1.017 \\
 & Mixed Sens/Spec & 1.025 & 1.033 & 1.020 & 1.027 & 1.028 & 1.019 & 1.004 \\
\midrule
OR Misspec. & ME in One Exposure & 1.023 & 1.008 & 1.021 & 1.012 & 1.006 & 1.007 & 0.998 \\
 & ME in Both Exposures & 1.003 & 1.004 & 1.003 & 0.999 & 1.006 & 0.998 & 0.999 \\
 & Low Specificity & 1.026 & 1.023 & 1.016 & 1.014 & 1.011 & 1.011 & 1.008 \\
 & Low Sensitivity & 1.031 & 1.022 & 1.016 & 1.023 & 1.010 & 1.011 & 1.008 \\
 & Mixed Sens/Spec & 1.041 & 1.032 & 1.024 & 1.013 & 1.024 & 1.007 & 1.005 \\
\bottomrule
\end{tabular}
\caption{Relative efficiency of the control variate estimator versus the gold-standard estimator, $\widehat{\mathrm{Var}}\left(\hat{\tau}_{\gs}^{(1,0)}\right)/\widehat{\mathrm{Var}}\left(\hat{\xi}^{(1,0)}\right)$, for the parameter $\tau^{(1,0)}$.  Values exceeding $1$ indicate efficiency gains from the control variate adjustment.}
\label{tab:sim-eff-10}
\end{table}

\begin{table}[H]
\centering
\small
\setlength{\tabcolsep}{6pt}
\begin{tabular}{llccccccc}
\toprule
Scenario & ME configuration & $\eta=0.2$ & $\eta=0.3$ & $\eta=0.4$ & $\eta=0.5$ & $\eta=0.6$ & $\eta=0.7$ & $\eta=0.8$ \\
\midrule
Correct & ME in One Exposure & 1.041 & 1.043 & 1.035 & 1.028 & 1.022 & 1.030 & 1.020 \\
 & ME in Both Exposures & 1.012 & 1.016 & 1.008 & 1.000 & 1.009 & 1.013 & 1.002 \\
 & Low Specificity & 1.048 & 1.032 & 1.032 & 1.029 & 1.032 & 1.035 & 1.025 \\
 & Low Sensitivity & 1.043 & 1.040 & 1.037 & 1.032 & 1.022 & 1.024 & 1.018 \\
 & Mixed Sens/Spec & 1.045 & 1.046 & 1.037 & 1.027 & 1.028 & 1.024 & 1.028 \\
\midrule
PS Misspec. & ME in One Exposure & 1.046 & 1.040 & 1.043 & 1.039 & 1.034 & 1.023 & 1.015 \\
 & ME in Both Exposures & 1.022 & 1.015 & 1.019 & 1.018 & 1.016 & 1.010 & 1.000 \\
 & Low Specificity & 1.043 & 1.048 & 1.037 & 1.038 & 1.024 & 1.023 & 1.029 \\
 & Low Sensitivity & 1.044 & 1.034 & 1.028 & 1.032 & 1.025 & 1.027 & 1.011 \\
 & Mixed Sens/Spec & 1.052 & 1.042 & 1.041 & 1.037 & 1.024 & 1.023 & 1.016 \\
\midrule
OR Misspec. & ME in One Exposure & 1.039 & 1.029 & 1.043 & 1.034 & 1.024 & 1.024 & 1.015 \\
 & ME in Both Exposures & 1.015 & 1.019 & 1.009 & 1.017 & 1.015 & 1.006 & 1.002 \\
 & Low Specificity & 1.045 & 1.043 & 1.047 & 1.034 & 1.021 & 1.030 & 1.024 \\
 & Low Sensitivity & 1.048 & 1.030 & 1.029 & 1.028 & 1.023 & 1.024 & 1.017 \\
 & Mixed Sens/Spec & 1.048 & 1.032 & 1.039 & 1.028 & 1.028 & 1.016 & 1.023 \\
\bottomrule
\end{tabular}
\caption{Relative efficiency of the control variate estimator versus the gold-standard estimator, $\widehat{\mathrm{Var}}\left(\hat{\tau}_{\gs}^{(1,1)}\right)/\widehat{\mathrm{Var}}\left(\hat{\xi}^{(1,1)}\right)$, for the parameter $\tau^{(1,1)}$.  Values exceeding $1$ indicate efficiency gains from the control variate adjustment.}
\label{tab:sim-eff-11}
\end{table}

\begin{table}[H]
\centering
\small
\setlength{\tabcolsep}{6pt}
\begin{tabular}{llccccccc}
\toprule
Scenario & ME configuration & $\eta=0.2$ & $\eta=0.3$ & $\eta=0.4$ & $\eta=0.5$ & $\eta=0.6$ & $\eta=0.7$ & $\eta=0.8$ \\
\midrule
Correct & ME in One Exposure & 1.013 & 1.013 & 1.010 & 1.008 & 1.007 & 1.003 & 1.001 \\
 & ME in Both Exposures & 1.001 & 1.002 & 1.000 & 1.005 & 1.001 & 1.005 & 0.997 \\
 & Low Specificity & 1.016 & 1.017 & 1.012 & 1.004 & 1.010 & 1.010 & 1.002 \\
 & Low Sensitivity & 1.008 & 1.010 & 1.007 & 1.013 & 1.015 & 1.007 & 1.004 \\
 & Mixed Sens/Spec & 1.015 & 1.008 & 1.016 & 1.016 & 1.009 & 1.007 & 1.006 \\
\midrule
PS Misspec. & ME in One Exposure & 1.015 & 1.009 & 1.007 & 1.007 & 1.010 & 1.008 & 1.007 \\
 & ME in Both Exposures & 1.001 & 1.003 & 1.005 & 1.002 & 1.002 & 1.003 & 1.001 \\
 & Low Specificity & 1.020 & 1.013 & 1.021 & 1.009 & 1.013 & 1.008 & 1.008 \\
 & Low Sensitivity & 1.018 & 1.013 & 1.015 & 1.015 & 1.017 & 1.014 & 1.008 \\
 & Mixed Sens/Spec & 1.014 & 1.015 & 1.014 & 1.024 & 1.014 & 1.010 & 1.011 \\
\midrule
OR Misspec. & ME in One Exposure & 1.018 & 1.004 & 1.012 & 1.011 & 1.009 & 1.007 & 1.001 \\
 & ME in Both Exposures & 1.000 & 0.999 & 0.999 & 0.999 & 0.998 & 1.000 & 0.998 \\
 & Low Specificity & 1.023 & 1.020 & 1.006 & 1.007 & 1.005 & 1.007 & 1.000 \\
 & Low Sensitivity & 1.017 & 1.012 & 1.010 & 1.011 & 1.003 & 1.009 & 1.009 \\
 & Mixed Sens/Spec & 1.026 & 1.022 & 1.021 & 1.007 & 1.008 & 1.008 & 0.999 \\
\bottomrule
\end{tabular}
\caption{Relative efficiency of the control variate estimator versus the gold-standard estimator, $\widehat{\mathrm{Var}}\left(\hat{\Delta}_{\gs}\right)/\widehat{\mathrm{Var}}\left(\hat{\xi}^{\Delta}\right)$, for the parameter $\Delta$.  Values exceeding $1$ indicate efficiency gains from the control variate adjustment.}
\label{tab:sim-eff-int}
\end{table}

\bibliographystyle{unsrtnat}
\bibliography{references}

\end{document}